# Steady vs. Dynamic Contributions of Different Doped Conducting Polymers in the Principal Components of an Electronic Nose's Response

Wiem Haj Ammar[a], Aicha Boujnah[a], Aimen Boubaker[a] Adel Kalboussi[a], Kamal Lmimouni[b] & Sébastien Pecqueur[b]*,

[a] Department of Physics, University of Monastir Tunisia.

[b] Univ. Lille, CNRS, Centrale Lille, Univ. Polytechnique Hauts-de-France, UMR 8520 - IEMN, F-59000 Lille, France.

Email: sebastien.pecqueur@*iemn*.fr

## Abstract

**Multivariate data analysis and machine-learning classification become popular tools to extract features without physical models for complex environments recognition. For electronic noses, time sampling over multiple sensors must be a fair compromise between a period sufficiently long to output a meaningful information pattern, and sufficiently short to minimize training time for practical applications. Particularly when reactivity's kinetics differ from thermodynamics' in sensitive materials, finding the best compromise to get the most from data is not obvious. Here, we investigate on the influence of data acquisition to improve or alter data clustering for molecular recognition on a conducting polymer electronic nose. We found out that waiting for the sensors to reach their steady state is not required for classification, and that reducing data acquisition down to the first dynamical information suffice to recognize molecular gases by principal component analysis with the same materials. Particularly for online inference, this study shows that a good sensing array is no array of good sensors, and that new figure-of-merits shall be defined for sensing hardware aiming machine-learning pattern-recognition rather than metrology.**

# Introduction

Current advances in artificial intelligence and internet-of-thing pushes for identifying new input layers as data generators to feed machine-learning (ML) for information classification. Before this, electronic noses (eNoses) have nourished the fantasy for more than 60 years of emulating a biological sense to perform classification tasks that sensors can barely make with molecular patterns.[**1-5**] It is therefore very appropriate to wonder whether conventional approaches to select sensing hardware are adapted to information processing requirements, with pattern recognition having more standards nowadays.

Chemistry is no trivial physics: it no 1D-continuum nor spectrum, but volatile molecules compose a vast group of hundred billions countable objects.[**6, 7**] Therefore, measuring all of them simultaneously with restricted and well-chosen set of ultimately selective sensors is unrealistic. If such complexity suits better ML-supported recognition than quantification using metrological sensors, proposing clear sensing figure-of-merits (FoM) that guide technologists to identify the right sets of materials is still needed in such context.[**8-11**] The main difficulty for identifying those lies in the fact that smart sensing arrays do not involve physical models at the opposite of metrological sensors: ML is purely mathematically-driven, which does not help for identifying device physical features to generate qualitative data. Each output signal conditions individual components of an input data, but the quality of the output data itself relates to features that affects the performances of the classifier as a group of device properties. The relationship between the data transformation as a result and the physical properties of a chosen sensing hardware is not straightforward and largely depends on the association of a software classifier with a sensing hardware. For an eNose, conductimetric sensing arrays are often used as a hardware,[**12-15**] in conjunction with principal component analysis (PCA)/k-mean clustering (k-means) as a classifier [**5, 16-18**]. In such systems, time-dependent current signals are inputted by environmentally sensitive materials in an array. Each current signal composes the feature of one single coordinate of a high-dimensional vector for classification. PCA is a linear classifier as it involves only one matrix transformation from a dataset of input vectors to a set of scores. The PCA projects data on a subspace defined by a basis of the covariance-matrix eigenvectors which have the highest eigenvalues. As a consequence, PCA aims at ignoring the contributions of “silent” sensing elements, or the ones which activities that are uncorrelated to the one another. Conventional sensors’ FoM do not consider information correlation between sensing elements, as it relates strictly at physical properties of individual elements to be standalone: for instance, selectivity, sensitivity or time response.[**19**] The latter is an important FoM that usually defines if a technology is suitable for online application: If information speed is essential for telecommunication, it is not necessarily the case of the velocity of its physical carriers of information themselves. The implementation of recurrent networks of sensing elements can illustrate such fact, where the group dynamics of different slow sensing elements enables recognizing frequency-modulated signals.[**20**] Homogenizing the physical properties of the population of sensing element can even yield to degrade the recognition performances of the classifier, so when the dynamics of sensing devices carries information, long-enough acquisition is crucial for classification.[**20**] However, time is also a resource to enrich a database to train a system: the slower the hardware the longer the training, which ultimately also conditions any classifier’s recognition performance. As a consequence, in case information lies in the dynamics of a hardware,[**21**] a fair compromise must be found for the time of data acquisition to optimize training with enough meaningful data but reasonable time for environmental pattern recognition.

In the framework of recognizing volatile solvents with doped conducting polymers, we propose in this study to quantify how relevant is the acquisition time as metrological FoM for effective molecular recognition with a conductimetric nose using PCA/k-means. By using both static and dynamic descriptors, we show that a sensor response does not necessarily require being steady for optimal classification, and that training time can greatly be shortened for practical inference.

# Experimental Section

**Device Fabrication:** The sensors fabrication has been reported elsewhere.[**22**] Concentric spiral-shaped Au microelectrodes (channel length L = 400 nm, spiral electrodes length W such as W/L > $10^3$ in a round cavity of 28 µm in diameter) were micro-fabricated by electron beam lithography in a cleanroom environment. Formulations of regio-regular P3HT and dopants solutions were subsequently deposited on top of the sensing arrays by drop casting.

**Electrical Characterization:** The current measurements were done with an Agilent 4155 parameter analyzer in an air blow, passing through a 5-mL vials that contains 1 mL of volatile solvents.[**22**] The system was calibrated to blow at a rate of 1 mL/s over the sensing array. The control of the vapor exposure was manually operated with pneumatic valves (asynchronous delays up to five seconds can be considered between the actual operation of the valves and the attributions of the labels on the current traces). Elementary sequences of gas exposures were set to three minutes to let the sensing elements reaching their steady-state. Each volatile solvent exposure is followed by a purging sequence, when the air flow passes through an empty vial, to let the response of the sensing arrays recover their response.

**Data Analysis:** Raw data currents were compiled and treated without filtering. Raw data curves have already been published,[**22**] and are available alongside on a public repository. Principal component analysis (PCA) was used as a linear classifier, by mean of the online open access tool Clustvis [**23**]. PCA data were scaled by unit variance and computed by singular value decomposition. All PCA data that are used in this experiment are available as supplementary information of this article. 95%-confidence ellipsoids were determined by k-means clustering with k=3, from random values for the initial centroids (in each case, k-means was run at least three times to ensure repeatability of the resulted clusters).

# Results

### Data Collection and Feature Extraction.

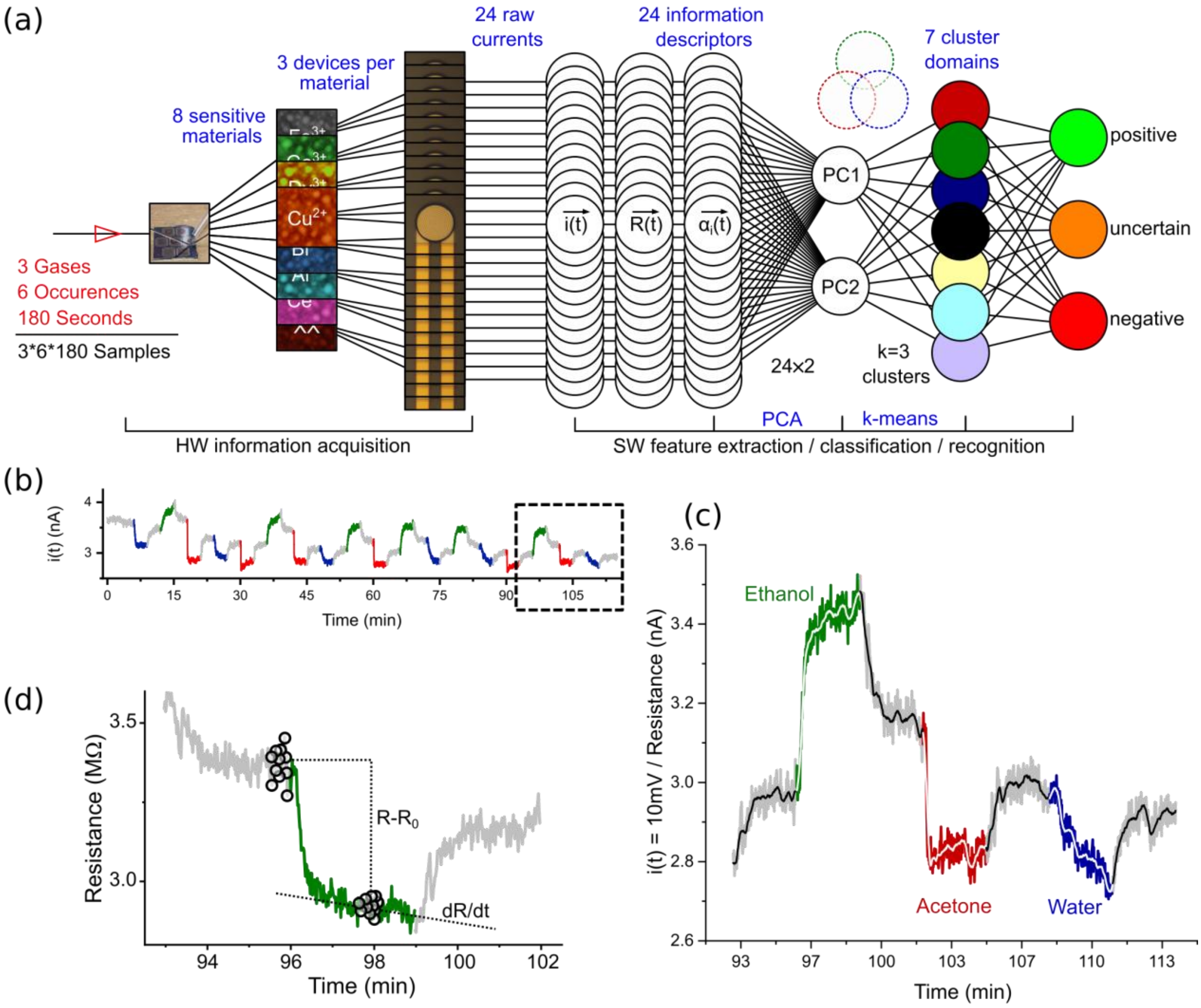

**Figure 1. Classifier Structure for Solvent Vapor Recognition | a,** Data classifier used in this study for unsupervised classification by using principal component analysis and k-means clustering on dynamical datasets generated by an array of sensors composed of different materials. **b,** Example of raw current signal output i(t) from one single sensor coated with one specific material, exposed to different sequences of solvent vapor exposures. **c,** Zoom-in on one time sequence when three different vapors affect the sensors conductance between purge sequences (zoom-in on fig.1b trace marked by a dashed square window). **b-c,** Blue, green and red sequences depict the sensor current output when water, acetone, ethanol vapors respectively are blown on the sensors with compressed air. Grey sequences depict the sensor current output when clean compressed air is blown on the sensors. **d,** Resistance value R(t) calculated from the raw current signal output i(t). Parameters $R\text{-}R_0$ and dR/dt for calculating the information features $\alpha_1(t)$ and $\alpha_2(t)$ are depicted as the absolute resistance modulation and the slope of the resistance variation overtime, extrapolated from the values of ten resistance data points in each case.

To interpret the physical significance of information features for the quality of a recognition, the machine-learning (ML) approach is preferred to the deep-learning (DL) one (see Fig.1 for the overall approach). DL algorithms differentiate from ML ones mostly by their complexity to allow the end-user not to have to identify information descriptors from raw data *a priori*. In our case, we chose specifically to compare two different features $\alpha_1$ and $\alpha_2$ that are extensively used in conductimetric eNoses as information descriptors: the relative resistance modulation $\alpha_1=R/R_0-1$ and the drift resistance $\alpha_2=\dot{R}/R$, with R the Ohmic resistance of a conductimetric sensing element measured at a given time during the exposure to an environment of a specific class, $\dot{R}$ its first-order derivative over time and $R_0$ its value right before it starts to be exposed to an environment of a specific class at $t_0$. As $\alpha_1$ compares two states before and after being exposed to a class, this information descriptor is qualified as a "thermodynamic feature" and relates to how much the signal's transport is affected by the presence of a given environment compared to a reference, through a given material. Instead, $\alpha_2$ uses a first-order derivative of time to quantify how fast or slow the signal decays or amplifies through a given material by the presence of a given environment: this information descriptor is qualified as a "kinetic feature". Each feature is calculated for different scenarios of crossed-coupled environments/devices parameters, with a population of 24 different sensitive devices (composed of eight populations of sensitive materials measured on three different devices), exposed at 18 occurrences to different environments (six alternate permutations of three different solvent vapor exposures). The experimental details are fully described in a first study,[**22**] where the PCA was used exclusively with the feature $\alpha_1$ calculated exclusively for the last ten seconds of each three-minute environment exposure. Although the steady-state response for these materials was relatively fast compared to other materials for comparable tests,[**24-27**] the dataset totalizes over 44 hours of time acquisition assisted by manual operations to expose the different materials on different devices with different environments repeatedly to train the system. Applying the same approach for exposing n! times the n different vapor classes for three minutes would yield to over a week, over a month and over seven months of total data acquisition, without interruption, to recognize respectively four, five and six different classes only. As reducing the number of materials, devices, or exposures would have consequences to degrade the recognition, it is straightforward to investigate on decreasing the elementary period for one single environment exposure to minimize the total acquisition duration. To assess on the classifier performances for a reduced acquisition time, a same dataset is considered, where data are pruned to ten seconds per acquisition sequences, after the exposure starts in order to make sure that the recognition performances do not depend on the acquisition itself, keeping the same number of datapoint to analyze in each case. Also, PCA/k-means was systematically used on all the different case scenario with the same normalization and initialization settings to not bias the clustering by user-dependent factors. The recognition after clustering uses the same rules and thresholds in each case, to assess a data sample as successfully, uncertainly or unsuccessfully recognized. All the individual PCA analysis are provided as supplementary information.

**Steady-Resistance Modulation as a Thermodynamic Information Feature.**

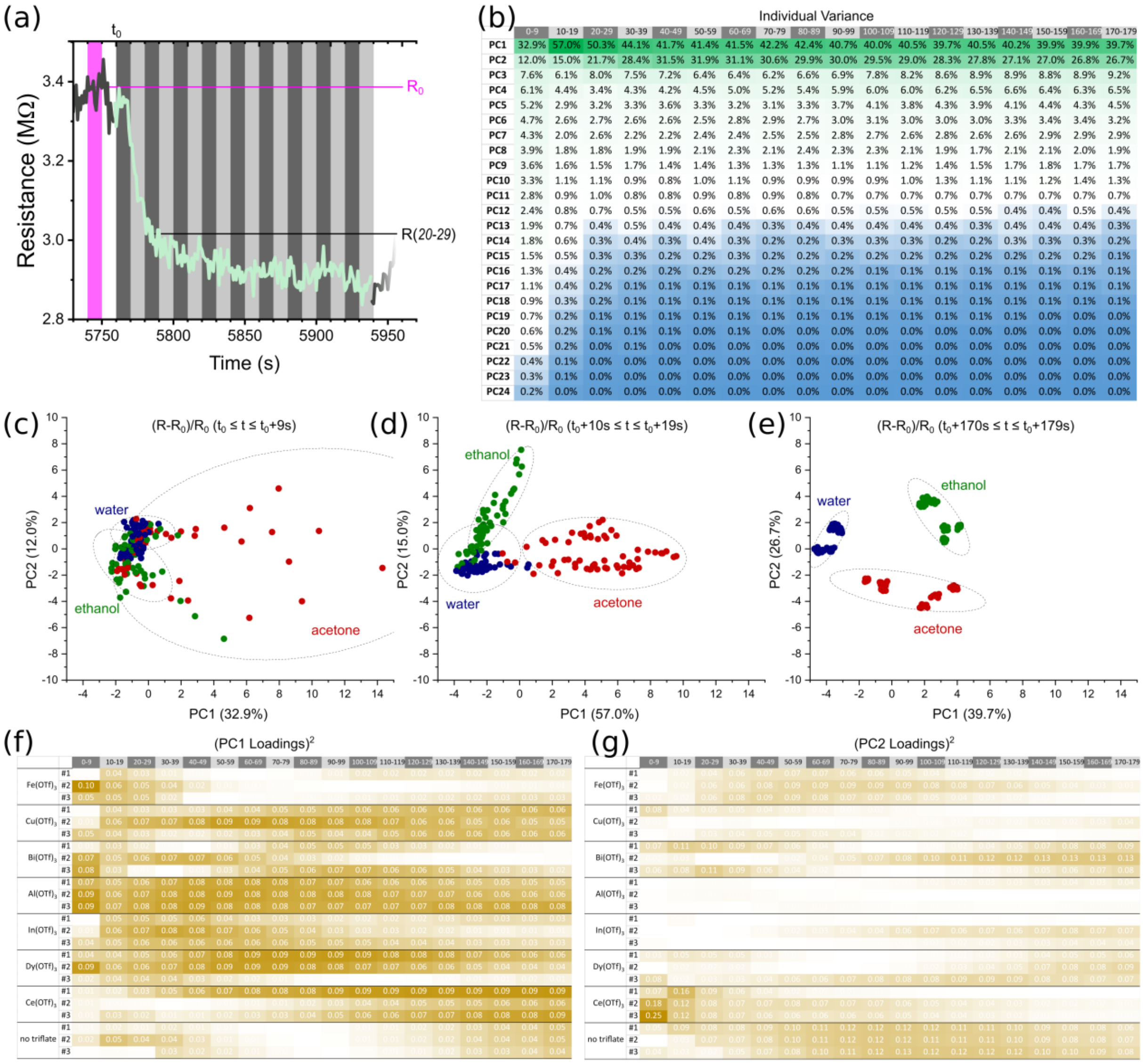

**Figure 2. Analysis of the Relative Resistance Modulation as Information Descriptor | a,** Extraction of the $R_0$ (as average for ten resistance values recorded between 11 and 20 seconds before vapor exposure, starting at $t_0$) and R (as average for ten resistance values recorded between t and t + 10 seconds every 10 seconds after $t_0$), to evaluate the feature $\alpha_1=(R-R_0)/R_0$. **b,** PC variance for different PCA on $\alpha_1$, where R is measured at different time interval [t;t+10s] after $t_0$ (the tricolor gradient shades linearly between the minimum and maximum value of the whole table). **c-e,** PCA scores on $\alpha_1$, where R is measured at different time interval [0s;9s] **(c)**, [10s;19s] **(d)** and [170s;179s] **(e)** after $t_0$ (confidence ellipsoids are set to 95% probability). **f-g,** Squared loadings of PC1 **(f)** and PC2 **(g)** for different PCA on $\alpha_1$, where R is measured at different time interval [t;t+10s] after $t_0$ (the bicolor gradient shades linearly between the minimum and maximum value of the whole tables).

The first analysis was performed using $\alpha_1$ as information descriptor (see Fig.2), as in the former studies using the same dataset. [**22, 28**] The choice for this specific feature was oriented by the hypothesis that the presence of volatile molecules modulates the polymer electrical conductance by interacting with its dopant shift its thermodynamic charge-transfer equilibrium. As a thermodynamic property, the acquisition was set to three minutes per exposure, to let enough time for all devices to reach a quasi-steady state of conductance (steady output currents at a constant applied voltage). Here from the whole dataset, 18 different sub-datasets were generated by segmenting the data to concatenation of sequences of ten seconds after each exposure starts at $t_0$ (ten datapoints per exposures, 180 datapoints per sub-datasets – see Fig.2a). For each exposure, the feature $\alpha_1$ takes into account two values: the resistance measured between the considered time interval and a reference resistance value $R_0$ measured while unexposed just before gas exposure. In each case, PCA was performed for the different sub-datasets, with the individual variance for each PC of each PCA for the different acquisition time compiled in Fig.2b. From these values, one can observe that most of the variance (>66%) is explained within the first two PC, except for data generated during the first ten seconds after $t_0$. It is also noticed that the variance for all 24 PC does not change significantly for acquisitions occurring 30 seconds after exposure (Fig.2b, the color shade encodes the variance values in linear scale), suggesting that the feature $\alpha_1$ is not enriched for the data separation in (PC1;PC2) if exposures last longer than 30 seconds. As shows their scores in the (PC1;PC2) projection (Fig.2c), the PCA on $\alpha_1$ during the first ten seconds acquisition, noted $PCA^{\alpha1}_{0\text{-}9}$, embeds most variability from acetone exposures on PC1 compared to two other clusters of data for water and ethanol, which are almost entirely contained in the acetone confidence ellipsoid. Already for $PCA^{\alpha1}_{10\text{-}19}$ (Fig.2d), acetone is completely separable from the other two gases, and some minor overlap remains between water and ethanol confidence ellipsoids. The final sequence $PCA^{\alpha1}_{170\text{-}179}$ in Fig.2e shows an ideal clustering in the (PC1;PC2) projection, that was used in the former study for gas recognition.[**22**] The loadings of PC1 and PC2 do not significantly depend on the acquisition sequences, except for the first ten seconds (Fig.2f and 2g): Excluding the first ten seconds, deviations overtime are however noticed: for instance the contribution of $Ce(OTf)_3$-doped devices increase monotonically in the $PC1^{\alpha1}$, such as the contribution of pristine P3HT in the $PC2^{\alpha1}$. This suggests that materials that help better for the (PC1;PC2) data separation but expresses slower than others do not necessarily enhances the clustering. The former study concluded on the exclusive basis of the $PCA^{\alpha1}_{170\text{-}179}$ analysis, two and three materials were preferred for recognizing ethanol from acetone from water using $\alpha_1$ as information descriptor.[**22**] Fig.2f and 2g show that such conclusion is highly acquisition-dependent (even for the same dataset considered), and that experiments performed with shorter time exposures might have concluded otherwise for the material selection. As such dependency on the sample acquisition time greatly conditions the preferred material choice for classification, it appears legitimate to investigate on a dynamical feature such as $\alpha_2$ as better information descriptor for gas recognition with the same raw current data.

**Resistance Drift as Kinetic Information Feature.**

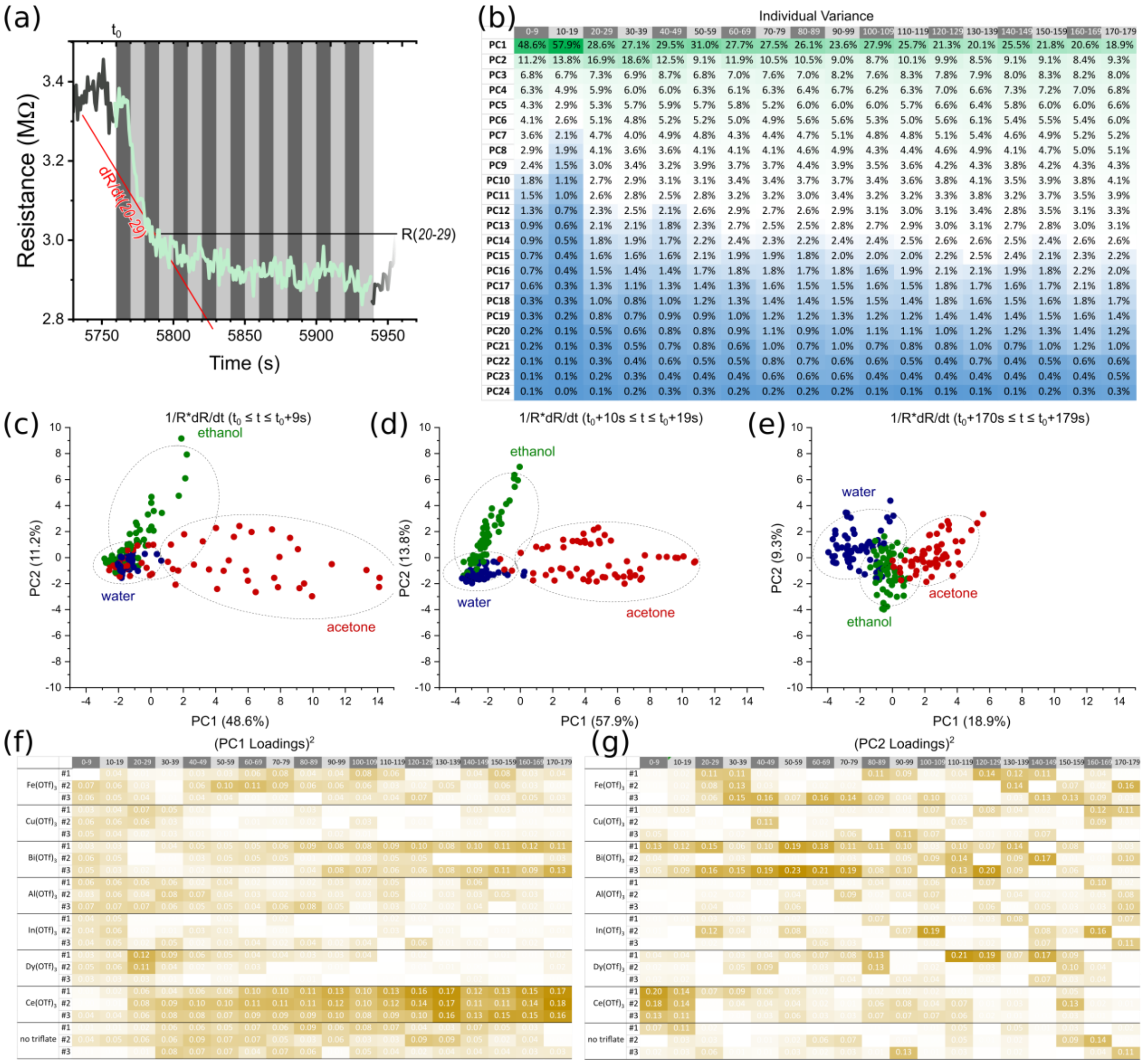

**Figure 3. Analysis of the Resistance Drift as Information Descriptor | a,** Extraction of dR/dt (linear regression over nine resistance values centered on R) and R, to evaluate the feature $\alpha_2$=1/R*dR/dt. **b,** PC variance for different PCA on $\alpha_2$, where R is measured at different time interval [t;t+10s] after $t_0$ (the tricolor gradient shades linearly between the minimum and maximum value of the whole table). **c-e,** PCA scores on $\alpha_2$, where R is measured at different time interval [0s;9s] **(c)**, [10s;19s] **(d)** and [170s;179s] **(e)** after $t_0$ (confidence ellipsoids are set to 95% probability). **f-g,** Squared loadings of PC1 **(f)** and PC2 **(g)** for different PCA on $\alpha_2$, where R is measured at different time interval [t;t+10s] after $t_0$ (the bicolor gradient shades linearly between the minimum and maximum value of the whole tables).

Drift resistance $\alpha_2$ in conductimetric eNoses is less often used as information descriptor. In a material, it can be associated to a memory effect that alter reversibly or irreversibly a structural property,[**29-33**] rather than the transduction of an environmental information in a device. Nevertheless, it is worth noticing that conventional techniques in analytical chemistry uses dynamical information features as information descriptors, such as chromatography (retention time of molecules on a substrate) or surface-plasmon resonance (absorption/desorption kinetic constants of molecules on a substrate). Supposed that part of the useful information to recognize gases lies in the diffusion kinetics of the volatile molecules through the doped polymers, or the adsorption/desorption kinetic constants of the molecular gases on the doped materials under flow, the drift-resistance may be a relevant information descriptor as it takes into account the first-order derivative of time (Fig.3a). Replicating the same methodology as previously for relative resistance modulation, PCA on $\alpha_2$ as information descriptor has been performed for the 18 different sub-datasets. As for $\alpha_1$, the evolution of the variance for the individual PC depends on the acquisition time (Fig.3b), mainly at the beginning of the transient response (typically the first 20 seconds). A singular difference with $\alpha_1$ is the impoverishing in data variance on the first two PC with the increase of the acquisition time (from >66% for $PCA^{\alpha1}_{170\text{-}179}$ down to <28% for $PCA^{\alpha2}_{170\text{-}179}$). This was expected as the information becomes more sensitive to the signal's noise at the reach of the steady-state, and therefore, with the increase of the acquisition time decreases the contribution of the molecularly-specific response compared in the overall information. We noticed also in Fig.3b that significantly less variance is contained in $PC2^{\alpha2}$ compared to $PC2^{\alpha1}$, and more importantly that variance for $PC2^{\alpha2}$ is far less dependent on the acquisition time than $PC2^{\alpha1}$. This suggests that despite any quality for the recognition, $\alpha_2$ seems to separate data mostly because of one single parameter (if a linear model is considered), while $\alpha_1$ seems to depend from mostly two parameters of a linear model. On the quality of the data separation (Fig.3c-e), similar trends for the (PC1;PC2) diagrams are observed depending on the acquisition time: At the first ten seconds of time acquisition after $t_0$, acetone data are better separated from the rest, followed by the ethanol data. Confidence ellipsoids seem to be always overlapping the one another, however, we observed that even until three minutes after time $t_0$ that the three volatile compounds are still recognizable, despite the residual dynamics of the sensors. It is also noticed that $PCA^{\alpha 2}_{170\text{-}179}$ (Fig.3e) clusters are more homogeneous than the ones of $PCA^{\alpha 1}_{170\text{-}179}$ (Fig.2e): smaller clusters within each ellipsoids can be distinguished in $PCA^{\alpha 1}_{170\text{-}179}$, which are attributed to experimental biases of the six repeated exposures for each classes. The dependency of $PCA^{\alpha 2}$ loadings with the acquisition time displays different trends than the one for $PCA^{\alpha 1}$ loadings (Fig.3g-h). For the first 20 s, all sensing elements except the ones coated with pristine P3HT and $Ce(OTf)_3$-doped P3HT seem to have a comparable contribution in the $PC1^{\alpha 2}$ and the $Ce(OTf)_3$-doped P3HT coated devices mostly constitute $PC2^{\alpha 2}$ if $t\text{-}t_0 < 20$ s. From 20 s after $t_0$, $Ce(OTf)_3$-doped P3HT coated devices become the most significant in the $PC1^{\alpha 2}$, while $PC2^{\alpha 2}$ loadings seem highly random if $t\text{-}t_0 < 20$ s. This closer look on the $PC1^{\alpha 2}$ and $PC2^{\alpha 2}$ loading highlights two different dynamics that successfully help for identifying the volatile compounds: before 20 seconds, a “fast” and a “slow” dynamics yields the array response to recognize all three classes from $PC1^{\alpha 2}$ and $PC2^{\alpha 2}$. After 20 seconds, the fast dynamics of all sensing elements has faded away, to leave a minor contribution on the $Ce(OTf)_3$-doped P3HT devices exclusively.

**Steady vs Dynamical Resistance for Gas Recognition.**

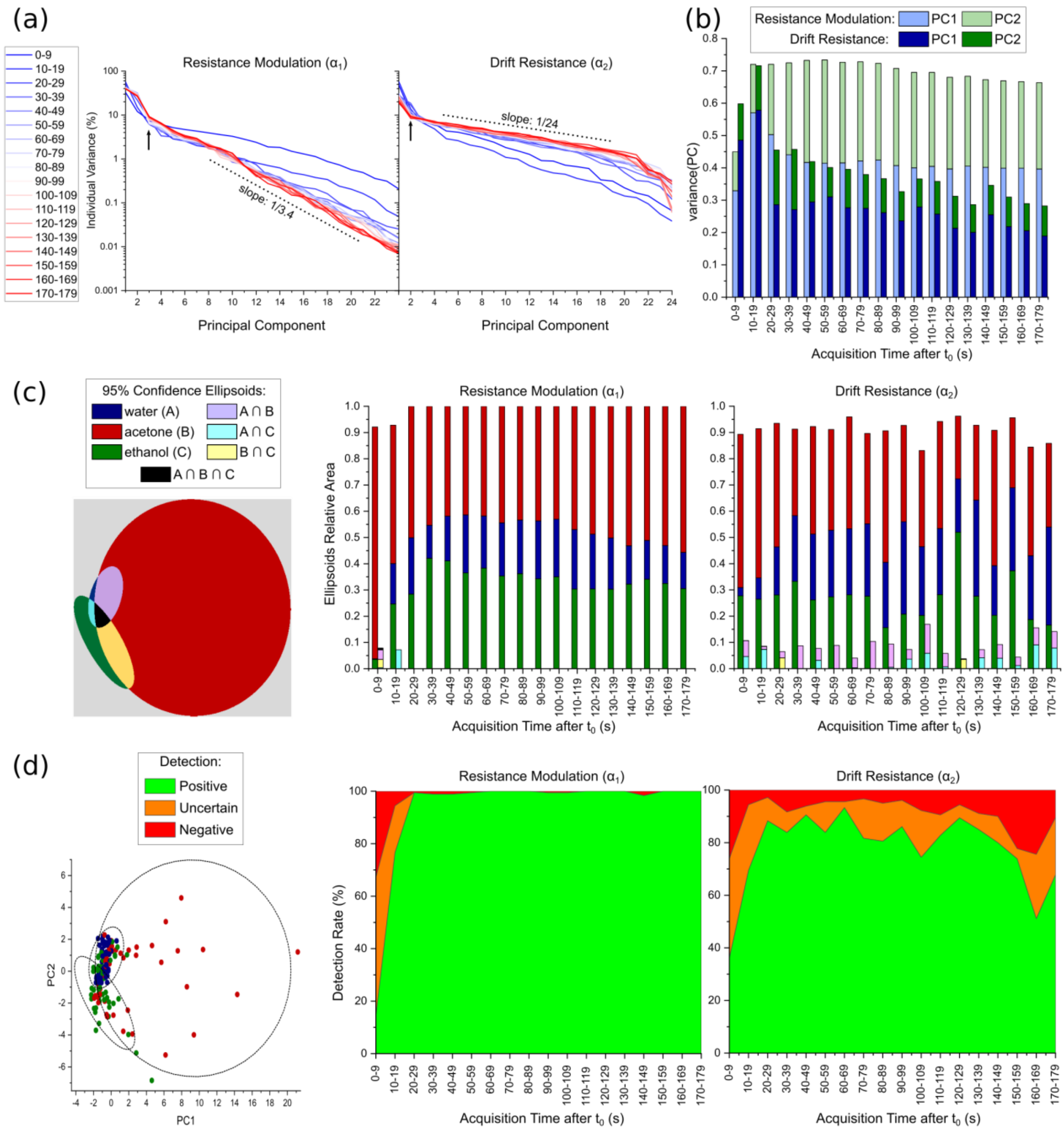

**Figure 4. Comparison of Both Descriptors | a,** Scree plots for the PCA performed with different acquisition times, from the first ten seconds after $t_0$ ("0-9" in blue) to the last ten seconds after $t_0$ ("170-179" in red), performed with either the relative resistance modulation ($\alpha_1$) or the drift resistance ($\alpha_2$) as information descriptor. **b,** Explained variance for PC1 and PC2 with PCA performed for different acquisition time with either $\alpha_1$ or $\alpha_2$. **c,** Overlap area of the three 95%-confidence ellipsoids in (PC1,PC2), for PCA performed for different acquisition time with either $\alpha_1$ or $\alpha_2$. The area were estimated by discretizing the different regions where the three ellipsoids are segmented in the smallest square of 500x500 pixels (on the left is displayed the example for the PCA with $\alpha_2$ for the "0-9" time acquisition). **d,** Detection tests of the PCA scores, for PCA performed for different acquisition time with either $\alpha_1$ or $\alpha_2$. The test is considered positive, uncertain and negative, respectively when scores are exclusively inside, not-exclusively inside and outside the corresponding 95%-confidence ellipsoid (on the left is displayed the example for the PCA with $\alpha_2$ for the "0-9" time acquisition).

To qualitatively compare the total variance distributions over the different PC with either $\alpha_1$ or $\alpha_2$ as information descriptor, the Scree plots are displayed on two different graphs and shows the effect of the acquisition time after $t_0$ for both descriptors (Fig.4a). Both of them confirms the trend of the variance with the acquisition time, for all principal components (not PC1 and PC2 exclusively). As Fig.4a shows, the increase of the acquisition time after exposure at $t_0$ tends to diminish the variance contained at the lowest PC in case of considering $\alpha_1$ as information descriptor, while it tends to increase the variance of the lower PC in case of considering $\alpha_2$ as information descriptor. We observed that this diminishing rate is different for both descriptors as displayed in Fig.4a, which might come from bias between the linear *ad hoc* model computed by the PCA and the actual physical model that governs the dependency of the information descriptors on the different sensing elements. The position of the elbows on both plots (indicated by arrows on the Scree plots in Fig.4a) confirms that most of the useful variance for $PCA^{\alpha1}$ is contained within the first three PC, while only the first two for $PCA^{\alpha2}$. Particularly for the drift resistance $\alpha_2$, this elbow is highly acquisition dependent, and one may appreciate a richer model with more PC to account for shorter acquisition time. The variance analysis on PC1 and PC2 for both $PCA^{\alpha1}$ and $PCA^{\alpha2}$ shows an acquisition time optimal to contain most of the data variance in both PC1 and PC2 independently from the chosen information descriptor (Fig.4b). The histogram displayed in Fig.4b shows that if the acquisition time is set at the transient between 10 and 19 s after $t_0$, both information features can gather at least 70% of the total variance in two different biparametric linear models. As both information features show opposed trends in the PC1+PC2 variance evolution with the acquisition time, it was expected to observe an acquisition-time optimum for both PCA, independently from the quality of the data clustering. To quantify the recognition performances of the sensing array with $PCA^{\alpha1}$ and $PCA^{\alpha2}$, the confidence ellipsoids determined by k-means clustering are used as a threshold to declare whether the given data belong to a given class of not, to define seven classes (as displayed in Fig.4c-left). The relative area of the overlap for the confidence ellipsoids has been evaluated for the different acquisition time and displayed in two different histograms for each descriptor in Fig.4c. For $\alpha_1$ (Fig.4c-histogramm in the middle), one can observe that ellipsoids do not overlap from $t-t_0 > 20s$, while in the case of $\alpha_2$ (Fig.4c-histogramm on the right), confidence ellipsoids always overlap the one another at any time for the acquisition. However, it should be noticed that the overlap in each case is only moderate: with less than 20% of the total area of the ellipsoids. This suggests that recognition tests may be higher in case $\alpha_1$ is used as information descriptor, than for $\alpha_2$. To verify this, recognition tests has been performed on the same sub-datasets, using the 95%-confidence ellipsoids as threshold for classification (Fig.4d). Detection is considered as “positive” if the actual class of the PCA score is exclusively contained within the corresponding confidence ellipsoid, “uncertain” if the actual class of the PCA score is not exclusively contained within the corresponding confidence ellipsoid, and “negative” if the actual class of the PCA score is excluded from the corresponding confidence ellipsoid. On the one hand for $\alpha_1$ (Fig.4d-graph in the middle), one can observe that the positive detection rate 99.4±0.6% for acquisition time higher than 20 s after $t_0$. On the one hand for $\alpha_2$ (Fig.4d-graph on the right), one can observe that the positive detection rate >80% for acquisition times higher between 20 s and 149 s after $t_0$. However, we noticed that using $\alpha_2$ as information descriptor enables substantially better recognition than $\alpha_2$ for short acquisition time below 10 s after $t_0$: typically, 36% positive and 38% uncertain using $\alpha_2$, while 14% positive and 53% uncertain using $\alpha_1$. Overall, this statistical study support that the different materials used in the array allow recognizing volatile organic compounds because of the relative resistance modulation ($\alpha_1$), rather than the resistance drift ($\alpha_2$), and points out the thermodynamics of the current modulations rather than the kinetics involved in the process.

# Discussion

Although this study focuses specifically on doped conducting polymers, the approach is generic to any conductimetric eNose and could have been address to different class of conducting materials used as transducers in an array. Despite the fact that $\alpha_1$ and $\alpha_2$ are frequently used as information descriptors for such application, our study proposes to investigate on which feature gather most of the useful information for an eNose. However, we want to highlight that for practical applications, such a choice may be conditioned rather by the application than the quality of the data *per se*. Specifically for the resistance modulation $\alpha_1$, the feature needs to routinely set a reference resistance $R_0$, when the environment is considered “neutral” between each data sampling. This requires therefore to have a hand on the sample exposure to periodically regenerate the device from a stimuli: it is therefore practically more convenient to use $\alpha_1$ as information descriptor in cases where analytes can be exposed to a system in a supervised manner (to evaluate fragrance batch quality in an analytical laboratory for instances). Such thermodynamic feature seems practically inadequate to extract for environment recognition in case the system is permanently exposed without periodic regeneration. For the drift resistance $\alpha_2$, the feature does not depend on a buffered reference. However, a steady value of the sampling period must be set to measure the dynamics: it is therefore a more adequate feature for online inference for environment recognition, at the condition that the sampling period is set *a posteriori* given the dynamic of the pattern to recognize. It is therefore a more adequate feature to recognize stimuli that are governed by a rhythm (for instance, pollution cycles in urban environments for instance). Here in this study, the selection of these features $\alpha_1$ and $\alpha_2$ is not based on the practical convenience to compute them in a conductimetric eNose application, but on the quality of the data they output for conducting polymers used in a recognition tests that conveniently enables computing both of them at will.

In such framework, PCA confirmed that the useful information in the current response of conducting polymer materials, doped with different metal triflates and exposed to different volatile organic compounds, lies specifically in the thermodynamics of the device resistance modulation and not its kinetics. The analysis of the relative resistance modulation $\alpha_1$ is more robust overtime when analyzed 20 seconds after each exposures at $t_0$: the first two PC explains at least two third of the data variance and allows an optimal clustering without overlap and enables 99.4±0.6% recognition. Considering that all devices have the same geometry, mechanisms responsible for the modulation of the electrical resistivity are pointed out. As different doping effects were assessed by varying the nature of the dopant,[**22**] modulation of the doping yield, by the interaction of different volatile molecules as electron-donating ligands on an electrophilic Lewis acid,[**34-36**] is not invalidated by this analysis. The statistical approach does not allow discriminating between the effect on the charge carrier density or the charge carrier mobility for the different materials exposed to different gases. Despite the fact that materials' conductivity is function of two independent physical properties (the charge carrier density and the charge carrier mobility), one cannot conclude that ant of these properties are purely marking $PC1^{\alpha1}$ and $PC2^{\alpha1}$, since conductivity is the product of both and PCA scores are linear combinations of PC. Such investigation could be carried out using the logarithmic value of the resistance modulation $\alpha_3=\log(R/R_0)$ as information descriptor for the PCA, in order to assess the individual contributions of the carrier mobility "μ" and carrier density "n" as principal components, since $\alpha_3 = \log(n_0/n) + \log(\mu_0/\mu)$. Such validation would have to be confirmed by correlating the $PC^{\alpha3}$ loadings with systematic study of carrier density and mobility assessment, in another study that couples this approach with further material characterizations.

# Conclusion

By statistical analysis, we evidenced that the relative resistance modulation of a doped conducting polymer array is a better information descriptor than the drift resistance, to recognize volatile organic compounds with different metal-triflate doped polythiophene sensing elements. This shows that the information carrier for the molecular recognition using these materials in an electronic nose is rather linked to the thermodynamic equilibrium that yields the doped conducting polymers' conductivity with its environment, rather than kinetic limitations which would select molecular targets' based on their drift diffusion through different doped polymers. Despite this result suggesting to wait for a complete steady state for each sensing element response, we evidenced that the thermodynamic properties to recognize molecular gases lies already in the transience of the material response. Despite materials reach a fast quasi-steady response within three minutes of environment exposures, principal component analysis and k-mean clustering show that the recognition is already higher than 99% after only 20 seconds of gas exposures. This suggests that training time to recognize molecular environment can greatly be optimized for practical applications, even in the case conducting materials in an electronics nose show slow response. Applied for practical recognition with an eNose, this approach to find a preliminary optimum for data acquisition before test shall greatly increase information throughputs for using eNose for chemical quality assessments by reducing the training duration while preserving most of the data quality to train an eNose to recognize various volatile molecular patterns.

# Supporting Information

Supplementary information is available in the online version of the paper.

# Acknowledgements

SP thanks the French Research Agency (ANR) for funding the “Sensation” project (Grant # ANR-22-CE24-0001-01). The authors thank the French Nanofabrication Network RENATECH for financial support of the IEMN cleanroom.

# Competing Interests

The authors declare no competing interests.

# Supporting Information for:

# Steady vs. Dynamic Contributions of Different Doped Conducting Polymers in the Principal Components of an Electronic Nose's Response

Wiem Haj Ammar[a], Aicha Boujnah[a], Aimen Boubaker[a] Adel Kalboussi[a], Kamal Lmimouni[b] & Sébastien Pecqueur[b]*,

[a] Department of Physics, University of Monastir Tunisia.

[b] Univ. Lille, CNRS, Centrale Lille, Univ. Polytechnique Hauts-de-France, UMR 8520 - IEMN, F-59000 Lille, France.

Email: sebastien.pecqueur@*iemn*.fr

**PCA scores (a), variance for the different PC (b) and their loadings (c) are organized in the different supporting figures by acquisition time for both information descriptors $\alpha_1$ and $\alpha_2$ such as:**

| | | Descriptors | |
|---|---|---|---|
| | | $\alpha_1$ | $\alpha_2$ |
| Acquisition Time Interval (in seconds) | **[0;9]** | Figure S1 | Figure S19 |
| | **[10;19]** | Figure S2 | Figure S20 |
| | **[20;29]** | Figure S3 | Figure S21 |
| | **[30;39]** | Figure S4 | Figure S22 |
| | **[40;49]** | Figure S5 | Figure S23 |
| | **[50;59]** | Figure S6 | Figure S24 |
| | **[60;69]** | Figure S7 | Figure S25 |
| | **[70;79]** | Figure S8 | Figure S26 |
| | **[80;89]** | Figure S9 | Figure S27 |
| | **[90;99]** | Figure S10 | Figure S28 |
| | **[100;109]** | Figure S11 | Figure S29 |
| | **[110;119]** | Figure S12 | Figure S30 |
| | **[120;129]** | Figure S13 | Figure S31 |
| | **[130;139]** | Figure S14 | Figure S32 |
| | **[140;149]** | Figure S15 | Figure S33 |
| | **[150;159]** | Figure S16 | Figure S34 |
| | **[160;169]** | Figure S17 | Figure S35 |
| | **[170;179]** | Figure S18 | Figure S36 |

a)

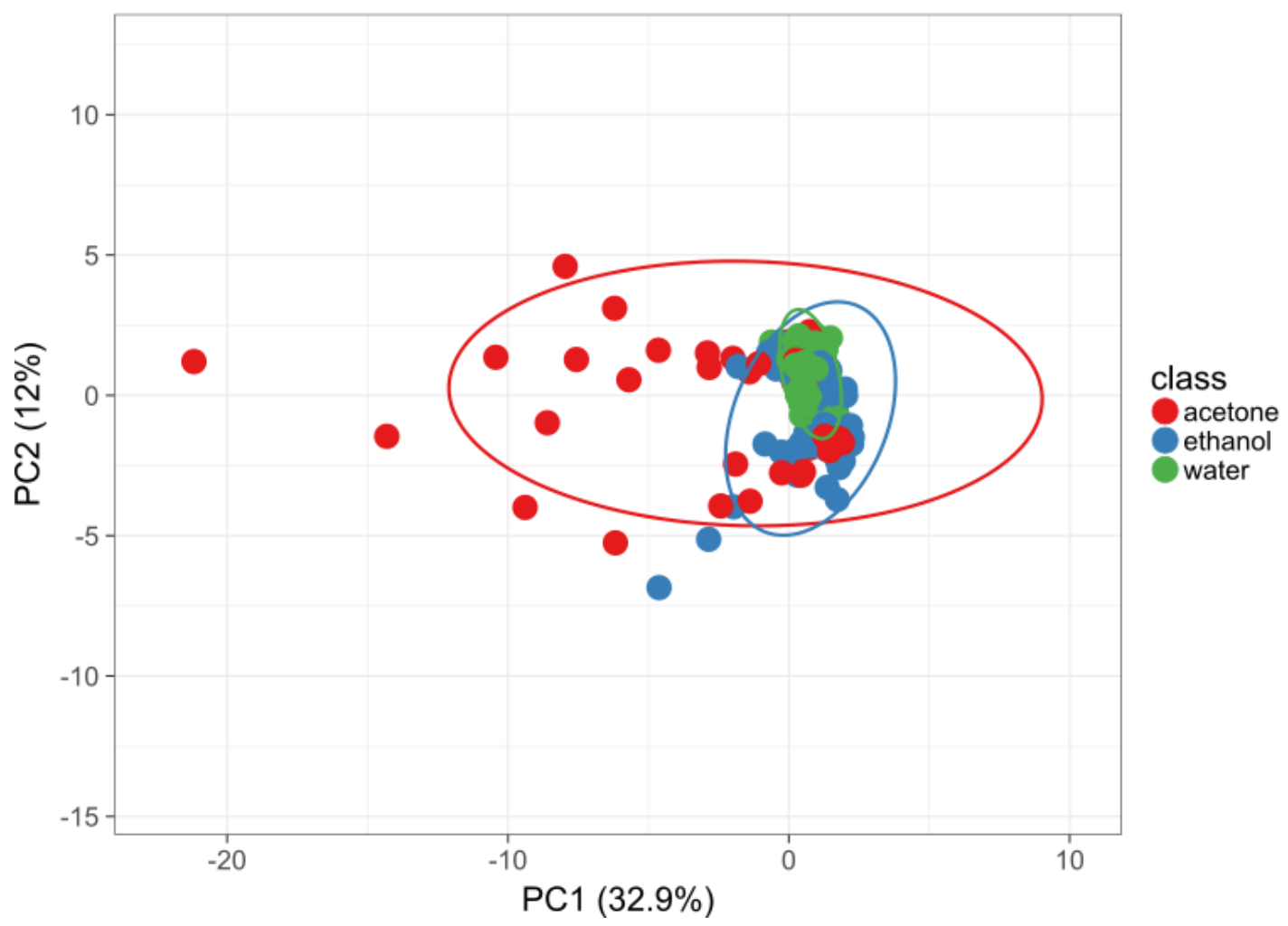

b)

| | Variance |
|---|---|
| **PC1** | 32.9% |
| **PC2** | 12.0% |
| **PC3** | 7.6% |
| **PC4** | 6.1% |
| **PC5** | 5.2% |
| **PC6** | 4.7% |
| **PC7** | 4.3% |
| **PC8** | 3.9% |
| **PC9** | 3.6% |
| **PC10** | 3.3% |
| **PC11** | 2.8% |
| **PC12** | 2.4% |
| **PC13** | 1.9% |
| **PC14** | 1.8% |
| **PC15** | 1.5% |
| **PC16** | 1.3% |
| **PC17** | 1.1% |
| **PC18** | 0.9% |
| **PC19** | 0.7% |
| **PC20** | 0.6% |
| **PC21** | 0.5% |
| **PC22** | 0.4% |
| **PC23** | 0.3% |
| **PC24** | 0.2% |

c)

| | | Loadings | |
|---|---|---|---|
| | | PC1 | PC2 |
| $Fe(OTf)_3$ | #1 | -0.32 | 0.08 |
| | #2 | -0.32 | 0.04 |
| | #3 | -0.23 | 0.15 |
| $Cu(OTf)_2$ | #1 | -0.05 | -0.28 |
| | #2 | -0.12 | -0.12 |
| | #3 | -0.22 | -0.05 |
| $Bi(OTf)_3$ | #1 | -0.11 | 0.27 |
| | #2 | -0.27 | -0.13 |
| | #3 | -0.28 | 0.24 |
| $Al(OTf)_3$ | #1 | -0.26 | 0.01 |
| | #2 | -0.30 | 0.06 |
| | #3 | -0.31 | 0.05 |
| $In(OTf)_3$ | #1 | -0.03 | -0.02 |
| | #2 | -0.11 | -0.03 |
| | #3 | -0.19 | -0.08 |
| $Dy(OTf)_3$ | #1 | -0.20 | -0.19 |
| | #2 | -0.30 | -0.05 |
| | #3 | -0.14 | 0.28 |
| $Ce(OTf)_3$ | #1 | -0.12 | -0.27 |
| | #2 | -0.10 | -0.43 |
| | #3 | -0.10 | -0.50 |
| no triflate | #1 | 0.08 | -0.21 |
| | #2 | 0.15 | 0.02 |
| | #3 | -0.03 | 0.20 |

**Figure S1. PCA on $\alpha_1$, for R is measured at different time interval [0s;9s] | a,** PCA scores with 95% confidence ellipsoids. **b,** Individual variance for the different PC. **c,** PCA loadings of the different sensing elements' response for PC1 and PC2.

a)

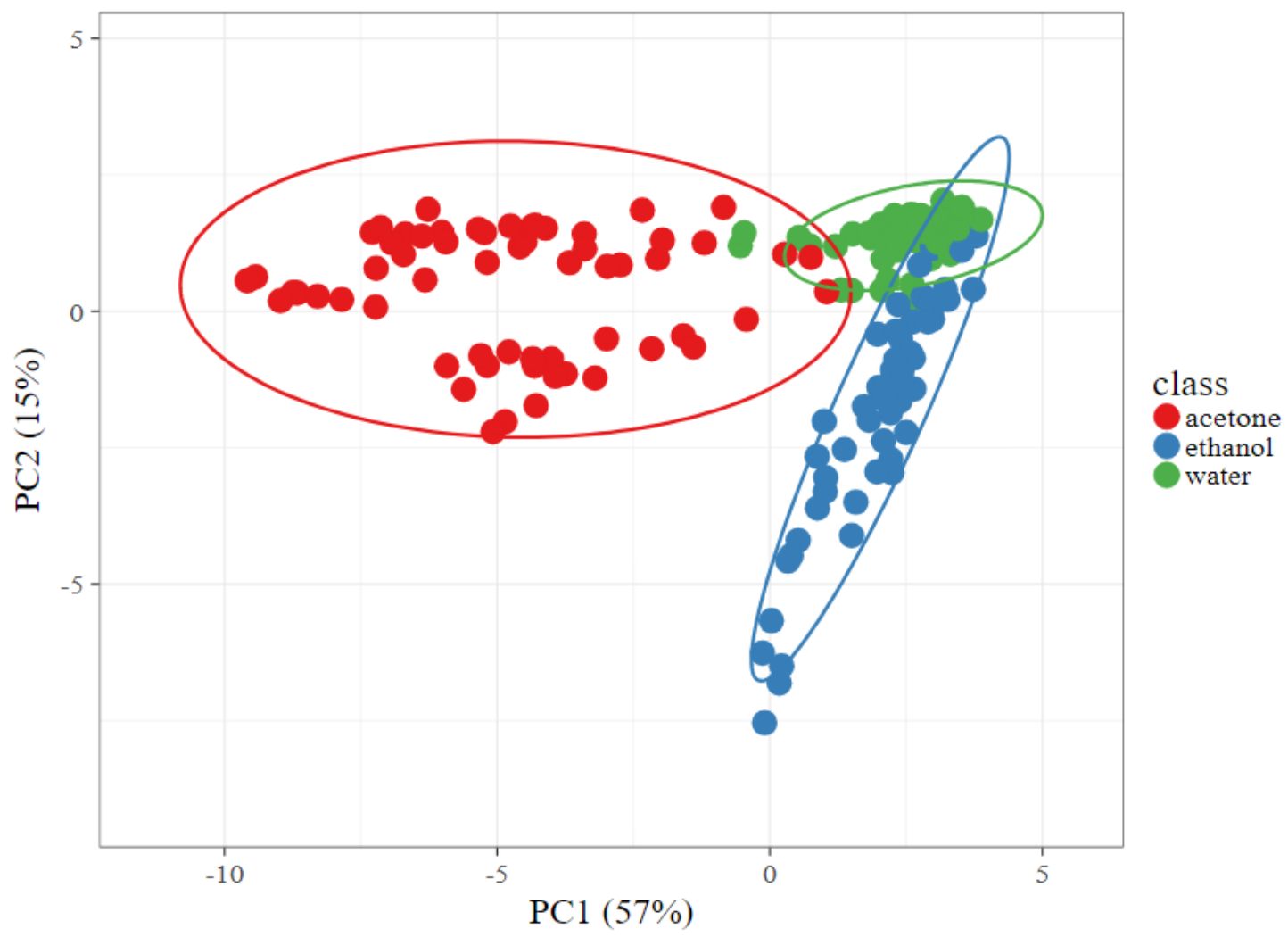

b)

| | Variance |
|---|---|
| **PC1** | 57.0% |
| **PC2** | 15.0% |
| **PC3** | 6.1% |
| **PC4** | 4.4% |
| **PC5** | 2.9% |
| **PC6** | 2.6% |
| **PC7** | 2.0% |
| **PC8** | 1.8% |
| **PC9** | 1.6% |
| **PC10** | 1.1% |
| **PC11** | 0.9% |
| **PC12** | 0.8% |
| **PC13** | 0.7% |
| **PC14** | 0.6% |
| **PC15** | 0.5% |
| **PC16** | 0.4% |
| **PC17** | 0.4% |
| **PC18** | 0.3% |
| **PC19** | 0.2% |
| **PC20** | 0.2% |
| **PC21** | 0.2% |
| **PC22** | 0.1% |
| **PC23** | 0.1% |
| **PC24** | 0.0% |

c)

| | | Loadings | |
|---|---|---|---|
| | | PC1 | PC2 |
| $Fe(OTf)_3$ | #1 | -0.20 | -0.16 |
| | #2 | -0.24 | -0.16 |
| | #3 | -0.23 | -0.14 |
| $Cu(OTf)_2$ | #1 | -0.19 | 0.19 |
| | #2 | -0.24 | -0.05 |
| | #3 | -0.21 | -0.11 |
| $Bi(OTf)_3$ | #1 | -0.18 | -0.33 |
| | #2 | -0.21 | 0.17 |
| | #3 | -0.18 | -0.27 |
| $Al(OTf)_3$ | #1 | -0.23 | 0.08 |
| | #2 | -0.25 | 0.09 |
| | #3 | -0.26 | -0.05 |
| $In(OTf)_3$ | #1 | -0.22 | 0.11 |
| | #2 | -0.25 | 0.00 |
| | #3 | -0.21 | -0.02 |
| $Dy(OTf)_3$ | #1 | -0.20 | 0.23 |
| | #2 | -0.24 | 0.15 |
| | #3 | -0.19 | -0.14 |
| $Ce(OTf)_3$ | #1 | -0.13 | 0.39 |
| | #2 | -0.07 | 0.35 |
| | #3 | -0.17 | 0.35 |
| no triflate | #1 | 0.13 | 0.31 |
| | #2 | 0.23 | 0.16 |
| | #3 | -0.06 | 0.16 |

**Figure S2. PCA on $\alpha_1$, for R is measured at different time interval [10s;19s] | a,** PCA scores with 95% confidence ellipsoids. **b,** Individual variance for the different PC. **c,** PCA loadings of the different sensing elements' response for PC1 and PC2.

a)

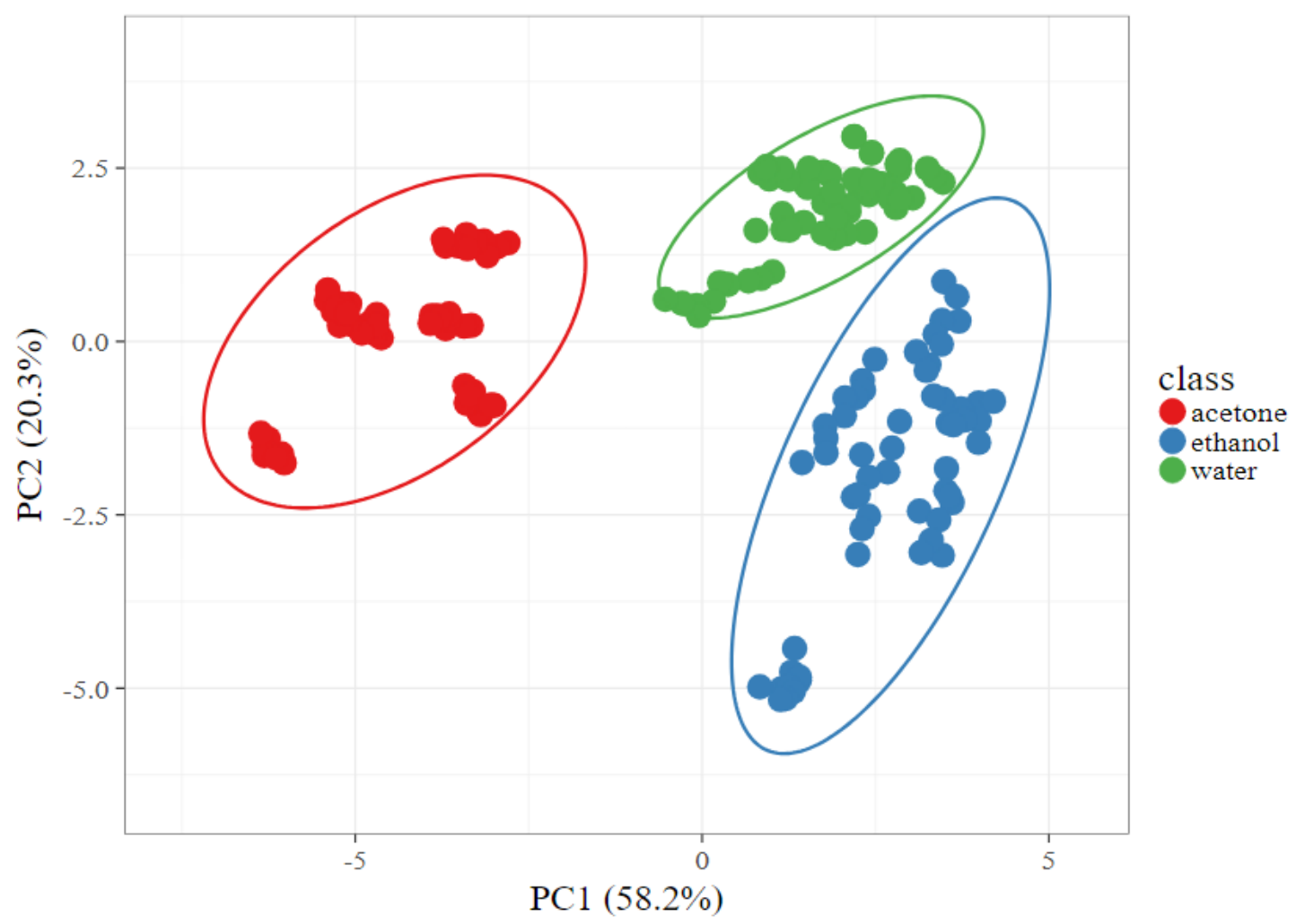

b)

| | Variance |
|---|---|
| **PC1** | 50.3% |
| **PC2** | 21.7% |
| **PC3** | 8.0% |
| **PC4** | 3.4% |
| **PC5** | 3.2% |
| **PC6** | 2.7% |
| **PC7** | 2.6% |
| **PC8** | 1.8% |
| **PC9** | 1.5% |
| **PC10** | 1.1% |
| **PC11** | 1.0% |
| **PC12** | 0.7% |
| **PC13** | 0.4% |
| **PC14** | 0.3% |
| **PC15** | 0.3% |
| **PC16** | 0.2% |
| **PC17** | 0.2% |
| **PC18** | 0.2% |
| **PC19** | 0.1% |
| **PC20** | 0.1% |
| **PC21** | 0.0% |
| **PC22** | 0.0% |
| **PC23** | 0.0% |
| **PC24** | 0.0% |

c)

| | | Loadings | |
|---|---|---|---|
| | | PC1 | PC2 |
| $Fe(OTf)_3$ | #1 | -0.16 | 0.21 |
| | #2 | -0.22 | 0.24 |
| | #3 | -0.22 | 0.24 |
| $Cu(OTf)_2$ | #1 | -0.17 | -0.22 |
| | #2 | -0.26 | -0.07 |
| | #3 | -0.17 | -0.19 |
| $Bi(OTf)_3$ | #1 | -0.14 | 0.32 |
| | #2 | -0.24 | -0.05 |
| | #3 | -0.08 | 0.33 |
| $Al(OTf)_3$ | #1 | -0.25 | -0.09 |
| | #2 | -0.27 | -0.10 |
| | #3 | -0.28 | 0.05 |
| $In(OTf)_3$ | #1 | -0.23 | -0.02 |
| | #2 | -0.27 | 0.02 |
| | #3 | -0.24 | 0.00 |
| $Dy(OTf)_3$ | #1 | -0.22 | -0.23 |
| | #2 | -0.25 | -0.14 |
| | #3 | -0.21 | 0.18 |
| $Ce(OTf)_3$ | #1 | -0.19 | -0.30 |
| | #2 | -0.07 | -0.29 |
| | #3 | -0.16 | -0.29 |
| no triflate | #1 | 0.11 | -0.28 |
| | #2 | 0.21 | -0.26 |
| | #3 | 0.04 | -0.14 |

**Figure S3. PCA on $\alpha_1$, for R is measured at different time interval [20s;29s] | a,** PCA scores with 95% confidence ellipsoids. **b,** Individual variance for the different PC. **c,** PCA loadings of the different sensing elements' response for PC1 and PC2.

a)

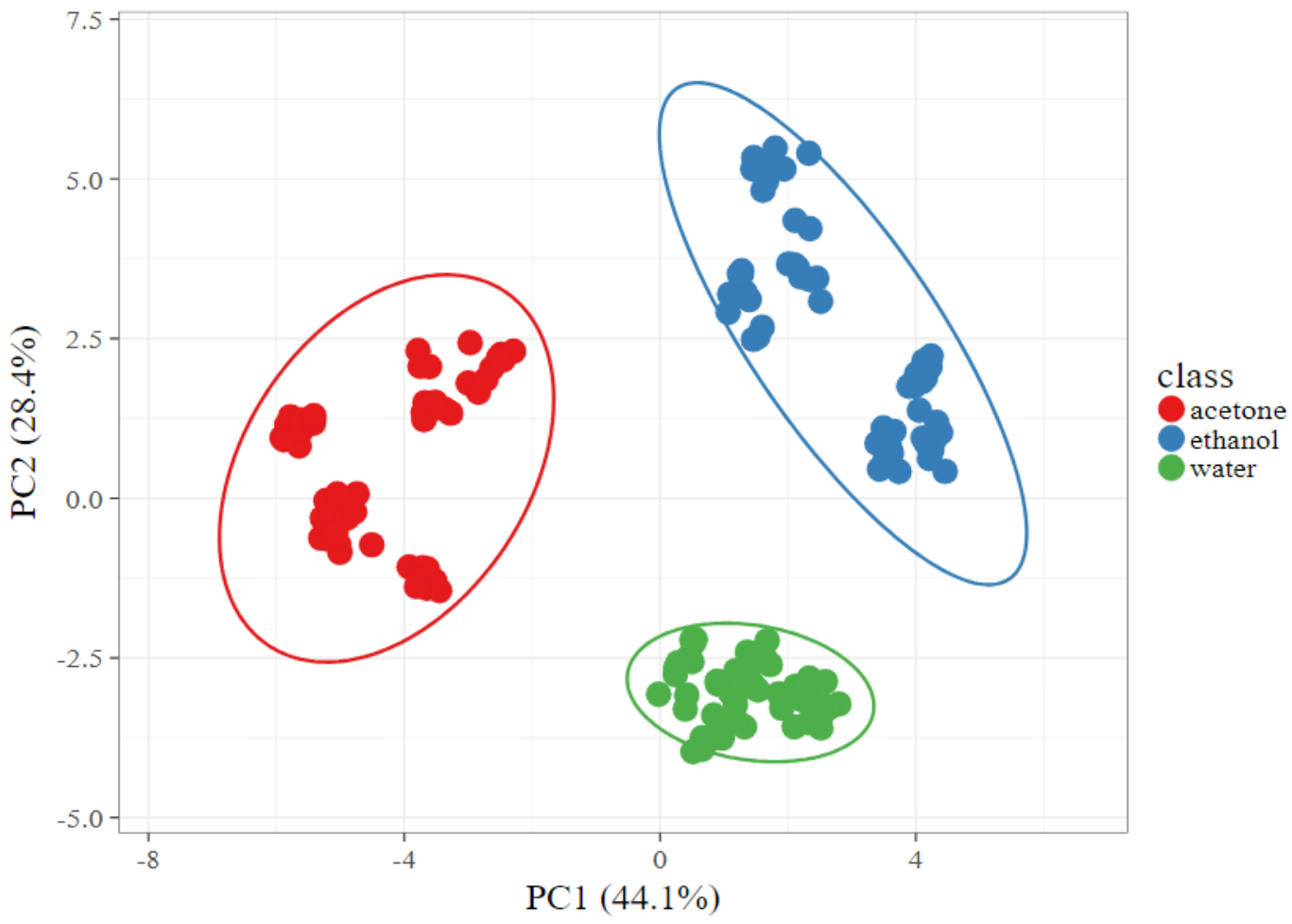

b)

| | Variance |
|---|---|
| **PC1** | 44.1% |
| **PC2** | 28.4% |
| **PC3** | 7.5% |
| **PC4** | 4.3% |
| **PC5** | 3.3% |
| **PC6** | 2.6% |
| **PC7** | 2.2% |
| **PC8** | 1.9% |
| **PC9** | 1.7% |
| **PC10** | 0.9% |
| **PC11** | 0.8% |
| **PC12** | 0.5% |
| **PC13** | 0.5% |
| **PC14** | 0.4% |
| **PC15** | 0.3% |
| **PC16** | 0.2% |
| **PC17** | 0.1% |
| **PC18** | 0.1% |
| **PC19** | 0.1% |
| **PC20** | 0.1% |
| **PC21** | 0.1% |
| **PC22** | 0.0% |
| **PC23** | 0.0% |
| **PC24** | 0.0% |

c)

| | | Loadings | |
|---|---|---|---|
| | | PC1 | PC2 |
| $Fe(OTf)_3$ | #1 | -0.12 | 0.25 |
| | #2 | -0.20 | 0.25 |
| | #3 | -0.15 | 0.29 |
| $Cu(OTf)_2$ | #1 | -0.16 | -0.21 |
| | #2 | -0.26 | -0.10 |
| | #3 | -0.13 | -0.21 |
| $Bi(OTf)_3$ | #1 | -0.06 | 0.30 |
| | #2 | -0.26 | -0.01 |
| | #3 | 0.01 | 0.29 |
| $Al(OTf)_3$ | #1 | -0.27 | -0.08 |
| | #2 | -0.28 | -0.11 |
| | #3 | -0.29 | 0.04 |
| $In(OTf)_3$ | #1 | -0.21 | -0.02 |
| | #2 | -0.28 | 0.06 |
| | #3 | -0.24 | -0.02 |
| $Dy(OTf)_3$ | #1 | -0.24 | -0.21 |
| | #2 | -0.26 | -0.14 |
| | #3 | -0.21 | 0.19 |
| $Ce(OTf)_3$ | #1 | -0.22 | -0.25 |
| | #2 | -0.03 | -0.27 |
| | #3 | -0.11 | -0.27 |
| no triflate | #1 | 0.14 | -0.28 |
| | #2 | 0.20 | -0.26 |
| | #3 | 0.17 | -0.21 |

**Figure S4. PCA on $\alpha_1$, for R is measured at different time interval [30s;39s] | a,** PCA scores with 95% confidence ellipsoids. **b,** Individual variance for the different PC. **c,** PCA loadings of the different sensing elements' response for PC1 and PC2.

a)

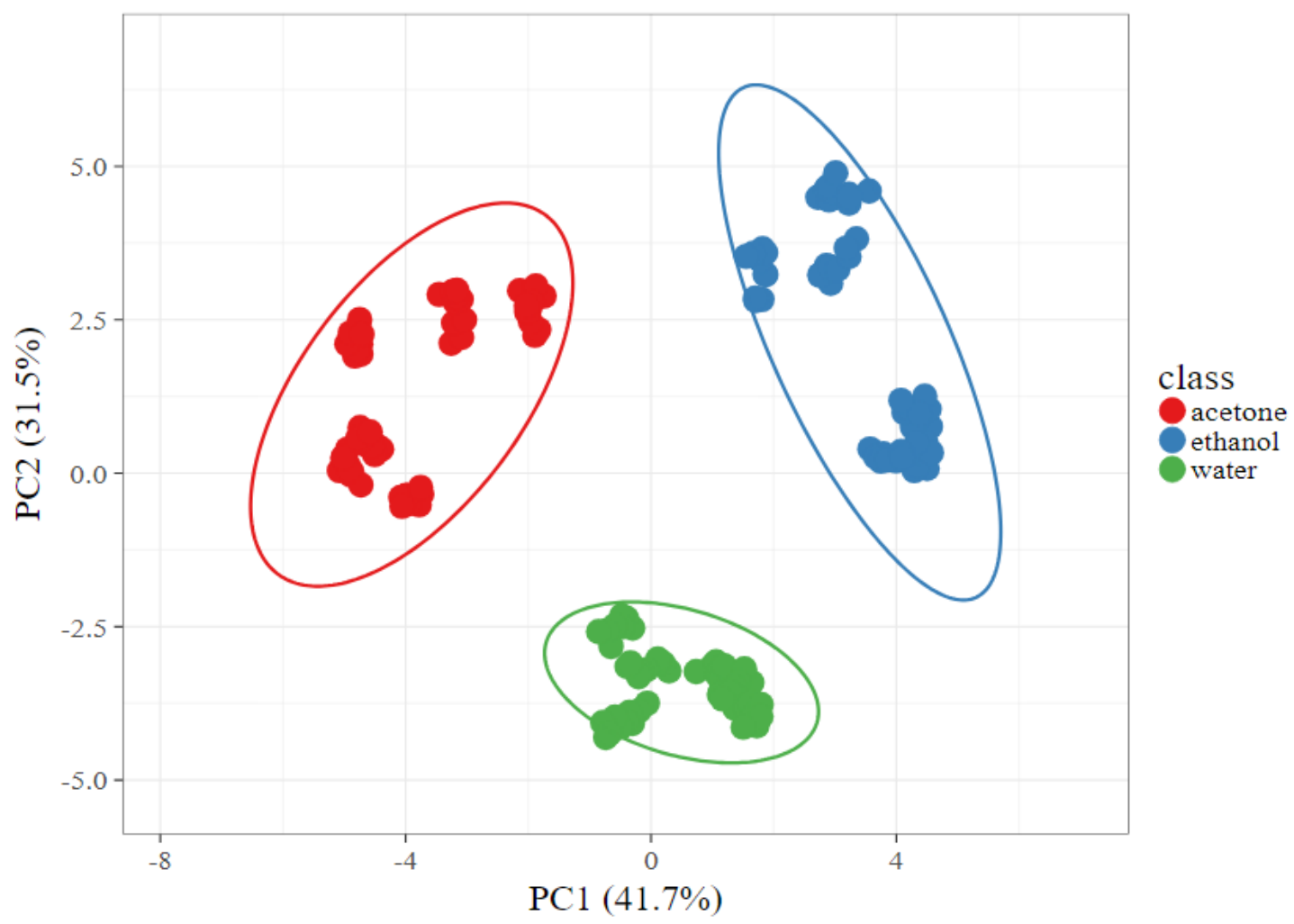

b)

| | Variance |
|---|---|
| **PC1** | 41.7% |
| **PC2** | 31.5% |
| **PC3** | 7.2% |
| **PC4** | 4.2% |
| **PC5** | 3.6% |
| **PC6** | 2.6% |
| **PC7** | 2.2% |
| **PC8** | 1.9% |
| **PC9** | 1.4% |
| **PC10** | 0.8% |
| **PC11** | 0.8% |
| **PC12** | 0.5% |
| **PC13** | 0.4% |
| **PC14** | 0.3% |
| **PC15** | 0.2% |
| **PC16** | 0.2% |
| **PC17** | 0.1% |
| **PC18** | 0.1% |
| **PC19** | 0.1% |
| **PC20** | 0.1% |
| **PC21** | 0.0% |
| **PC22** | 0.0% |
| **PC23** | 0.0% |
| **PC24** | 0.0% |

c)

| | | Loadings | |
|---|---|---|---|
| | | PC1 | PC2 |
| $Fe(OTf)_3$ | #1 | -0.07 | 0.26 |
| | #2 | -0.15 | 0.28 |
| | #3 | -0.05 | 0.31 |
| $Cu(OTf)_2$ | #1 | -0.17 | -0.21 |
| | #2 | -0.28 | -0.07 |
| | #3 | -0.13 | -0.21 |
| $Bi(OTf)_3$ | #1 | 0.03 | 0.27 |
| | #2 | -0.26 | 0.08 |
| | #3 | 0.10 | 0.25 |
| $Al(OTf)_3$ | #1 | -0.28 | -0.04 |
| | #2 | -0.29 | -0.09 |
| | #3 | -0.29 | 0.07 |
| $In(OTf)_3$ | #1 | -0.24 | 0.01 |
| | #2 | -0.28 | 0.11 |
| | #3 | -0.25 | 0.00 |
| $Dy(OTf)_3$ | #1 | -0.27 | -0.16 |
| | #2 | -0.28 | -0.09 |
| | #3 | -0.18 | 0.21 |
| $Ce(OTf)_3$ | #1 | -0.25 | -0.19 |
| | #2 | -0.05 | -0.27 |
| | #3 | -0.12 | -0.25 |
| no triflate | #1 | 0.12 | -0.30 |
| | #2 | 0.16 | -0.29 |
| | #3 | 0.16 | -0.26 |

**Figure S5. PCA on $\alpha_1$, for R is measured at different time interval [40s;49s] | a,** PCA scores with 95% confidence ellipsoids. **b,** Individual variance for the different PC. **c,** PCA loadings of the different sensing elements' response for PC1 and PC2.

a)

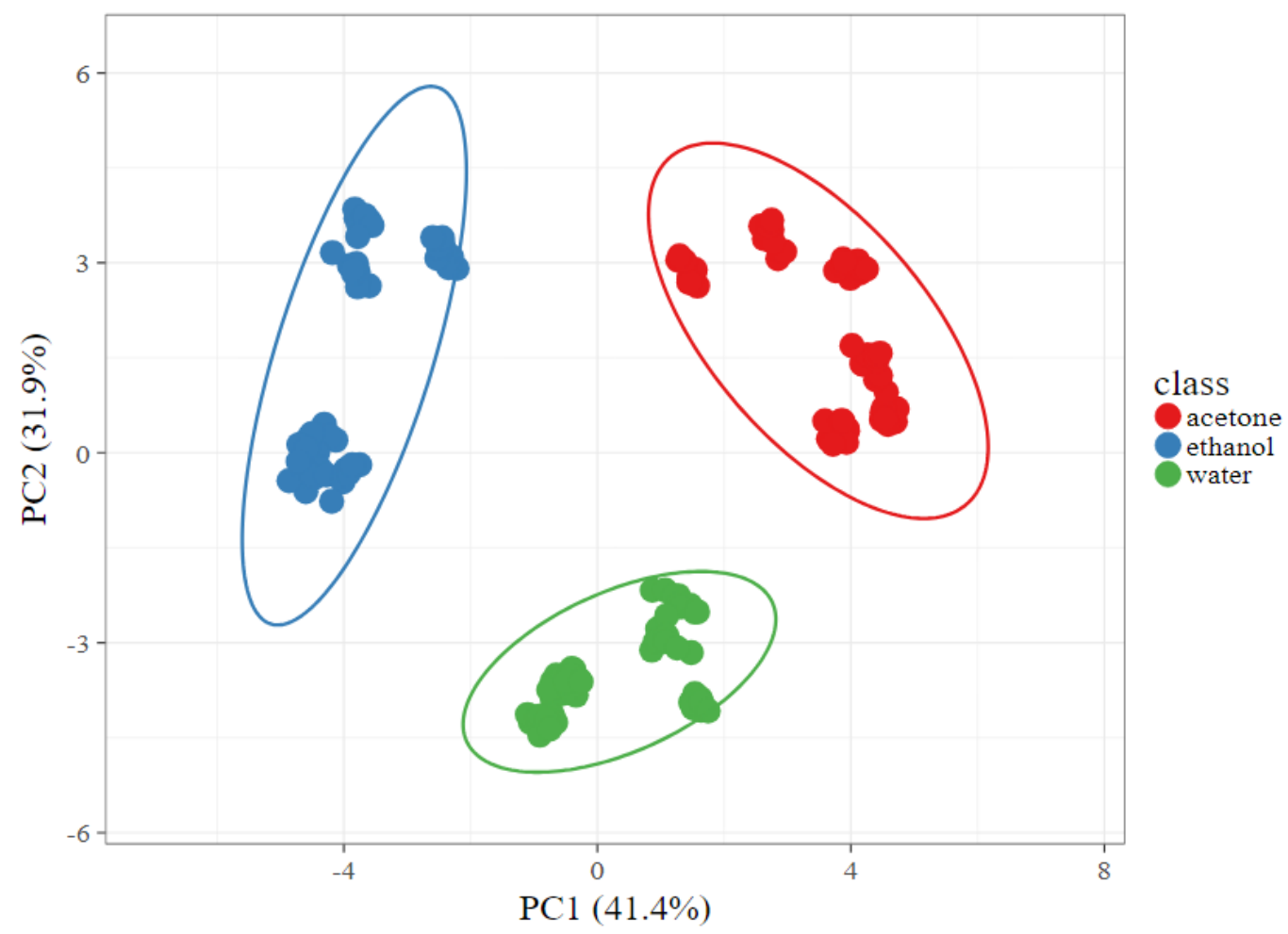

b)

| | Variance |
|---|---|
| **PC1** | 41.4% |
| **PC2** | 31.9% |
| **PC3** | 6.4% |
| **PC4** | 4.5% |
| **PC5** | 3.3% |
| **PC6** | 2.5% |
| **PC7** | 2.4% |
| **PC8** | 2.1% |
| **PC9** | 1.4% |
| **PC10** | 1.0% |
| **PC11** | 0.9% |
| **PC12** | 0.6% |
| **PC13** | 0.4% |
| **PC14** | 0.4% |
| **PC15** | 0.2% |
| **PC16** | 0.2% |
| **PC17** | 0.1% |
| **PC18** | 0.1% |
| **PC19** | 0.1% |
| **PC20** | 0.0% |
| **PC21** | 0.0% |
| **PC22** | 0.0% |
| **PC23** | 0.0% |
| **PC24** | 0.0% |

c)

| | | Loadings | |
|---|---|---|---|
| | | PC1 | PC2 |
| $Fe(OTf)_3$ | #1 | -0.03 | 0.26 |
| | #2 | -0.09 | 0.30 |
| | #3 | 0.02 | 0.30 |
| $Cu(OTf)_2$ | #1 | -0.19 | -0.18 |
| | #2 | -0.30 | -0.01 |
| | #3 | -0.16 | -0.22 |
| $Bi(OTf)_3$ | #1 | 0.10 | 0.24 |
| | #2 | -0.25 | 0.14 |
| | #3 | 0.17 | 0.20 |
| $Al(OTf)_3$ | #1 | -0.28 | 0.01 |
| | #2 | -0.29 | -0.07 |
| | #3 | -0.28 | 0.10 |
| $In(OTf)_3$ | #1 | -0.20 | 0.05 |
| | #2 | -0.26 | 0.14 |
| | #3 | -0.25 | 0.03 |
| $Dy(OTf)_3$ | #1 | -0.29 | -0.11 |
| | #2 | -0.29 | -0.04 |
| | #3 | -0.15 | 0.24 |
| $Ce(OTf)_3$ | #1 | -0.27 | -0.15 |
| | #2 | -0.08 | -0.28 |
| | #3 | -0.14 | -0.24 |
| no triflate | #1 | 0.08 | -0.32 |
| | #2 | 0.11 | -0.32 |
| | #3 | 0.14 | -0.29 |

**Figure S6. PCA on $\alpha_1$, for R is measured at different time interval [50s;59s] | a,** PCA scores with 95% confidence ellipsoids. **b,** Individual variance for the different PC. **c,** PCA loadings of the different sensing elements' response for PC1 and PC2.

a)

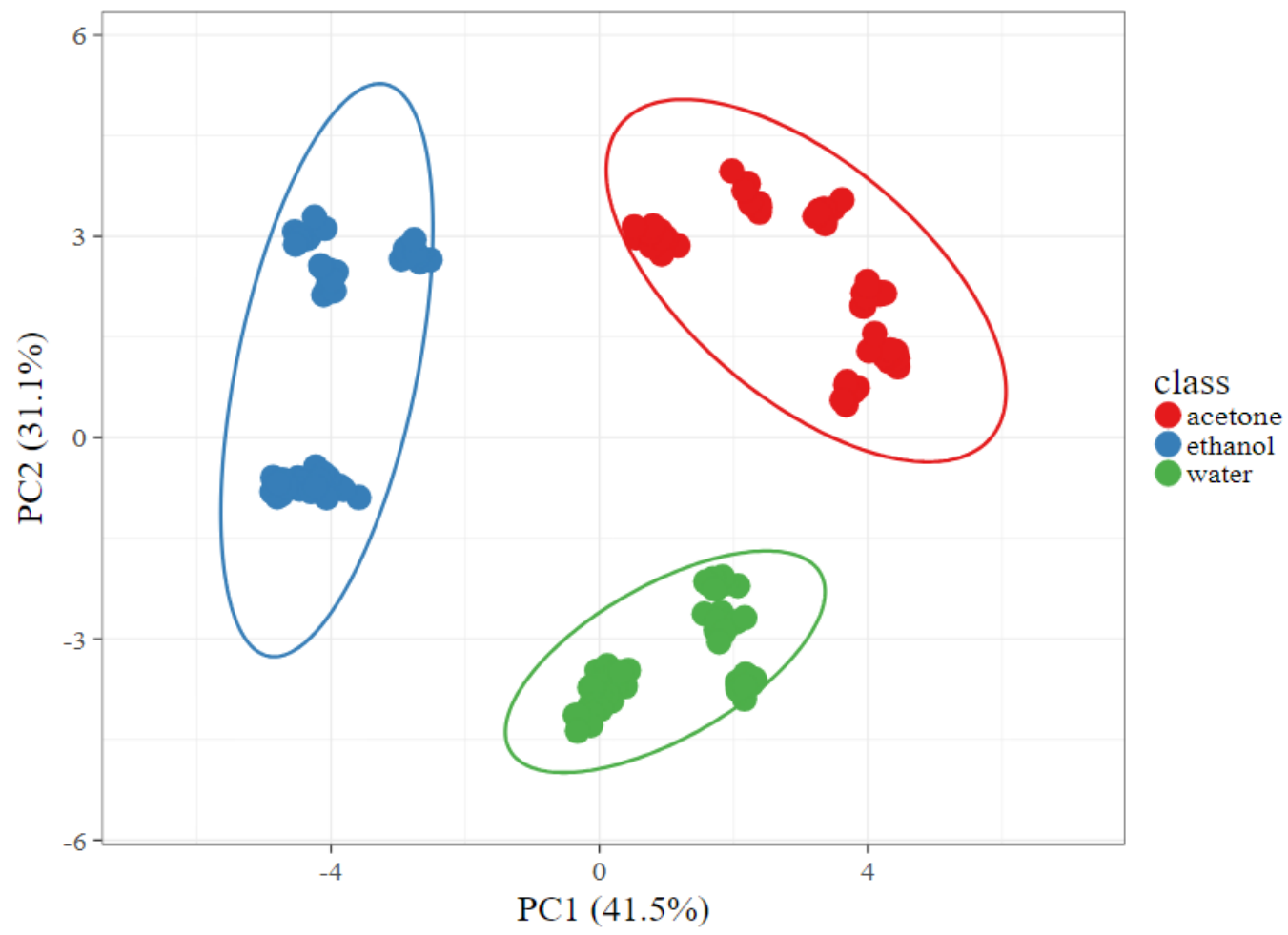

b)

| | Variance |
|---|---|
| **PC1** | 41.5% |
| **PC2** | 31.1% |
| **PC3** | 6.4% |
| **PC4** | 5.0% |
| **PC5** | 3.2% |
| **PC6** | 2.8% |
| **PC7** | 2.4% |
| **PC8** | 2.3% |
| **PC9** | 1.3% |
| **PC10** | 1.1% |
| **PC11** | 0.8% |
| **PC12** | 0.5% |
| **PC13** | 0.4% |
| **PC14** | 0.3% |
| **PC15** | 0.3% |
| **PC16** | 0.2% |
| **PC17** | 0.1% |
| **PC18** | 0.1% |
| **PC19** | 0.1% |
| **PC20** | 0.1% |
| **PC21** | 0.0% |
| **PC22** | 0.0% |
| **PC23** | 0.0% |
| **PC24** | 0.0% |

c)

| | | Loadings | |
|---|---|---|---|
| | | PC1 | PC2 |
| $Fe(OTf)_3$ | #1 | -0.01 | -0.26 |
| | #2 | 0.07 | -0.30 |
| | #3 | -0.05 | -0.29 |
| $Cu(OTf)_2$ | #1 | 0.20 | 0.15 |
| | #2 | 0.29 | -0.05 |
| | #3 | 0.17 | 0.21 |
| $Bi(OTf)_3$ | #1 | -0.16 | -0.19 |
| | #2 | 0.22 | -0.20 |
| | #3 | -0.21 | -0.15 |
| $Al(OTf)_3$ | #1 | 0.28 | -0.04 |
| | #2 | 0.29 | 0.07 |
| | #3 | 0.28 | -0.11 |
| $In(OTf)_3$ | #1 | 0.18 | -0.08 |
| | #2 | 0.25 | -0.17 |
| | #3 | 0.25 | -0.05 |
| $Dy(OTf)_3$ | #1 | 0.30 | 0.06 |
| | #2 | 0.29 | 0.01 |
| | #3 | 0.13 | -0.25 |
| $Ce(OTf)_3$ | #1 | 0.28 | 0.13 |
| | #2 | 0.10 | 0.27 |
| | #3 | 0.16 | 0.24 |
| no triflate | #1 | -0.04 | 0.34 |
| | #2 | -0.08 | 0.33 |
| | #3 | -0.10 | 0.32 |

**Figure S7. PCA on $\alpha_1$, for R is measured at different time interval [60s;69s] | a,** PCA scores with 95% confidence ellipsoids. **b,** Individual variance for the different PC. **c,** PCA loadings of the different sensing elements' response for PC1 and PC2.

a)

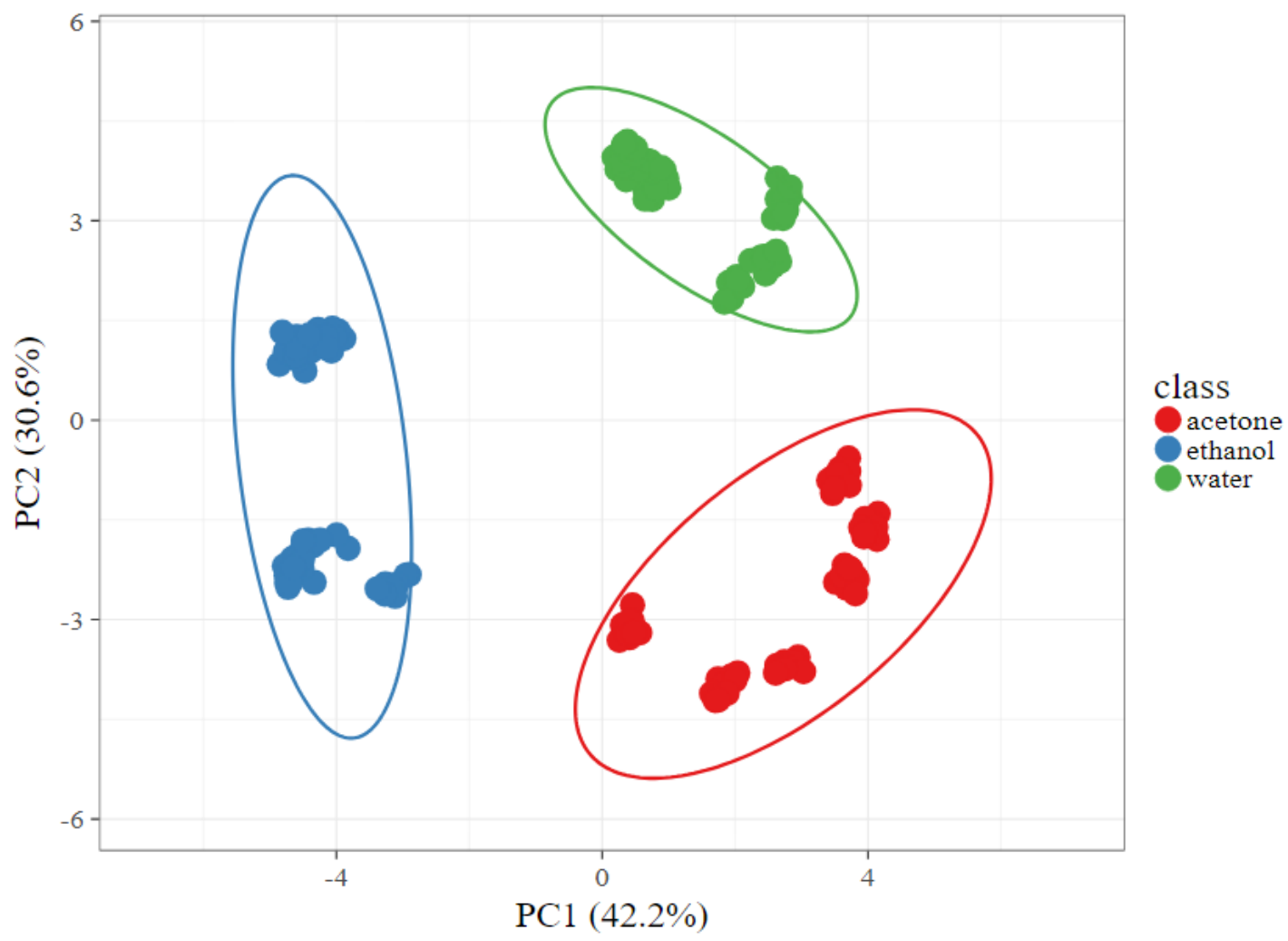

b)

| | Variance |
|---|---|
| **PC1** | 42.2% |
| **PC2** | 30.6% |
| **PC3** | 6.2% |
| **PC4** | 5.2% |
| **PC5** | 3.1% |
| **PC6** | 2.9% |
| **PC7** | 2.5% |
| **PC8** | 2.1% |
| **PC9** | 1.3% |
| **PC10** | 0.9% |
| **PC11** | 0.9% |
| **PC12** | 0.6% |
| **PC13** | 0.3% |
| **PC14** | 0.2% |
| **PC15** | 0.2% |
| **PC16** | 0.2% |
| **PC17** | 0.1% |
| **PC18** | 0.1% |
| **PC19** | 0.1% |
| **PC20** | 0.0% |
| **PC21** | 0.0% |
| **PC22** | 0.0% |
| **PC23** | 0.0% |
| **PC24** | 0.0% |

c)

| | | Loadings | |
|---|---|---|---|
| | | PC1 | PC2 |
| $Fe(OTf)_3$ | #1 | -0.04 | -0.25 |
| | #2 | 0.06 | -0.30 |
| | #3 | -0.06 | -0.27 |
| $Cu(OTf)_2$ | #1 | 0.22 | 0.12 |
| | #2 | 0.29 | -0.06 |
| | #3 | 0.19 | 0.20 |
| $Bi(OTf)_3$ | #1 | -0.20 | -0.12 |
| | #2 | 0.19 | -0.23 |
| | #3 | -0.23 | -0.11 |
| $Al(OTf)_3$ | #1 | 0.28 | -0.05 |
| | #2 | 0.28 | 0.06 |
| | #3 | 0.27 | -0.10 |
| $In(OTf)_3$ | #1 | 0.18 | -0.06 |
| | #2 | 0.23 | -0.19 |
| | #3 | 0.24 | -0.07 |
| $Dy(OTf)_3$ | #1 | 0.30 | 0.03 |
| | #2 | 0.29 | -0.02 |
| | #3 | 0.11 | -0.27 |
| $Ce(OTf)_3$ | #1 | 0.28 | 0.11 |
| | #2 | 0.12 | 0.27 |
| | #3 | 0.18 | 0.25 |
| no triflate | #1 | -0.01 | 0.34 |
| | #2 | -0.04 | 0.34 |
| | #3 | -0.07 | 0.33 |

**Figure S8. PCA on $\alpha_1$, for R is measured at different time interval [70s;79s] | a,** PCA scores with 95% confidence ellipsoids. **b,** Individual variance for the different PC. **c,** PCA loadings of the different sensing elements' response for PC1 and PC2.

a)

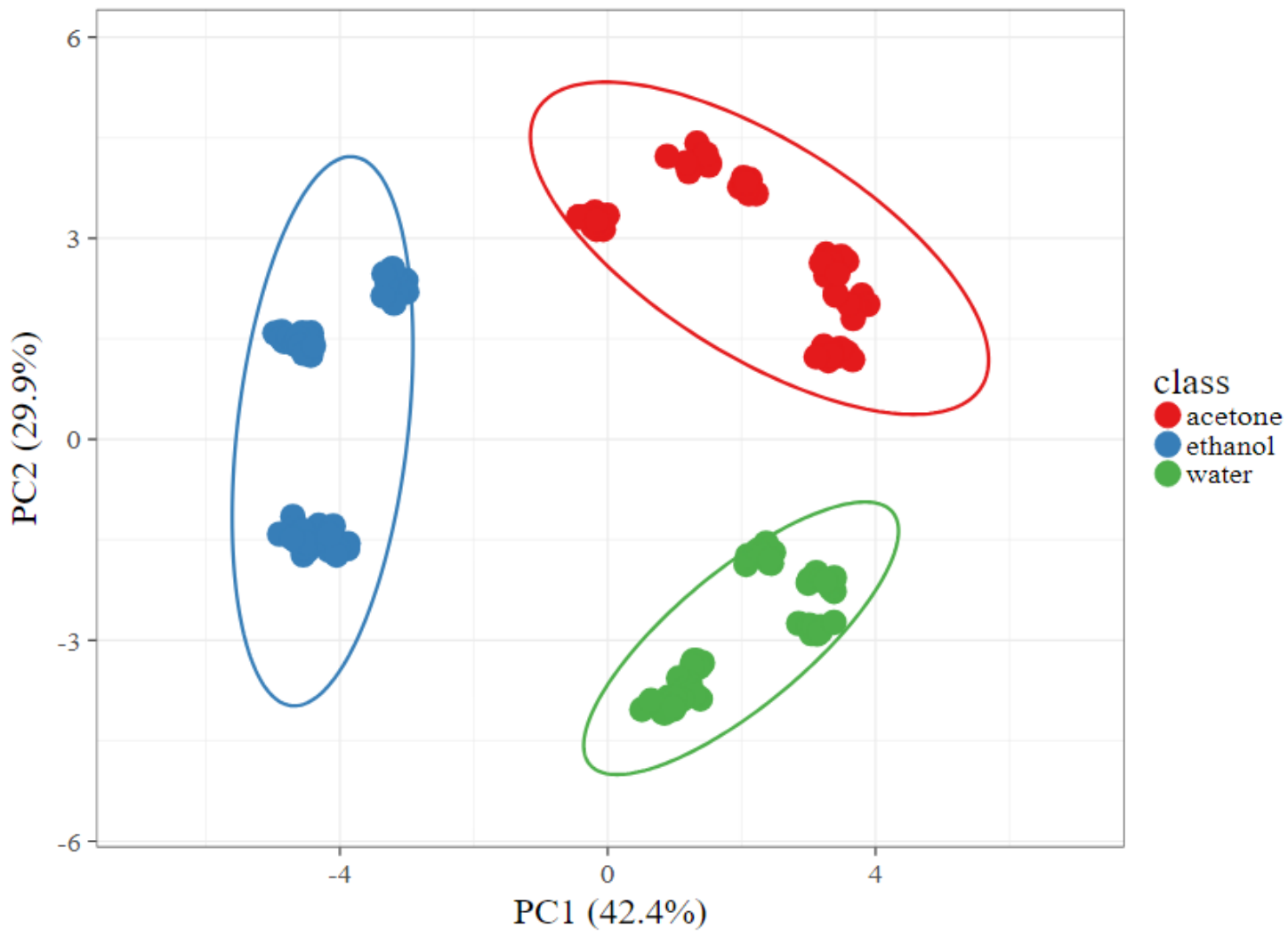

b)

| | Variance |
|---|---|
| **PC1** | 42.4% |
| **PC2** | 29.9% |
| **PC3** | 6.6% |
| **PC4** | 5.4% |
| **PC5** | 3.3% |
| **PC6** | 2.7% |
| **PC7** | 2.5% |
| **PC8** | 2.4% |
| **PC9** | 1.3% |
| **PC10** | 0.9% |
| **PC11** | 0.8% |
| **PC12** | 0.6% |
| **PC13** | 0.4% |
| **PC14** | 0.2% |
| **PC15** | 0.2% |
| **PC16** | 0.2% |
| **PC17** | 0.1% |
| **PC18** | 0.1% |
| **PC19** | 0.1% |
| **PC20** | 0.0% |
| **PC21** | 0.0% |
| **PC22** | 0.0% |
| **PC23** | 0.0% |
| **PC24** | 0.0% |

c)

| | | Loadings | |
|---|---|---|---|
| | | PC1 | PC2 |
| $Fe(OTf)_3$ | #1 | -0.07 | 0.24 |
| | #2 | 0.04 | 0.31 |
| | #3 | -0.07 | 0.26 |
| $Cu(OTf)_2$ | #1 | 0.23 | -0.14 |
| | #2 | 0.29 | 0.07 |
| | #3 | 0.22 | -0.14 |
| $Bi(OTf)_3$ | #1 | -0.22 | 0.04 |
| | #2 | 0.16 | 0.27 |
| | #3 | -0.25 | 0.06 |
| $Al(OTf)_3$ | #1 | 0.27 | 0.09 |
| | #2 | 0.28 | -0.03 |
| | #3 | 0.27 | 0.09 |
| $In(OTf)_3$ | #1 | 0.17 | 0.07 |
| | #2 | 0.21 | 0.21 |
| | #3 | 0.23 | 0.09 |
| $Dy(OTf)_3$ | #1 | 0.30 | 0.01 |
| | #2 | 0.29 | 0.05 |
| | #3 | 0.06 | 0.28 |
| $Ce(OTf)_3$ | #1 | 0.28 | -0.10 |
| | #2 | 0.14 | -0.26 |
| | #3 | 0.19 | -0.25 |
| no triflate | #1 | 0.03 | -0.35 |
| | #2 | -0.01 | -0.35 |
| | #3 | -0.04 | -0.35 |

**Figure S9. PCA on $\alpha_1$, for R is measured at different time interval [80s;89s] | a,** PCA scores with 95% confidence ellipsoids. **b,** Individual variance for the different PC. **c,** PCA loadings of the different sensing elements' response for PC1 and PC2.

a)

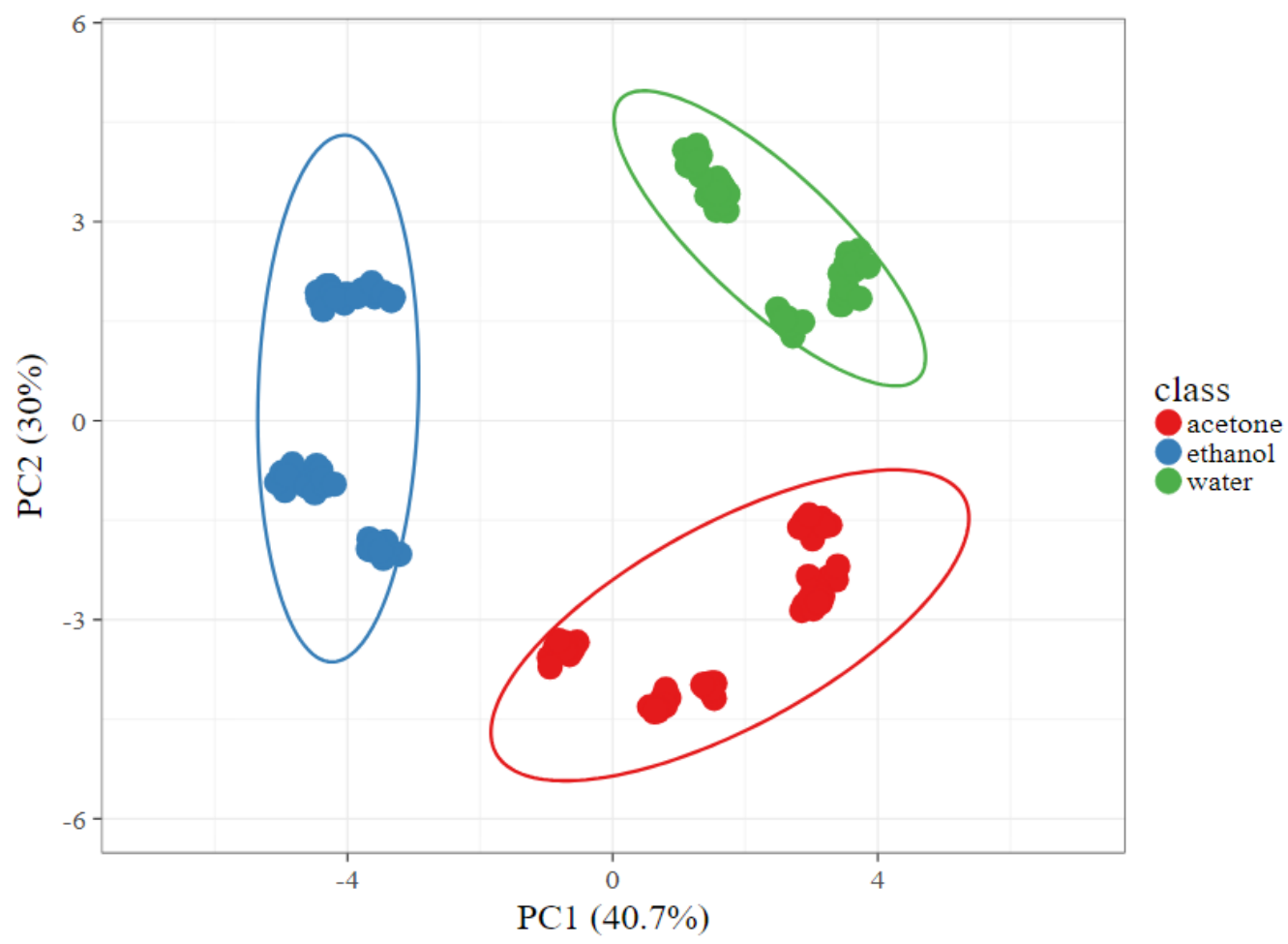

b)

| | Variance |
|---|---|
| **PC1** | 40.7% |
| **PC2** | 30.0% |
| **PC3** | 6.9% |
| **PC4** | 5.9% |
| **PC5** | 3.7% |
| **PC6** | 3.0% |
| **PC7** | 2.8% |
| **PC8** | 2.3% |
| **PC9** | 1.1% |
| **PC10** | 0.9% |
| **PC11** | 0.7% |
| **PC12** | 0.5% |
| **PC13** | 0.4% |
| **PC14** | 0.3% |
| **PC15** | 0.2% |
| **PC16** | 0.2% |
| **PC17** | 0.1% |
| **PC18** | 0.1% |
| **PC19** | 0.1% |
| **PC20** | 0.0% |
| **PC21** | 0.0% |
| **PC22** | 0.0% |
| **PC23** | 0.0% |
| **PC24** | 0.0% |

c)

| | | Loadings | |
|---|---|---|---|
| | | PC1 | PC2 |
| $Fe(OTf)_3$ | #1 | -0.11 | 0.22 |
| | #2 | 0.02 | 0.30 |
| | #3 | -0.08 | 0.24 |
| $Cu(OTf)_2$ | #1 | 0.23 | -0.11 |
| | #2 | 0.28 | 0.10 |
| | #3 | 0.20 | -0.16 |
| $Bi(OTf)_3$ | #1 | -0.22 | -0.04 |
| | #2 | 0.14 | 0.29 |
| | #3 | -0.26 | 0.01 |
| $Al(OTf)_3$ | #1 | 0.26 | 0.11 |
| | #2 | 0.28 | -0.02 |
| | #3 | 0.27 | 0.08 |
| $In(OTf)_3$ | #1 | 0.15 | 0.06 |
| | #2 | 0.18 | 0.24 |
| | #3 | 0.22 | 0.11 |
| $Dy(OTf)_3$ | #1 | 0.30 | 0.05 |
| | #2 | 0.29 | 0.09 |
| | #3 | 0.03 | 0.29 |
| $Ce(OTf)_3$ | #1 | 0.29 | -0.08 |
| | #2 | 0.16 | -0.24 |
| | #3 | 0.22 | -0.23 |
| no triflate | #1 | 0.05 | -0.35 |
| | #2 | 0.04 | -0.35 |
| | #3 | 0.01 | -0.35 |

**Figure S10. PCA on $\alpha_1$, for R is measured at different time interval [90s;99s] | a,** PCA scores with 95% confidence ellipsoids. **b,** Individual variance for the different PC. **c,** PCA loadings of the different sensing elements' response for PC1 and PC2.

a)

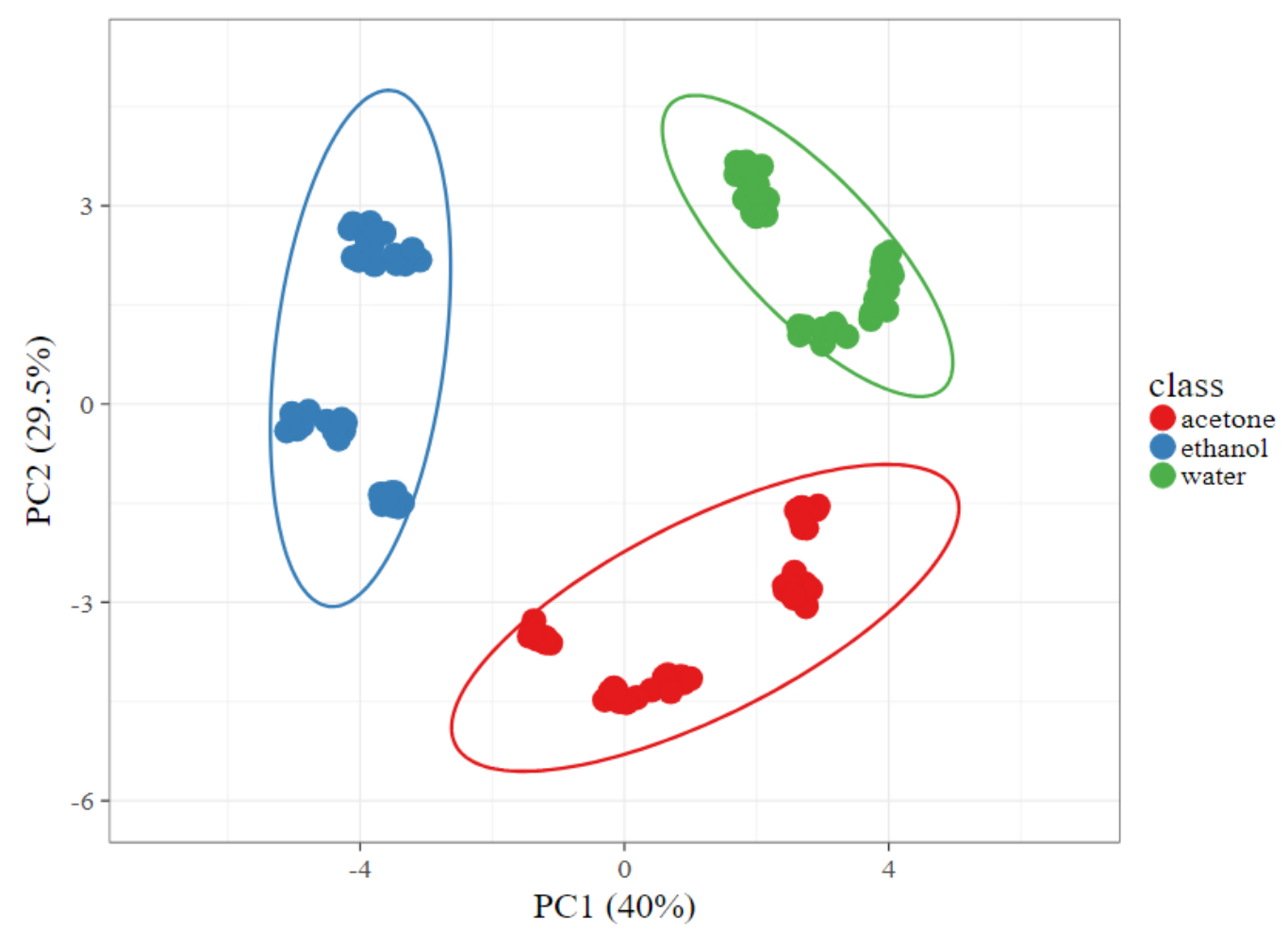

b)

| | Variance |
|---|---|
| **PC1** | 40.0% |
| **PC2** | 29.5% |
| **PC3** | 7.8% |
| **PC4** | 6.0% |
| **PC5** | 4.1% |
| **PC6** | 3.1% |
| **PC7** | 2.7% |
| **PC8** | 2.3% |
| **PC9** | 1.1% |
| **PC10** | 0.9% |
| **PC11** | 0.7% |
| **PC12** | 0.5% |
| **PC13** | 0.4% |
| **PC14** | 0.3% |
| **PC15** | 0.2% |
| **PC16** | 0.1% |
| **PC17** | 0.1% |
| **PC18** | 0.1% |
| **PC19** | 0.1% |
| **PC20** | 0.0% |
| **PC21** | 0.0% |
| **PC22** | 0.0% |
| **PC23** | 0.0% |
| **PC24** | 0.0% |

c)

| | | Loadings | |
|---|---|---|---|
| | | PC1 | PC2 |
| $Fe(OTf)_3$ | #1 | -0.14 | 0.19 |
| | #2 | -0.01 | 0.29 |
| | #3 | -0.11 | 0.22 |
| $Cu(OTf)_2$ | #1 | 0.24 | -0.09 |
| | #2 | 0.28 | 0.11 |
| | #3 | 0.20 | -0.14 |
| $Bi(OTf)_3$ | #1 | -0.21 | -0.11 |
| | #2 | 0.10 | 0.32 |
| | #3 | -0.26 | -0.05 |
| $Al(OTf)_3$ | #1 | 0.25 | 0.14 |
| | #2 | 0.27 | -0.02 |
| | #3 | 0.27 | 0.09 |
| $In(OTf)_3$ | #1 | 0.14 | 0.04 |
| | #2 | 0.16 | 0.26 |
| | #3 | 0.21 | 0.13 |
| $Dy(OTf)_3$ | #1 | 0.30 | 0.09 |
| | #2 | 0.27 | 0.14 |
| | #3 | -0.01 | 0.29 |
| $Ce(OTf)_3$ | #1 | 0.30 | -0.05 |
| | #2 | 0.19 | -0.22 |
| | #3 | 0.24 | -0.21 |
| no triflate | #1 | 0.09 | -0.34 |
| | #2 | 0.09 | -0.34 |
| | #3 | 0.06 | -0.35 |

**Figure S11. PCA on $\alpha_1$, for R is measured at different time interval [100s;109s] | a,** PCA scores with 95% confidence ellipsoids. **b,** Individual variance for the different PC. **c,** PCA loadings of the different sensing elements' response for PC1 and PC2.

a)

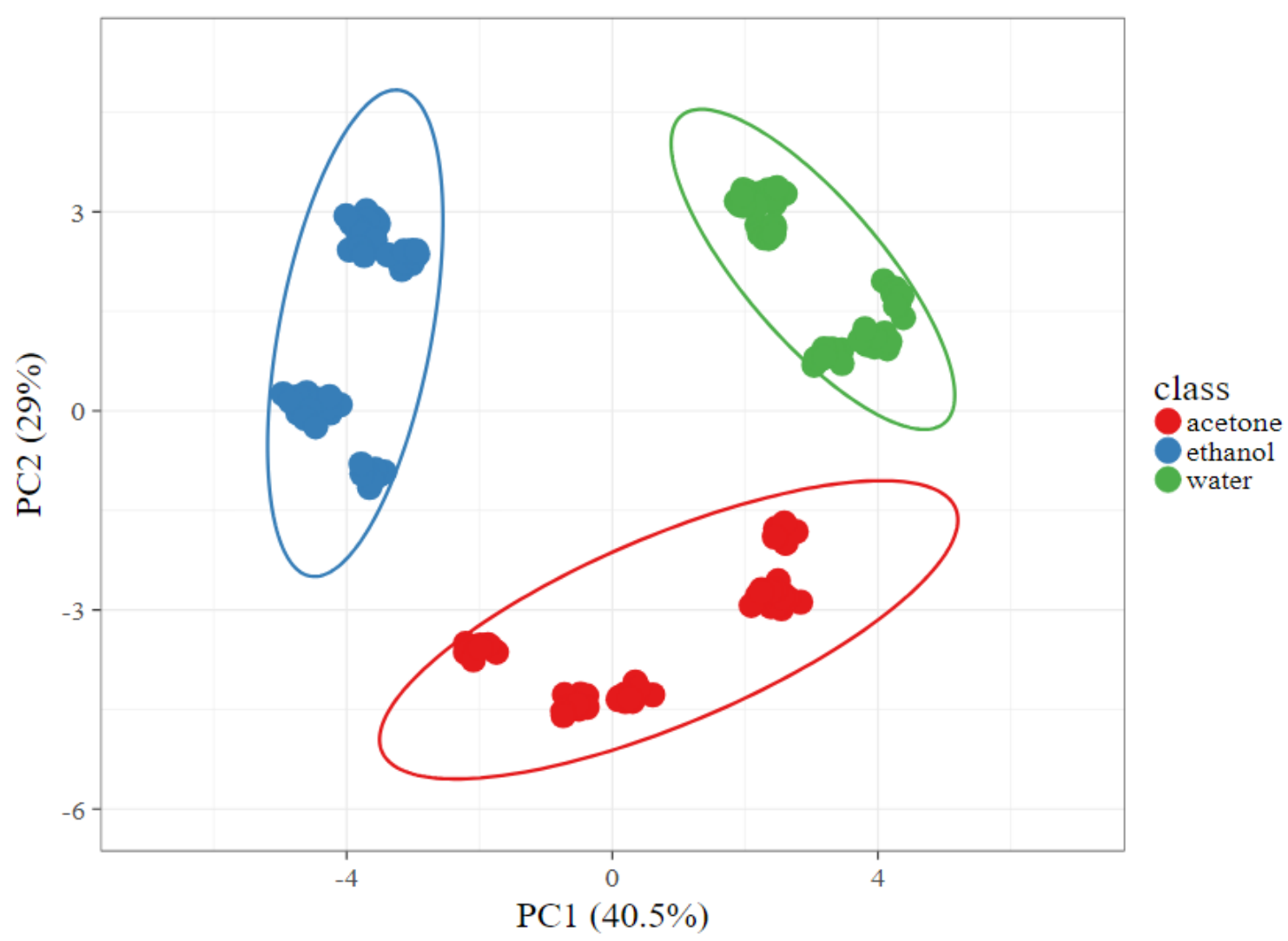

b)

| | Variance |
|---|---|
| **PC1** | 40.5% |
| **PC2** | 29.0% |
| **PC3** | 8.2% |
| **PC4** | 6.0% |
| **PC5** | 3.8% |
| **PC6** | 3.0% |
| **PC7** | 2.6% |
| **PC8** | 2.1% |
| **PC9** | 1.2% |
| **PC10** | 1.0% |
| **PC11** | 0.7% |
| **PC12** | 0.5% |
| **PC13** | 0.4% |
| **PC14** | 0.3% |
| **PC15** | 0.2% |
| **PC16** | 0.1% |
| **PC17** | 0.1% |
| **PC18** | 0.1% |
| **PC19** | 0.1% |
| **PC20** | 0.0% |
| **PC21** | 0.0% |
| **PC22** | 0.0% |
| **PC23** | 0.0% |
| **PC24** | 0.0% |

c)

| | | Loadings | |
|---|---|---|---|
| | | PC1 | PC2 |
| $Fe(OTf)_3$ | #1 | -0.15 | 0.16 |
| | #2 | -0.02 | 0.28 |
| | #3 | -0.12 | 0.19 |
| $Cu(OTf)_2$ | #1 | 0.24 | -0.07 |
| | #2 | 0.27 | 0.12 |
| | #3 | 0.23 | -0.13 |
| $Bi(OTf)_3$ | #1 | -0.19 | -0.16 |
| | #2 | 0.07 | 0.33 |
| | #3 | -0.25 | -0.10 |
| $Al(OTf)_3$ | #1 | 0.24 | 0.15 |
| | #2 | 0.27 | -0.01 |
| | #3 | 0.27 | 0.09 |
| $In(OTf)_3$ | #1 | 0.16 | -0.01 |
| | #2 | 0.16 | 0.25 |
| | #3 | 0.20 | 0.15 |
| $Dy(OTf)_3$ | #1 | 0.29 | 0.12 |
| | #2 | 0.26 | 0.17 |
| | #3 | -0.03 | 0.30 |
| $Ce(OTf)_3$ | #1 | 0.30 | -0.04 |
| | #2 | 0.20 | -0.21 |
| | #3 | 0.25 | -0.19 |
| no triflate | #1 | 0.11 | -0.33 |
| | #2 | 0.11 | -0.33 |
| | #3 | 0.08 | -0.35 |

**Figure S12. PCA on $\alpha_1$, for R is measured at different time interval [110s;119s] | a,** PCA scores with 95% confidence ellipsoids. **b,** Individual variance for the different PC. **c,** PCA loadings of the different sensing elements' response for PC1 and PC2.

a)

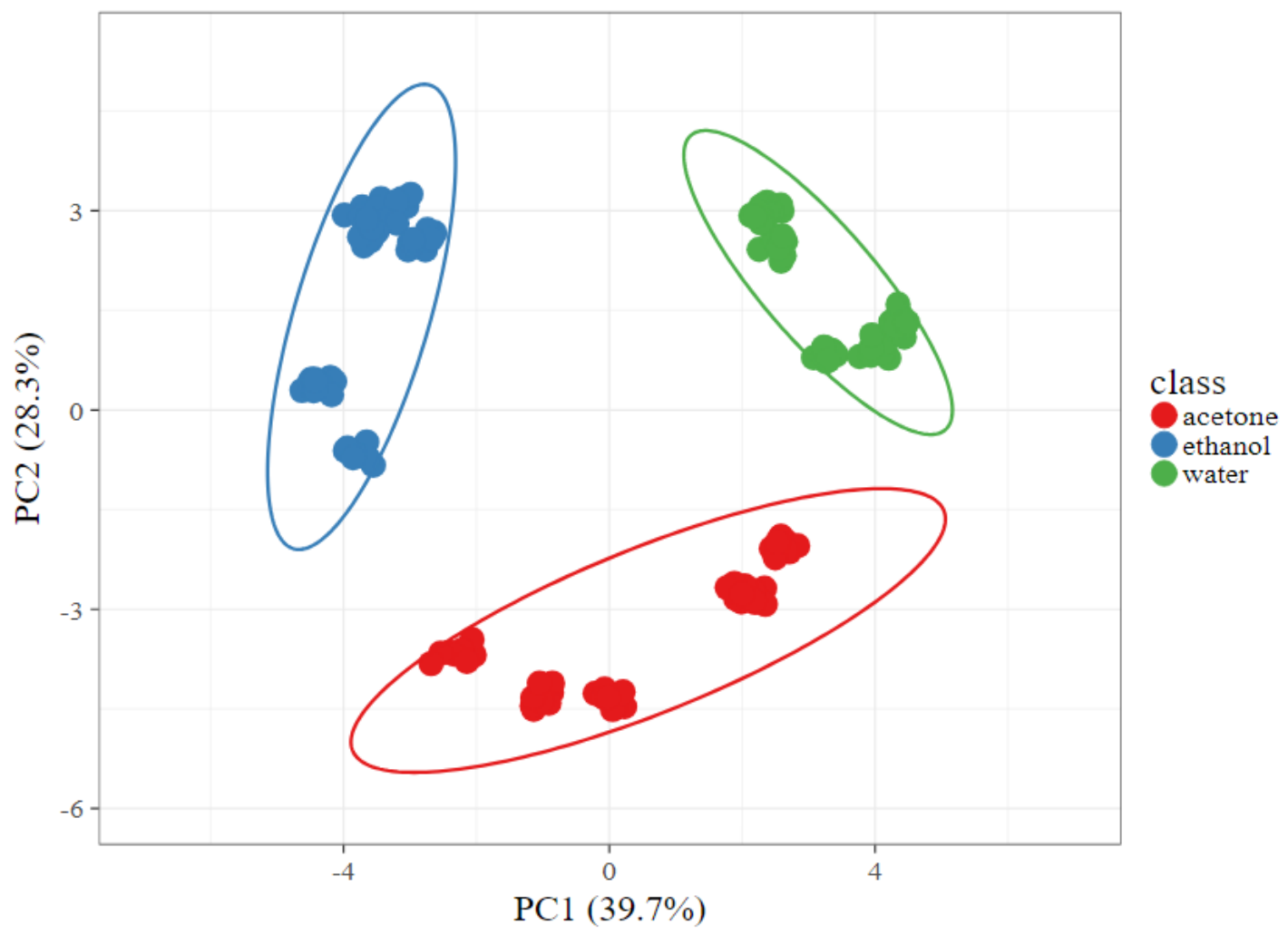

b)

| | Variance |
|---|---|
| **PC1** | 39.7% |
| **PC2** | 28.3% |
| **PC3** | 8.6% |
| **PC4** | 6.2% |
| **PC5** | 4.3% |
| **PC6** | 3.0% |
| **PC7** | 2.8% |
| **PC8** | 1.9% |
| **PC9** | 1.4% |
| **PC10** | 1.3% |
| **PC11** | 0.7% |
| **PC12** | 0.5% |
| **PC13** | 0.4% |
| **PC14** | 0.2% |
| **PC15** | 0.2% |
| **PC16** | 0.1% |
| **PC17** | 0.1% |
| **PC18** | 0.1% |
| **PC19** | 0.1% |
| **PC20** | 0.0% |
| **PC21** | 0.0% |
| **PC22** | 0.0% |
| **PC23** | 0.0% |
| **PC24** | 0.0% |

c)

| | | Loadings | |
|---|---|---|---|
| | | PC1 | PC2 |
| $Fe(OTf)_3$ | #1 | -0.14 | 0.14 |
| | #2 | -0.03 | 0.27 |
| | #3 | -0.13 | 0.16 |
| $Cu(OTf)_2$ | #1 | 0.25 | -0.06 |
| | #2 | 0.25 | 0.14 |
| | #3 | 0.24 | -0.08 |
| $Bi(OTf)_3$ | #1 | -0.17 | -0.20 |
| | #2 | 0.05 | 0.34 |
| | #3 | -0.24 | -0.14 |
| $Al(OTf)_3$ | #1 | 0.24 | 0.16 |
| | #2 | 0.26 | -0.03 |
| | #3 | 0.27 | 0.08 |
| $In(OTf)_3$ | #1 | 0.15 | -0.05 |
| | #2 | 0.15 | 0.26 |
| | #3 | 0.20 | 0.15 |
| $Dy(OTf)_3$ | #1 | 0.29 | 0.14 |
| | #2 | 0.25 | 0.20 |
| | #3 | -0.04 | 0.30 |
| $Ce(OTf)_3$ | #1 | 0.31 | -0.04 |
| | #2 | 0.22 | -0.20 |
| | #3 | 0.26 | -0.18 |
| no triflate | #1 | 0.13 | -0.33 |
| | #2 | 0.13 | -0.33 |
| | #3 | 0.10 | -0.35 |

**Figure S13. PCA on $\alpha_1$, for R is measured at different time interval [120s;129s] | a,** PCA scores with 95% confidence ellipsoids. **b,** Individual variance for the different PC. **c,** PCA loadings of the different sensing elements' response for PC1 and PC2.

a)

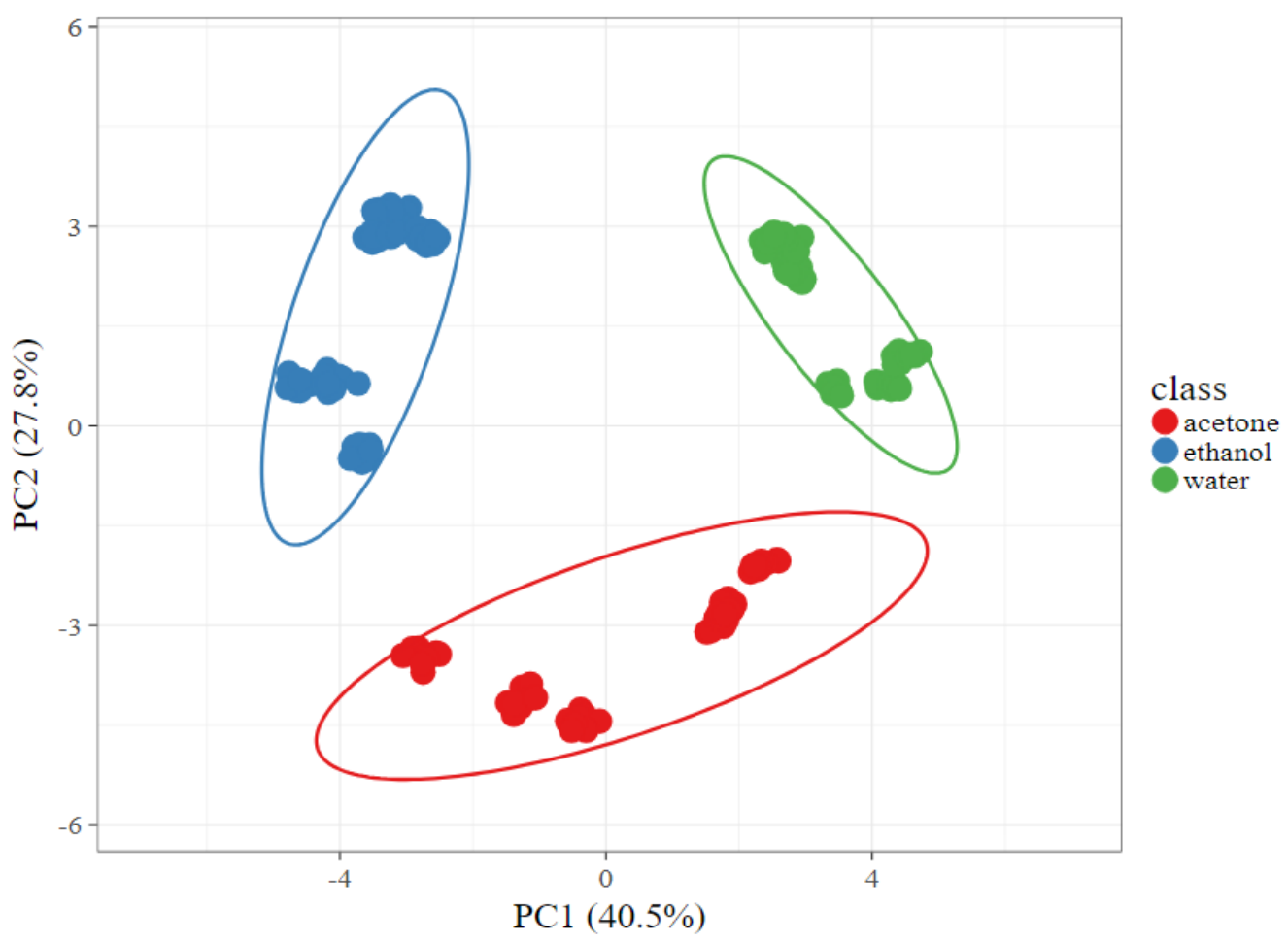

b)

| | Variance |
|---|---|
| **PC1** | 40.5% |
| **PC2** | 27.8% |
| **PC3** | 8.9% |
| **PC4** | 6.5% |
| **PC5** | 3.9% |
| **PC6** | 3.0% |
| **PC7** | 2.6% |
| **PC8** | 1.7% |
| **PC9** | 1.5% |
| **PC10** | 1.1% |
| **PC11** | 0.7% |
| **PC12** | 0.5% |
| **PC13** | 0.3% |
| **PC14** | 0.2% |
| **PC15** | 0.2% |
| **PC16** | 0.1% |
| **PC17** | 0.1% |
| **PC18** | 0.1% |
| **PC19** | 0.1% |
| **PC20** | 0.0% |
| **PC21** | 0.0% |
| **PC22** | 0.0% |
| **PC23** | 0.0% |
| **PC24** | 0.0% |

c)

| | | Loadings | |
|---|---|---|---|
| | | PC1 | PC2 |
| Fe(OTf)$_3$ | #1 | -0.14 | 0.14 |
| | #2 | -0.05 | 0.25 |
| | #3 | -0.14 | 0.15 |
| Cu(OTf)$_2$ | #1 | 0.25 | -0.03 |
| | #2 | 0.25 | 0.13 |
| | #3 | 0.25 | -0.06 |
| Bi(OTf)$_3$ | #1 | -0.14 | -0.23 |
| | #2 | 0.03 | 0.35 |
| | #3 | -0.23 | -0.17 |
| Al(OTf)$_3$ | #1 | 0.24 | 0.18 |
| | #2 | 0.26 | -0.03 |
| | #3 | 0.28 | 0.07 |
| In(OTf)$_3$ | #1 | 0.16 | -0.03 |
| | #2 | 0.14 | 0.25 |
| | #3 | 0.19 | 0.17 |
| Dy(OTf)$_3$ | #1 | 0.28 | 0.16 |
| | #2 | 0.24 | 0.22 |
| | #3 | -0.07 | 0.30 |
| Ce(OTf)$_3$ | #1 | 0.30 | -0.03 |
| | #2 | 0.22 | -0.18 |
| | #3 | 0.27 | -0.16 |
| no triflate | #1 | 0.16 | -0.31 |
| | #2 | 0.16 | -0.32 |
| | #3 | 0.13 | -0.34 |

**Figure S14. PCA on $\alpha_1$, for R is measured at different time interval [130s;139s] | a,** PCA scores with 95% confidence ellipsoids. **b,** Individual variance for the different PC. **c,** PCA loadings of the different sensing elements' response for PC1 and PC2.

a)

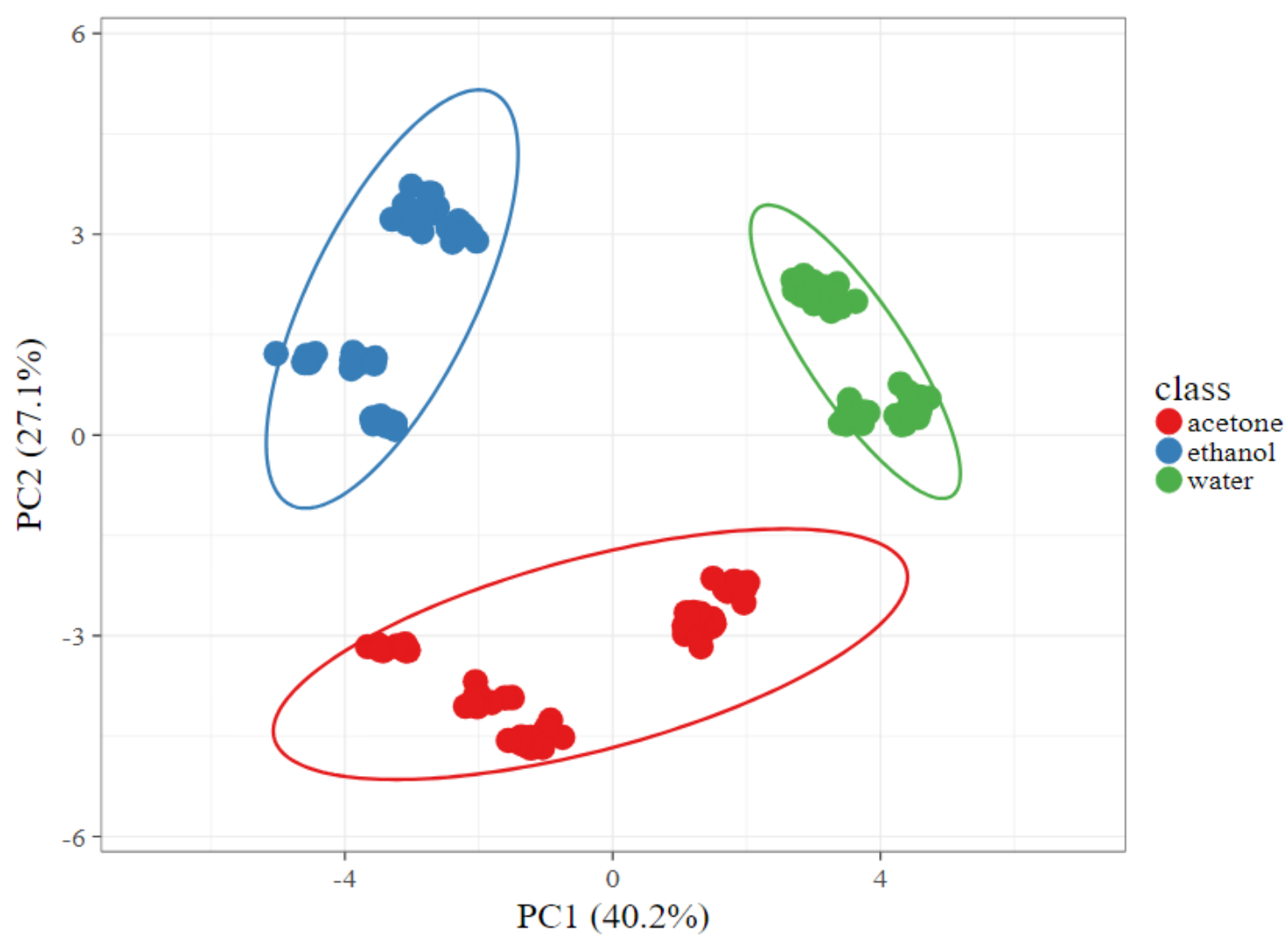

b)

| | Variance |
|---|---|
| **PC1** | 40.2% |
| **PC2** | 27.1% |
| **PC3** | 8.9% |
| **PC4** | 6.6% |
| **PC5** | 4.1% |
| **PC6** | 3.3% |
| **PC7** | 2.6% |
| **PC8** | 2.1% |
| **PC9** | 1.7% |
| **PC10** | 1.1% |
| **PC11** | 0.7% |
| **PC12** | 0.4% |
| **PC13** | 0.4% |
| **PC14** | 0.3% |
| **PC15** | 0.2% |
| **PC16** | 0.1% |
| **PC17** | 0.1% |
| **PC18** | 0.1% |
| **PC19** | 0.1% |
| **PC20** | 0.0% |
| **PC21** | 0.0% |
| **PC22** | 0.0% |
| **PC23** | 0.0% |
| **PC24** | 0.0% |

c)

| | | Loadings | |
|---|---|---|---|
| | | PC1 | PC2 |
| $Fe(OTf)_3$ | #1 | -0.15 | 0.11 |
| | #2 | -0.07 | 0.22 |
| | #3 | -0.16 | 0.11 |
| $Cu(OTf)_2$ | #1 | 0.25 | -0.01 |
| | #2 | 0.23 | 0.16 |
| | #3 | 0.25 | -0.04 |
| $Bi(OTf)_3$ | #1 | -0.11 | -0.26 |
| | #2 | -0.01 | 0.36 |
| | #3 | -0.20 | -0.22 |
| $Al(OTf)_3$ | #1 | 0.23 | 0.19 |
| | #2 | 0.25 | -0.01 |
| | #3 | 0.27 | 0.08 |
| $In(OTf)_3$ | #1 | 0.18 | -0.07 |
| | #2 | 0.12 | 0.26 |
| | #3 | 0.18 | 0.18 |
| $Dy(OTf)_3$ | #1 | 0.26 | 0.21 |
| | #2 | 0.22 | 0.26 |
| | #3 | -0.11 | 0.28 |
| $Ce(OTf)_3$ | #1 | 0.30 | 0.00 |
| | #2 | 0.23 | -0.15 |
| | #3 | 0.28 | -0.12 |
| no triflate | #1 | 0.19 | -0.29 |
| | #2 | 0.19 | -0.30 |
| | #3 | 0.17 | -0.32 |

**Figure S15. PCA on $\alpha_1$, for R is measured at different time interval [140s;149s] | a,** PCA scores with 95% confidence ellipsoids. **b,** Individual variance for the different PC. **c,** PCA loadings of the different sensing elements' response for PC1 and PC2.

a)

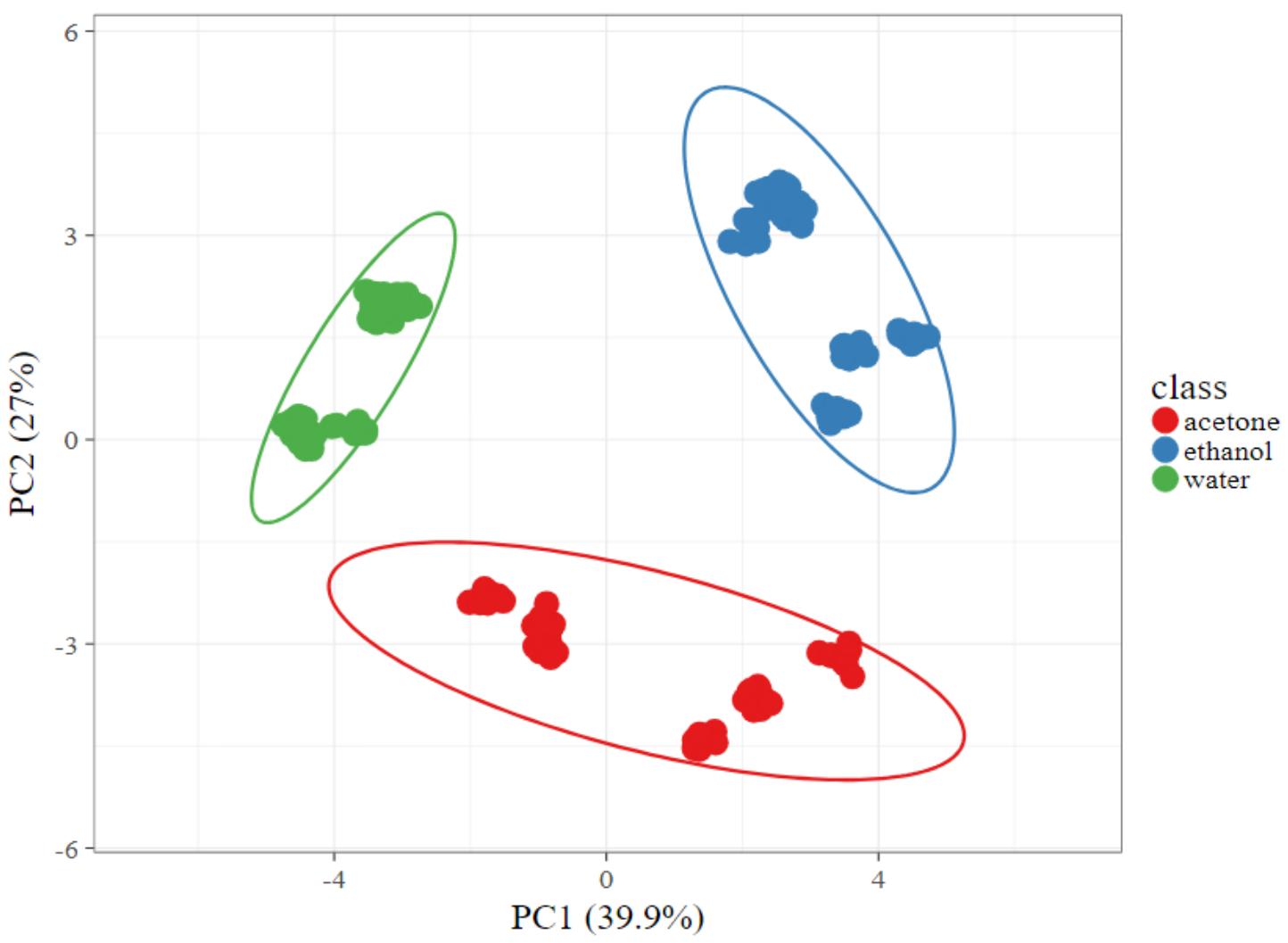

b)

| | Variance |
|---|---|
| **PC1** | 39.9% |
| **PC2** | 27.0% |
| **PC3** | 8.8% |
| **PC4** | 6.4% |
| **PC5** | 4.4% |
| **PC6** | 3.4% |
| **PC7** | 2.9% |
| **PC8** | 2.1% |
| **PC9** | 1.8% |
| **PC10** | 1.2% |
| **PC11** | 0.7% |
| **PC12** | 0.4% |
| **PC13** | 0.4% |
| **PC14** | 0.3% |
| **PC15** | 0.2% |
| **PC16** | 0.1% |
| **PC17** | 0.1% |
| **PC18** | 0.1% |
| **PC19** | 0.0% |
| **PC20** | 0.0% |
| **PC21** | 0.0% |
| **PC22** | 0.0% |
| **PC23** | 0.0% |
| **PC24** | 0.0% |

c)

| | | Loadings | |
|---|---|---|---|
| | | PC1 | PC2 |
| $Fe(OTf)_3$ | #1 | -0.14 | 0.10 |
| | #2 | -0.09 | 0.19 |
| | #3 | -0.16 | 0.09 |
| $Cu(OTf)_2$ | #1 | 0.25 | -0.01 |
| | #2 | 0.24 | 0.15 |
| | #3 | 0.24 | -0.08 |
| $Bi(OTf)_3$ | #1 | -0.09 | -0.28 |
| | #2 | -0.03 | 0.35 |
| | #3 | -0.18 | -0.25 |
| $Al(OTf)_3$ | #1 | 0.22 | 0.19 |
| | #2 | 0.25 | -0.02 |
| | #3 | 0.28 | 0.08 |
| $In(OTf)_3$ | #1 | 0.18 | -0.04 |
| | #2 | 0.11 | 0.28 |
| | #3 | 0.18 | 0.20 |
| $Dy(OTf)_3$ | #1 | 0.25 | 0.22 |
| | #2 | 0.20 | 0.28 |
| | #3 | -0.10 | 0.27 |
| $Ce(OTf)_3$ | #1 | 0.31 | 0.00 |
| | #2 | 0.24 | -0.14 |
| | #3 | 0.29 | -0.11 |
| no triflate | #1 | 0.21 | -0.28 |
| | #2 | 0.20 | -0.28 |
| | #3 | 0.19 | -0.31 |

**Figure S16. PCA on $\alpha_1$, for R is measured at different time interval [150s;159s] | a,** PCA scores with 95% confidence ellipsoids. **b,** Individual variance for the different PC. **c,** PCA loadings of the different sensing elements' response for PC1 and PC2.

a)

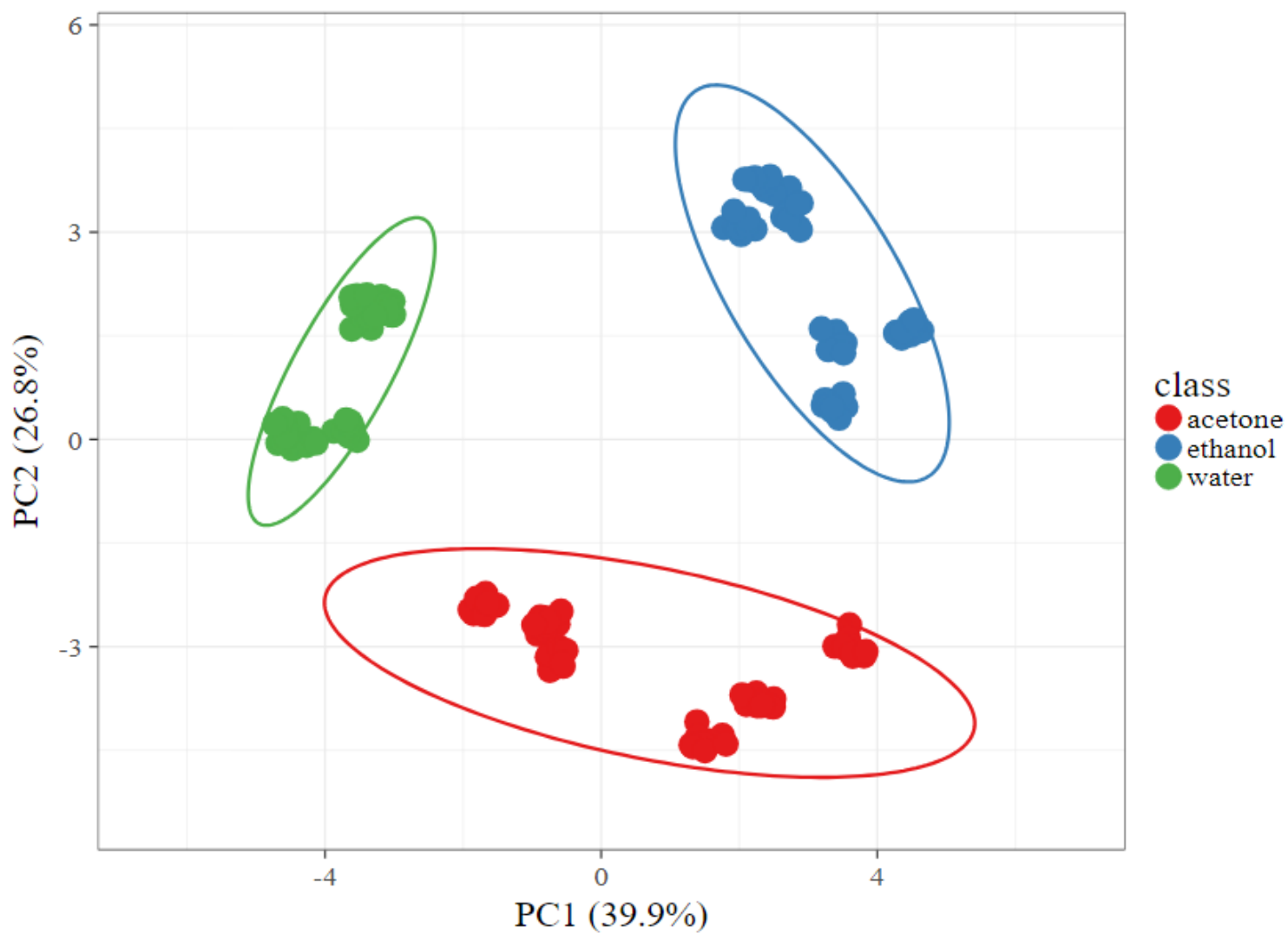

b)

| | Variance |
|---|---|
| **PC1** | 39.9% |
| **PC2** | 26.8% |
| **PC3** | 8.9% |
| **PC4** | 6.3% |
| **PC5** | 4.3% |
| **PC6** | 3.4% |
| **PC7** | 2.9% |
| **PC8** | 2.0% |
| **PC9** | 1.7% |
| **PC10** | 1.4% |
| **PC11** | 0.7% |
| **PC12** | 0.5% |
| **PC13** | 0.4% |
| **PC14** | 0.3% |
| **PC15** | 0.2% |
| **PC16** | 0.1% |
| **PC17** | 0.1% |
| **PC18** | 0.1% |
| **PC19** | 0.0% |
| **PC20** | 0.0% |
| **PC21** | 0.0% |
| **PC22** | 0.0% |
| **PC23** | 0.0% |
| **PC24** | 0.0% |

c)

| | | Loadings | |
|---|---|---|---|
| | | PC1 | PC2 |
| $Fe(OTf)_3$ | #1 | -0.14 | 0.07 |
| | #2 | -0.09 | 0.18 |
| | #3 | -0.17 | 0.07 |
| $Cu(OTf)_2$ | #1 | 0.25 | -0.01 |
| | #2 | 0.23 | 0.17 |
| | #3 | 0.24 | -0.09 |
| $Bi(OTf)_3$ | #1 | -0.07 | -0.29 |
| | #2 | -0.03 | 0.36 |
| | #3 | -0.17 | -0.27 |
| $Al(OTf)_3$ | #1 | 0.22 | 0.18 |
| | #2 | 0.24 | -0.02 |
| | #3 | 0.28 | 0.10 |
| $In(OTf)_3$ | #1 | 0.17 | -0.03 |
| | #2 | 0.12 | 0.27 |
| | #3 | 0.18 | 0.20 |
| $Dy(OTf)_3$ | #1 | 0.25 | 0.23 |
| | #2 | 0.20 | 0.28 |
| | #3 | -0.12 | 0.26 |
| $Ce(OTf)_3$ | #1 | 0.31 | -0.01 |
| | #2 | 0.24 | -0.13 |
| | #3 | 0.29 | -0.11 |
| no triflate | #1 | 0.21 | -0.28 |
| | #2 | 0.21 | -0.28 |
| | #3 | 0.19 | -0.31 |

**Figure S17. PCA on $\alpha_1$, for R is measured at different time interval [160s;169s] | a,** PCA scores with 95% confidence ellipsoids. **b,** Individual variance for the different PC. **c,** PCA loadings of the different sensing elements' response for PC1 and PC2.

a)

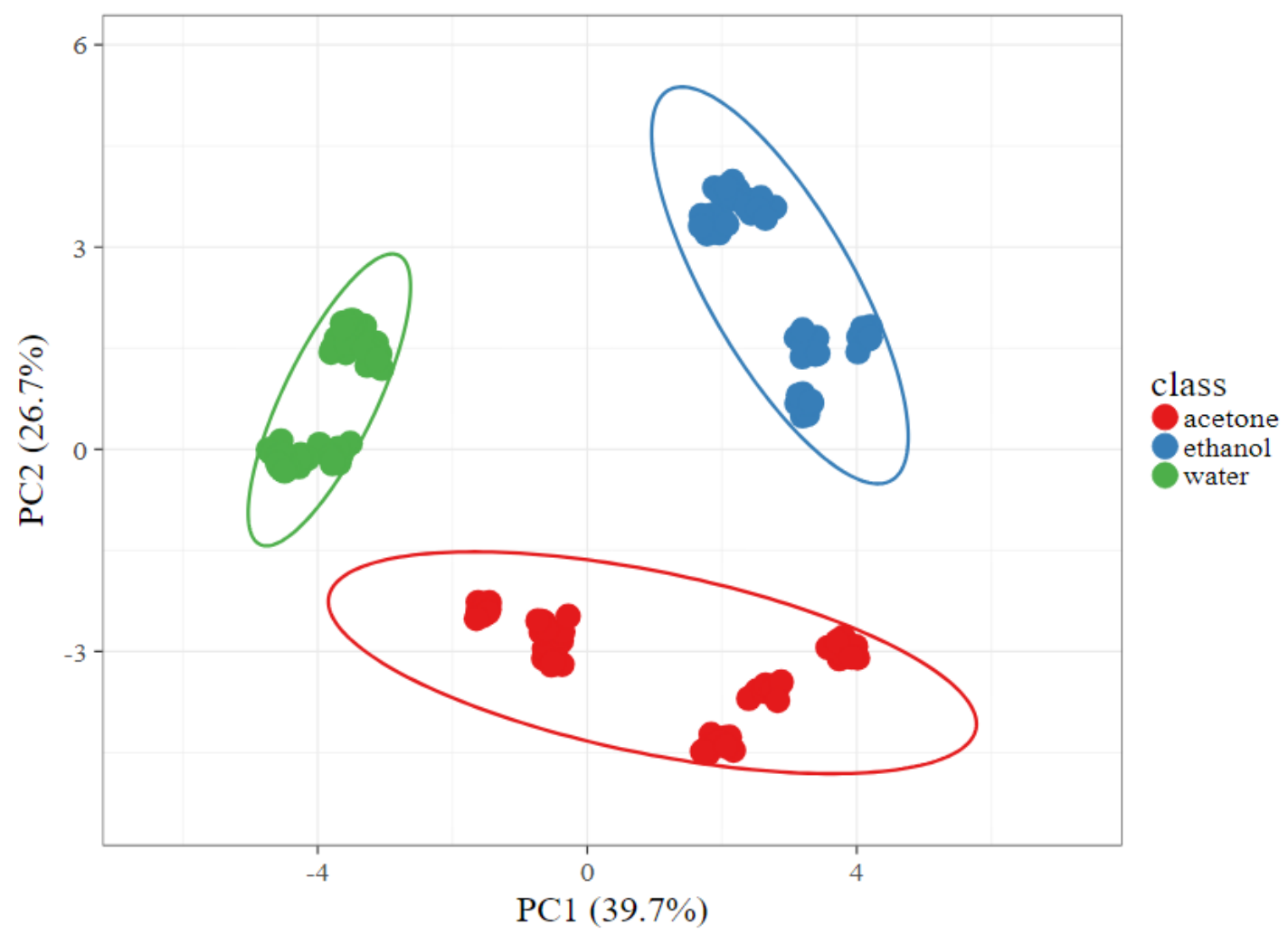

b)

| | Variance |
|---|---|
| **PC1** | 39.7% |
| **PC2** | 26.7% |
| **PC3** | 9.2% |
| **PC4** | 6.5% |
| **PC5** | 4.5% |
| **PC6** | 3.2% |
| **PC7** | 2.9% |
| **PC8** | 1.9% |
| **PC9** | 1.7% |
| **PC10** | 1.3% |
| **PC11** | 0.7% |
| **PC12** | 0.4% |
| **PC13** | 0.3% |
| **PC14** | 0.2% |
| **PC15** | 0.1% |
| **PC16** | 0.1% |
| **PC17** | 0.1% |
| **PC18** | 0.1% |
| **PC19** | 0.0% |
| **PC20** | 0.0% |
| **PC21** | 0.0% |
| **PC22** | 0.0% |
| **PC23** | 0.0% |
| **PC24** | 0.0% |

c)

| | | Loadings | |
|---|---|---|---|
| | | PC1 | PC2 |
| $Fe(OTf)_3$ | #1 | -0.13 | 0.04 |
| | #2 | -0.09 | 0.16 |
| | #3 | -0.16 | 0.05 |
| $Cu(OTf)_2$ | #1 | 0.25 | 0.00 |
| | #2 | 0.22 | 0.21 |
| | #3 | 0.24 | -0.10 |
| $Bi(OTf)_3$ | #1 | -0.05 | -0.30 |
| | #2 | -0.05 | 0.35 |
| | #3 | -0.15 | -0.28 |
| $Al(OTf)_3$ | #1 | 0.21 | 0.20 |
| | #2 | 0.24 | -0.03 |
| | #3 | 0.28 | 0.10 |
| $In(OTf)_3$ | #1 | 0.18 | -0.06 |
| | #2 | 0.11 | 0.26 |
| | #3 | 0.18 | 0.21 |
| $Dy(OTf)_3$ | #1 | 0.24 | 0.25 |
| | #2 | 0.19 | 0.30 |
| | #3 | -0.11 | 0.26 |
| $Ce(OTf)_3$ | #1 | 0.31 | 0.00 |
| | #2 | 0.25 | -0.12 |
| | #3 | 0.29 | -0.10 |
| no triflate | #1 | 0.23 | -0.25 |
| | #2 | 0.23 | -0.26 |
| | #3 | 0.21 | -0.29 |

**Figure S18. PCA on $\alpha_1$, for R is measured at different time interval [170s;179s] | a,** PCA scores with 95% confidence ellipsoids. **b,** Individual variance for the different PC. **c,** PCA loadings of the different sensing elements' response for PC1 and PC2.

a)

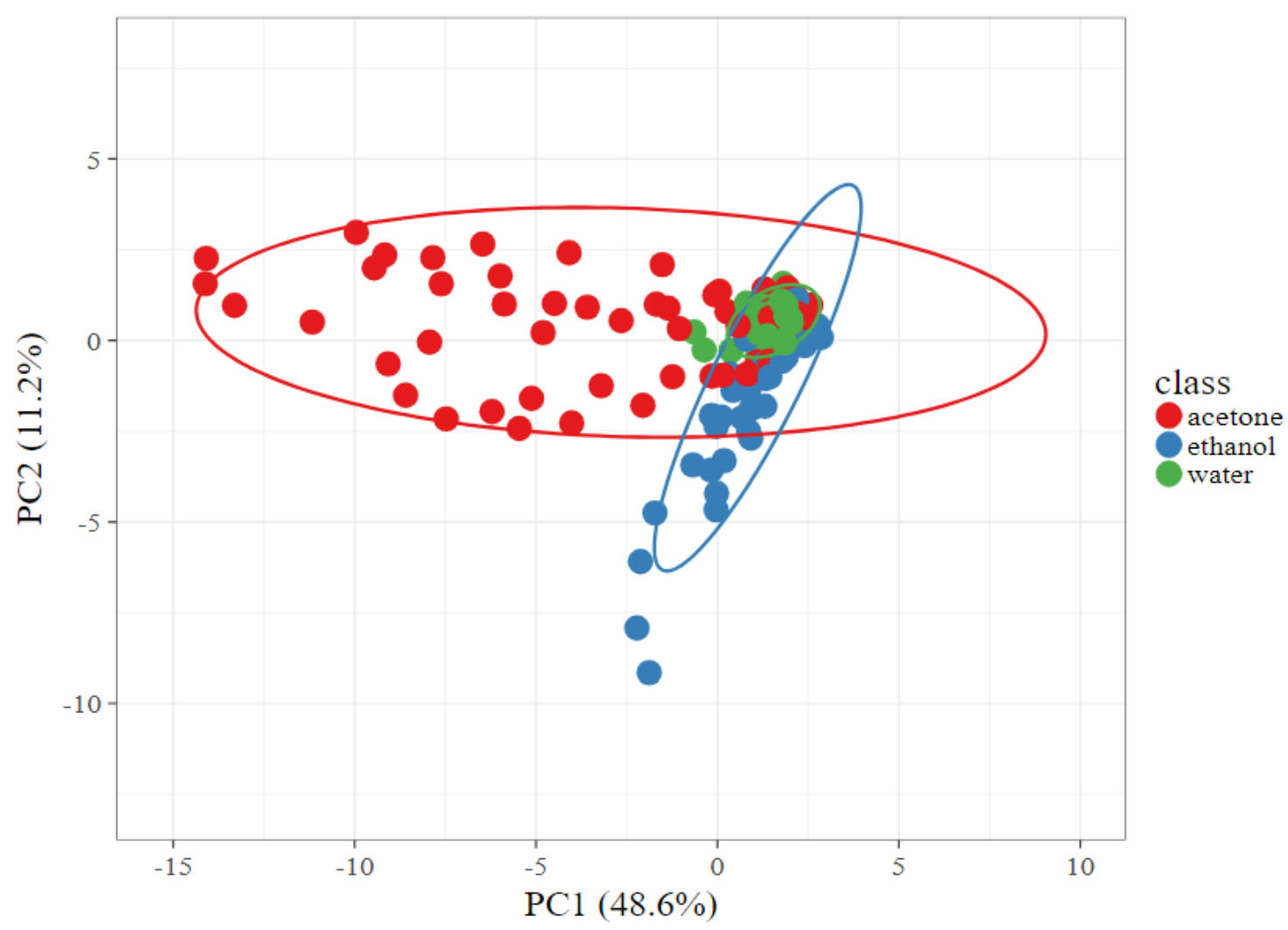

b)

| | Variance |
|---|---|
| **PC1** | 48.6% |
| **PC2** | 11.2% |
| **PC3** | 6.8% |
| **PC4** | 6.3% |
| **PC5** | 4.3% |
| **PC6** | 4.1% |
| **PC7** | 3.6% |
| **PC8** | 2.9% |
| **PC9** | 2.4% |
| **PC10** | 1.8% |
| **PC11** | 1.5% |
| **PC12** | 1.3% |
| **PC13** | 0.9% |
| **PC14** | 0.9% |
| **PC15** | 0.7% |
| **PC16** | 0.7% |
| **PC17** | 0.6% |
| **PC18** | 0.3% |
| **PC19** | 0.3% |
| **PC20** | 0.2% |
| **PC21** | 0.2% |
| **PC22** | 0.1% |
| **PC23** | 0.1% |
| **PC24** | 0.1% |

c)

| | | Loadings | |
|---|---|---|---|
| | | PC1 | PC2 |
| $Fe(OTf)_3$ | #1 | 0.24 | 0.10 |
| | #2 | 0.27 | 0.07 |
| | #3 | 0.25 | 0.03 |
| $Cu(OTf)_2$ | #1 | 0.19 | -0.13 |
| | #2 | 0.24 | 0.08 |
| | #3 | 0.23 | 0.21 |
| $Bi(OTf)_3$ | #1 | 0.18 | 0.36 |
| | #2 | 0.25 | -0.19 |
| | #3 | 0.23 | 0.22 |
| $Al(OTf)_3$ | #1 | 0.25 | -0.08 |
| | #2 | 0.18 | -0.03 |
| | #3 | 0.26 | 0.06 |
| $In(OTf)_3$ | #1 | 0.21 | -0.09 |
| | #2 | 0.21 | 0.09 |
| | #3 | 0.21 | 0.01 |
| $Dy(OTf)_3$ | #1 | 0.18 | -0.19 |
| | #2 | 0.23 | -0.13 |
| | #3 | 0.18 | 0.14 |
| $Ce(OTf)_3$ | #1 | 0.09 | -0.44 |
| | #2 | 0.07 | -0.43 |
| | #3 | 0.20 | -0.35 |
| no triflate | #1 | -0.12 | -0.27 |
| | #2 | -0.20 | -0.01 |
| | #3 | 0.07 | -0.15 |

**Figure S19. PCA on $\alpha_2$, for R is measured at different time interval [1s;9s] | a,** PCA scores with 95% confidence ellipsoids. **b,** Individual variance for the different PC. **c,** PCA loadings of the different sensing elements' response for PC1 and PC2.

a)

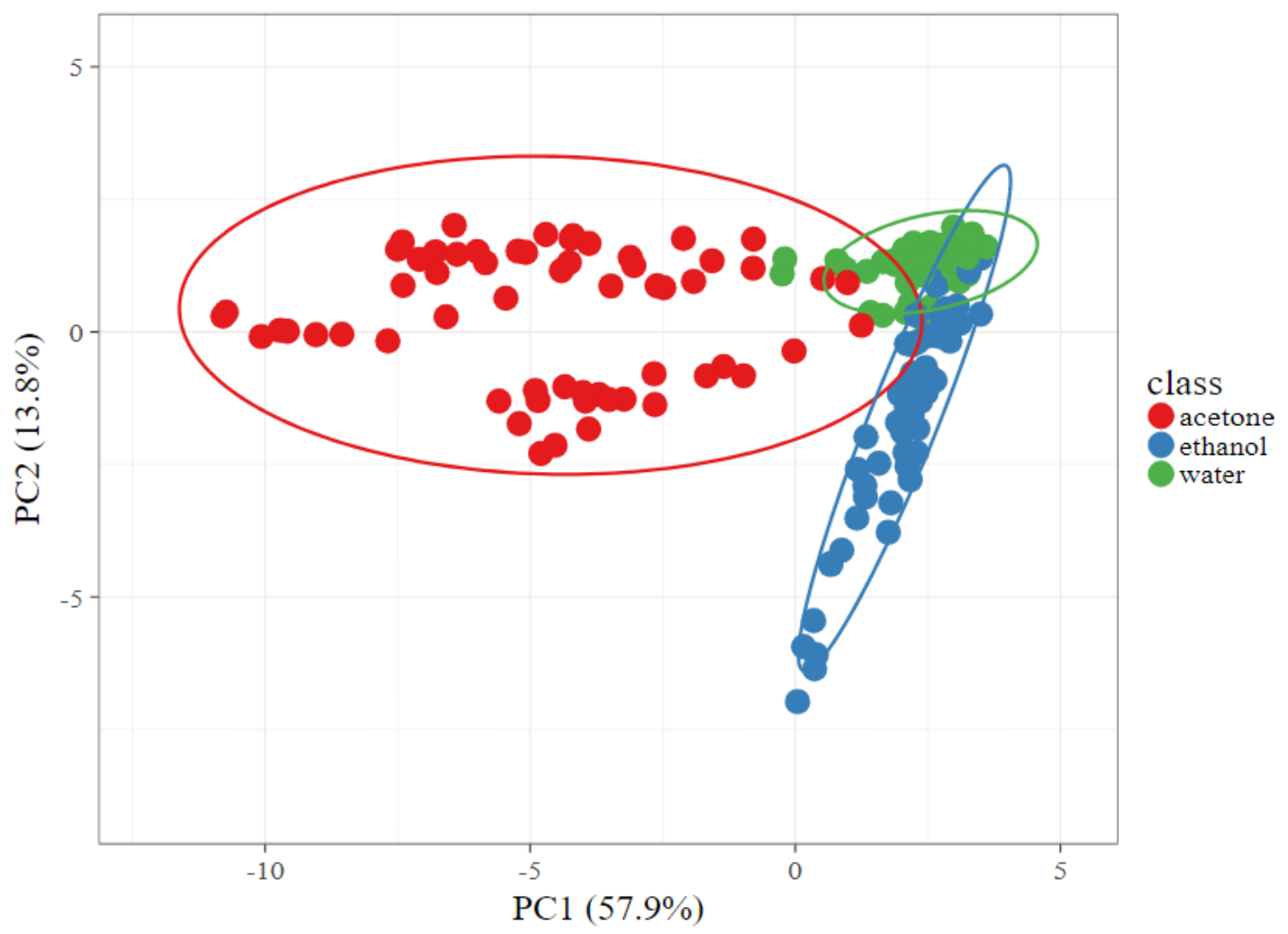

b)

| | Variance |
|---|---|
| **PC1** | 57.9% |
| **PC2** | 13.8% |
| **PC3** | 6.7% |
| **PC4** | 4.9% |
| **PC5** | 2.9% |
| **PC6** | 2.6% |
| **PC7** | 2.1% |
| **PC8** | 1.9% |
| **PC9** | 1.5% |
| **PC10** | 1.1% |
| **PC11** | 1.0% |
| **PC12** | 0.7% |
| **PC13** | 0.6% |
| **PC14** | 0.5% |
| **PC15** | 0.4% |
| **PC16** | 0.4% |
| **PC17** | 0.3% |
| **PC18** | 0.3% |
| **PC19** | 0.2% |
| **PC20** | 0.1% |
| **PC21** | 0.1% |
| **PC22** | 0.1% |
| **PC23** | 0.1% |
| **PC24** | 0.0% |

c)

| | | Loadings | |
|---|---|---|---|
| | | PC1 | PC2 |
| $Fe(OTf)_3$ | #1 | 0.20 | 0.15 |
| | #2 | 0.24 | 0.14 |
| | #3 | 0.23 | 0.14 |
| $Cu(OTf)_2$ | #1 | 0.20 | -0.18 |
| | #2 | 0.24 | 0.02 |
| | #3 | 0.21 | 0.12 |
| $Bi(OTf)_3$ | #1 | 0.18 | 0.35 |
| | #2 | 0.21 | -0.15 |
| | #3 | 0.18 | 0.30 |
| $Al(OTf)_3$ | #1 | 0.25 | -0.06 |
| | #2 | 0.24 | -0.06 |
| | #3 | 0.26 | 0.04 |
| $In(OTf)_3$ | #1 | 0.21 | -0.11 |
| | #2 | 0.25 | 0.00 |
| | #3 | 0.21 | 0.02 |
| $Dy(OTf)_3$ | #1 | 0.20 | -0.20 |
| | #2 | 0.24 | -0.12 |
| | #3 | 0.17 | 0.15 |
| $Ce(OTf)_3$ | #1 | 0.15 | -0.38 |
| | #2 | 0.09 | -0.37 |
| | #3 | 0.19 | -0.33 |
| no triflate | #1 | -0.12 | -0.33 |
| | #2 | -0.23 | -0.18 |
| | #3 | 0.06 | -0.17 |

**Figure S20. PCA on $\alpha_2$, for R is measured at different time interval [10s;19s] | a,** PCA scores with 95% confidence ellipsoids. **b,** Individual variance for the different PC. **c,** PCA loadings of the different sensing elements' response for PC1 and PC2.

a)

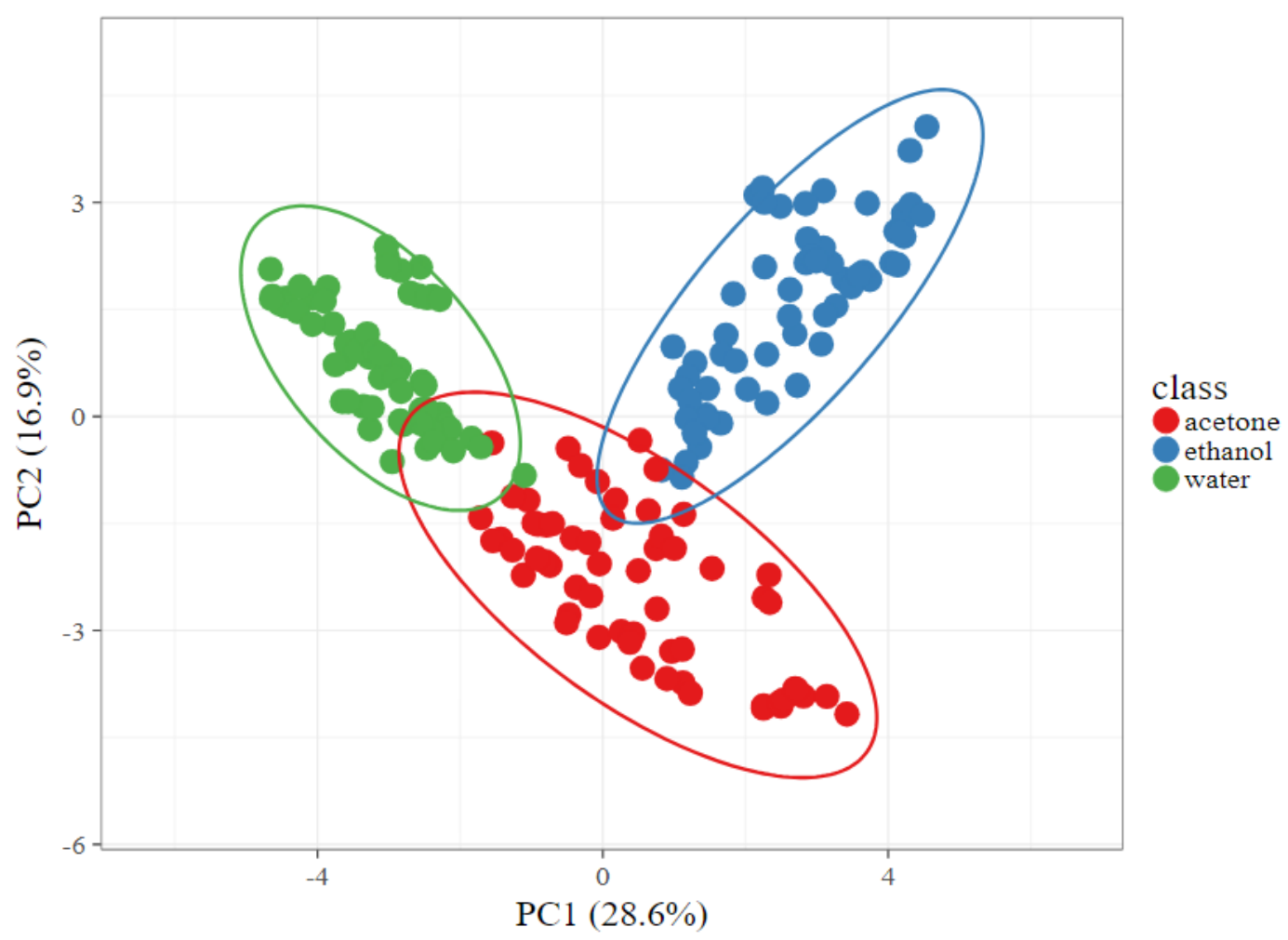

b)

| | Variance |
|---|---|
| **PC1** | 28.6% |
| **PC2** | 16.9% |
| **PC3** | 7.3% |
| **PC4** | 5.9% |
| **PC5** | 5.3% |
| **PC6** | 5.1% |
| **PC7** | 4.7% |
| **PC8** | 4.1% |
| **PC9** | 3.0% |
| **PC10** | 2.7% |
| **PC11** | 2.6% |
| **PC12** | 2.3% |
| **PC13** | 2.1% |
| **PC14** | 1.8% |
| **PC15** | 1.6% |
| **PC16** | 1.5% |
| **PC17** | 1.3% |
| **PC18** | 1.0% |
| **PC19** | 0.8% |
| **PC20** | 0.5% |
| **PC21** | 0.3% |
| **PC22** | 0.3% |
| **PC23** | 0.2% |
| **PC24** | 0.1% |

c)

| | | Loadings | |
|---|---|---|---|
| | | PC1 | PC2 |
| $Fe(OTf)_3$ | #1 | -0.11 | 0.33 |
| | #2 | -0.17 | 0.29 |
| | #3 | -0.21 | 0.25 |
| $Cu(OTf)_2$ | #1 | 0.26 | -0.01 |
| | #2 | 0.24 | 0.09 |
| | #3 | 0.14 | 0.06 |
| $Bi(OTf)_3$ | #1 | 0.04 | 0.39 |
| | #2 | 0.05 | -0.06 |
| | #3 | -0.13 | 0.40 |
| $Al(OTf)_3$ | #1 | 0.24 | 0.09 |
| | #2 | 0.19 | 0.13 |
| | #3 | 0.26 | 0.20 |
| $In(OTf)_3$ | #1 | 0.07 | 0.18 |
| | #2 | 0.07 | 0.34 |
| | #3 | 0.13 | -0.12 |
| $Dy(OTf)_3$ | #1 | 0.35 | 0.05 |
| | #2 | 0.33 | -0.01 |
| | #3 | 0.18 | 0.06 |
| $Ce(OTf)_3$ | #1 | 0.24 | -0.27 |
| | #2 | 0.28 | -0.05 |
| | #3 | 0.24 | -0.01 |
| no triflate | #1 | 0.16 | 0.13 |
| | #2 | 0.24 | 0.16 |
| | #3 | 0.12 | 0.24 |

**Figure S21. PCA on $\alpha_2$, for R is measured at different time interval [20s;29s] | a,** PCA scores with 95% confidence ellipsoids. **b,** Individual variance for the different PC. **c,** PCA loadings of the different sensing elements' response for PC1 and PC2.

a)

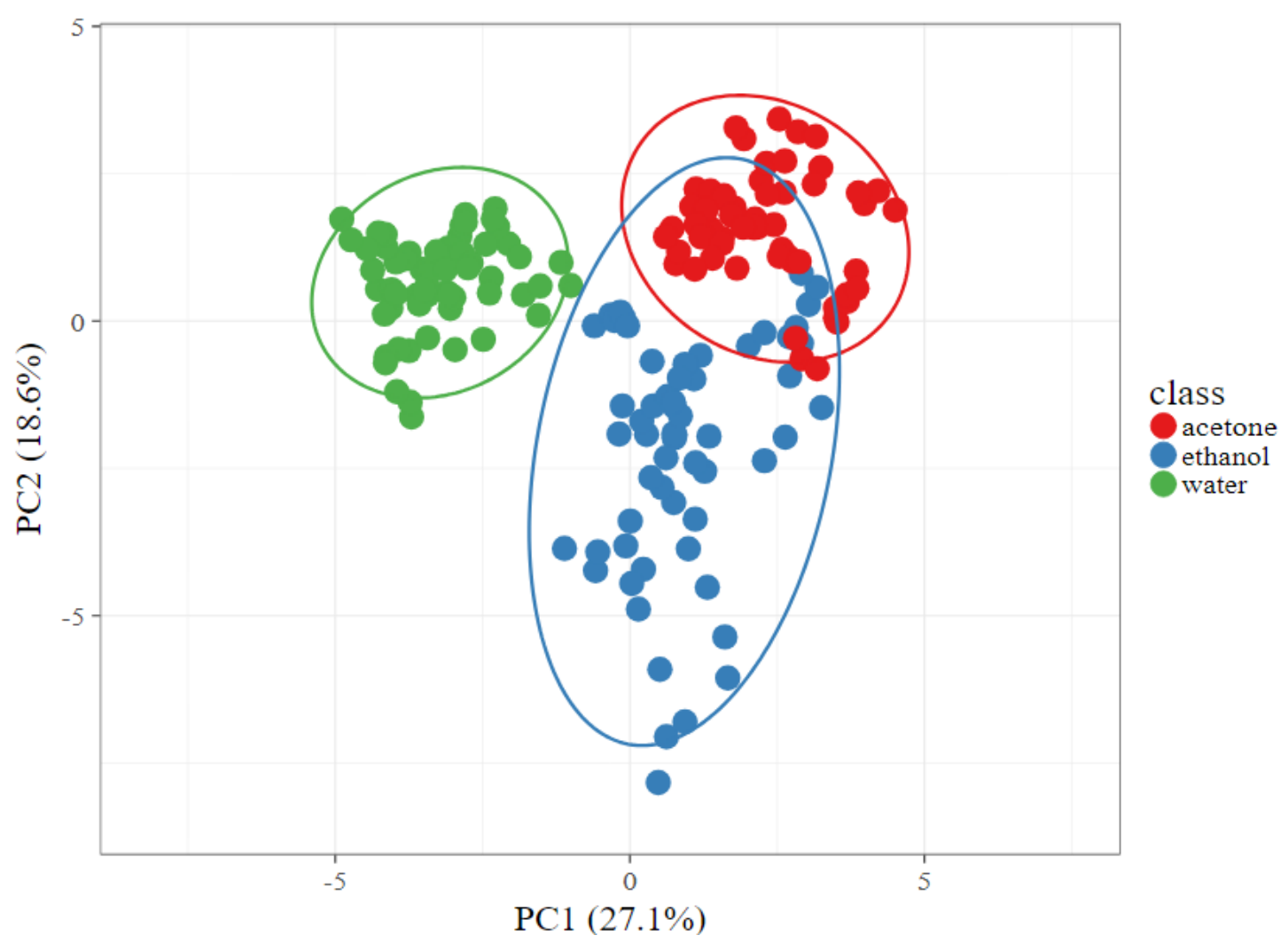

b)

| | Variance |
|---|---|
| **PC1** | 27.1% |
| **PC2** | 18.6% |
| **PC3** | 6.9% |
| **PC4** | 6.0% |
| **PC5** | 5.7% |
| **PC6** | 4.8% |
| **PC7** | 4.0% |
| **PC8** | 3.6% |
| **PC9** | 3.4% |
| **PC10** | 2.9% |
| **PC11** | 2.8% |
| **PC12** | 2.5% |
| **PC13** | 2.1% |
| **PC14** | 1.9% |
| **PC15** | 1.6% |
| **PC16** | 1.4% |
| **PC17** | 1.1% |
| **PC18** | 0.8% |
| **PC19** | 0.7% |
| **PC20** | 0.6% |
| **PC21** | 0.5% |
| **PC22** | 0.4% |
| **PC23** | 0.3% |
| **PC24** | 0.2% |

c)

| | | Loadings | |
|---|---|---|---|
| | | PC1 | PC2 |
| $Fe(OTf)_3$ | #1 | -0.10 | 0.34 |
| | #2 | -0.05 | 0.36 |
| | #3 | -0.08 | 0.39 |
| $Cu(OTf)_2$ | #1 | -0.23 | -0.09 |
| | #2 | -0.17 | 0.03 |
| | #3 | -0.16 | 0.13 |
| $Bi(OTf)_3$ | #1 | -0.19 | 0.24 |
| | #2 | 0.07 | 0.17 |
| | #3 | -0.09 | 0.39 |
| $Al(OTf)_3$ | #1 | -0.24 | 0.04 |
| | #2 | -0.28 | 0.13 |
| | #3 | -0.24 | -0.08 |
| $In(OTf)_3$ | #1 | -0.06 | 0.16 |
| | #2 | -0.11 | 0.21 |
| | #3 | -0.23 | -0.02 |
| $Dy(OTf)_3$ | #1 | -0.30 | -0.18 |
| | #2 | -0.21 | -0.23 |
| | #3 | -0.12 | 0.03 |
| $Ce(OTf)_3$ | #1 | -0.20 | -0.31 |
| | #2 | -0.30 | -0.20 |
| | #3 | -0.28 | -0.08 |
| no triflate | #1 | -0.24 | 0.03 |
| | #2 | -0.25 | -0.02 |
| | #3 | -0.28 | 0.11 |

**Figure S22. PCA on $\alpha_2$, for R is measured at different time interval [30s;39s] | a,** PCA scores with 95% confidence ellipsoids. **b,** Individual variance for the different PC. **c,** PCA loadings of the different sensing elements' response for PC1 and PC2.

a)

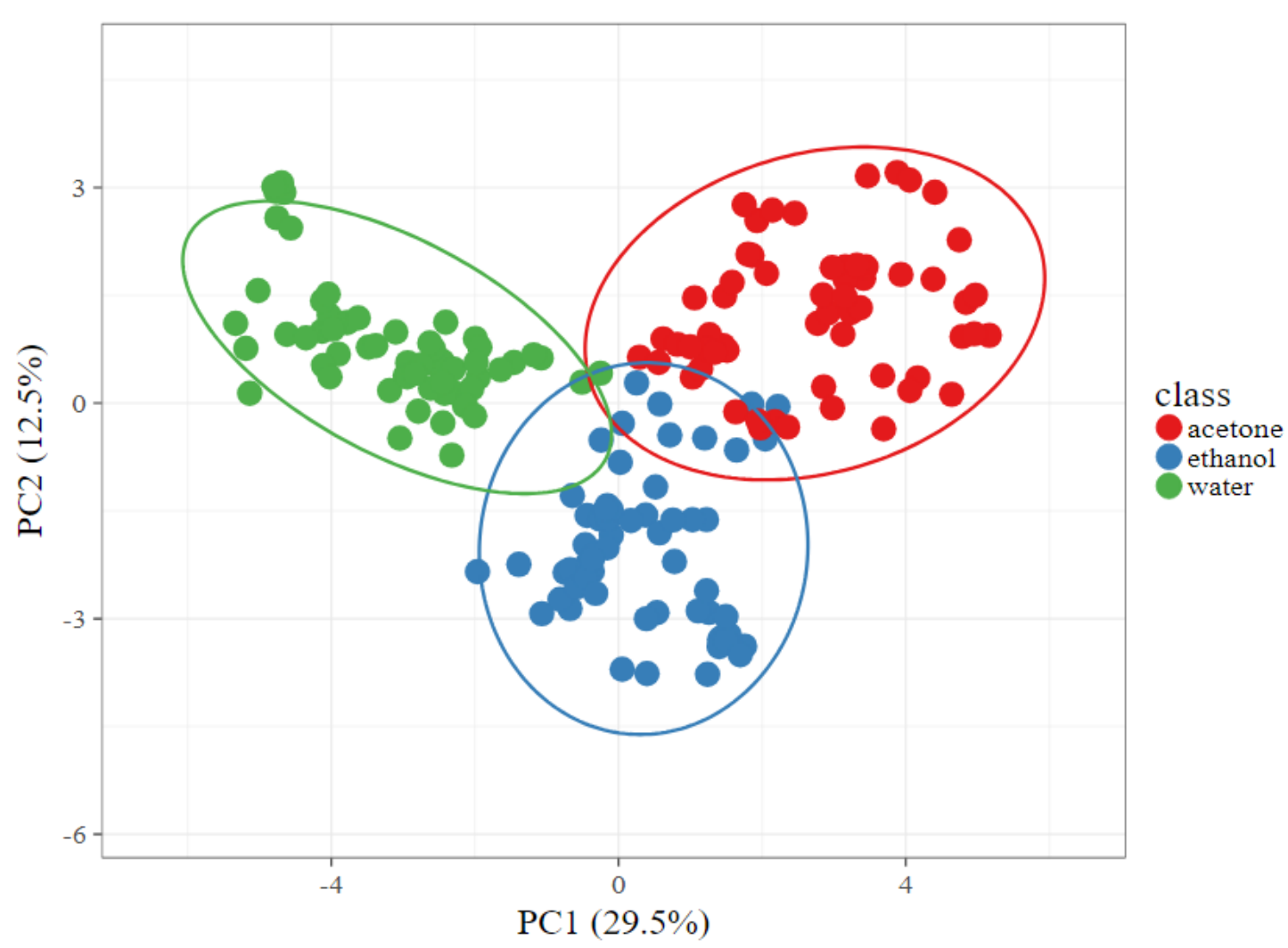

b)

| | Variance |
|---|---|
| **PC1** | 29.5% |
| **PC2** | 12.5% |
| **PC3** | 8.7% |
| **PC4** | 6.0% |
| **PC5** | 5.9% |
| **PC6** | 5.2% |
| **PC7** | 4.9% |
| **PC8** | 3.6% |
| **PC9** | 3.2% |
| **PC10** | 3.1% |
| **PC11** | 2.5% |
| **PC12** | 2.1% |
| **PC13** | 1.8% |
| **PC14** | 1.7% |
| **PC15** | 1.6% |
| **PC16** | 1.4% |
| **PC17** | 1.3% |
| **PC18** | 1.0% |
| **PC19** | 0.9% |
| **PC20** | 0.8% |
| **PC21** | 0.7% |
| **PC22** | 0.6% |
| **PC23** | 0.4% |
| **PC24** | 0.3% |

c)

| | | Loadings | |
|---|---|---|---|
| | | PC1 | PC2 |
| $Fe(OTf)_3$ | #1 | -0.18 | 0.14 |
| | #2 | -0.24 | -0.17 |
| | #3 | -0.20 | -0.40 |
| $Cu(OTf)_2$ | #1 | -0.14 | 0.16 |
| | #2 | -0.09 | 0.34 |
| | #3 | -0.12 | 0.13 |
| $Bi(OTf)_3$ | #1 | -0.23 | -0.31 |
| | #2 | 0.17 | -0.08 |
| | #3 | -0.13 | -0.44 |
| $Al(OTf)_3$ | #1 | -0.16 | -0.14 |
| | #2 | -0.26 | 0.01 |
| | #3 | -0.22 | 0.03 |
| $In(OTf)_3$ | #1 | -0.01 | 0.13 |
| | #2 | -0.18 | -0.10 |
| | #3 | -0.12 | -0.09 |
| $Dy(OTf)_3$ | #1 | -0.25 | 0.20 |
| | #2 | -0.14 | 0.30 |
| | #3 | -0.01 | -0.05 |
| $Ce(OTf)_3$ | #1 | -0.24 | 0.25 |
| | #2 | -0.31 | 0.20 |
| | #3 | -0.28 | 0.17 |
| no triflate | #1 | -0.25 | -0.05 |
| | #2 | -0.29 | -0.12 |
| | #3 | -0.27 | -0.10 |

**Figure S23. PCA on $\alpha_2$, for R is measured at different time interval [40s;49s] | a,** PCA scores with 95% confidence ellipsoids. **b,** Individual variance for the different PC. **c,** PCA loadings of the different sensing elements' response for PC1 and PC2.

a)

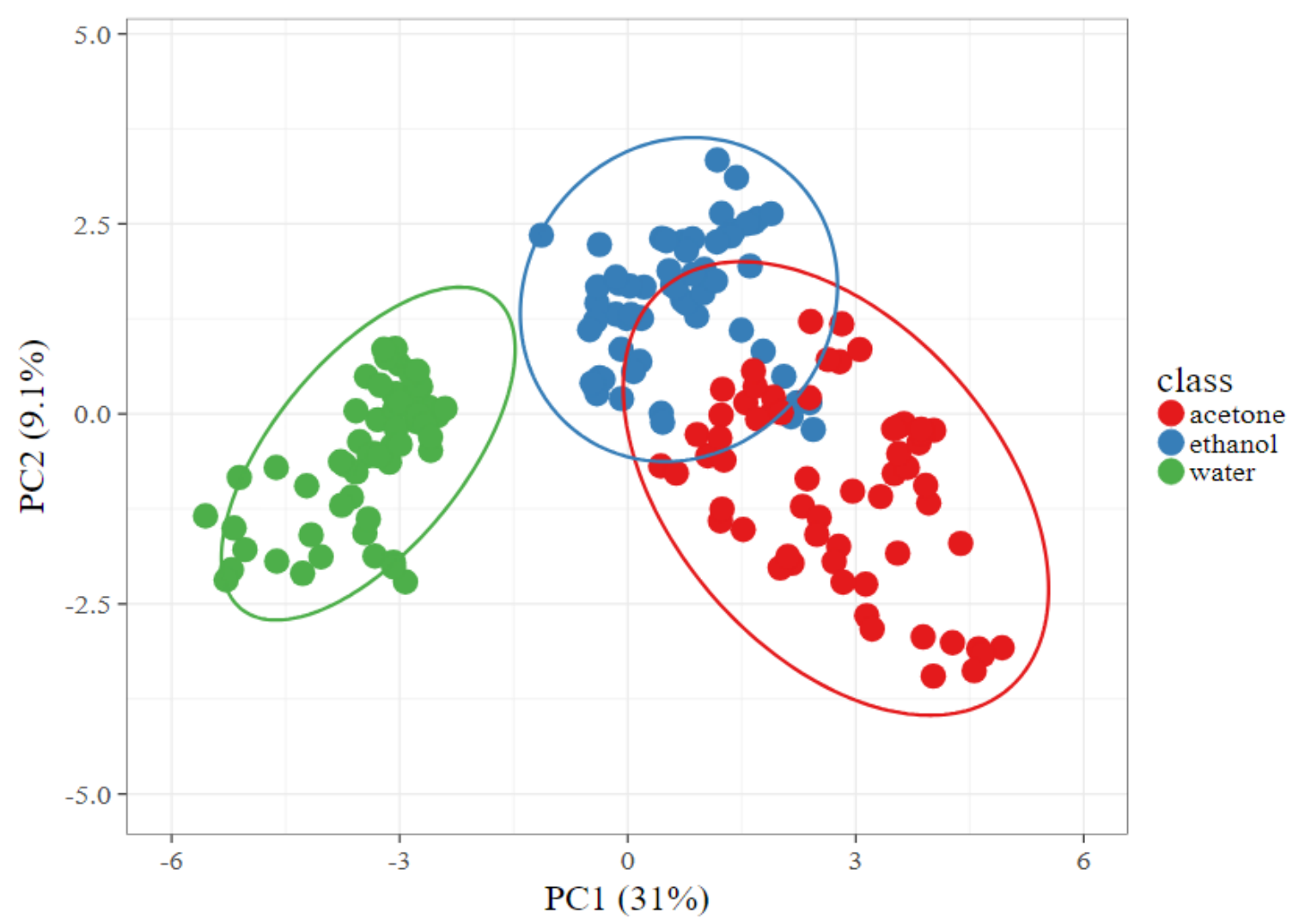

b)

| | Variance |
|---|---|
| **PC1** | 31.0% |
| **PC2** | 9.1% |
| **PC3** | 6.8% |
| **PC4** | 6.3% |
| **PC5** | 5.7% |
| **PC6** | 5.2% |
| **PC7** | 4.8% |
| **PC8** | 4.1% |
| **PC9** | 3.9% |
| **PC10** | 3.1% |
| **PC11** | 2.8% |
| **PC12** | 2.6% |
| **PC13** | 2.3% |
| **PC14** | 2.2% |
| **PC15** | 2.1% |
| **PC16** | 1.7% |
| **PC17** | 1.4% |
| **PC18** | 1.2% |
| **PC19** | 0.9% |
| **PC20** | 0.8% |
| **PC21** | 0.8% |
| **PC22** | 0.5% |
| **PC23** | 0.4% |
| **PC24** | 0.3% |

c)

| | | Loadings | |
|---|---|---|---|
| | | PC1 | PC2 |
| $Fe(OTf)_3$ | #1 | -0.18 | 0.09 |
| | #2 | -0.32 | -0.05 |
| | #3 | -0.19 | -0.26 |
| $Cu(OTf)_2$ | #1 | -0.07 | 0.08 |
| | #2 | 0.01 | -0.04 |
| | #3 | -0.09 | -0.03 |
| $Bi(OTf)_3$ | #1 | -0.21 | 0.44 |
| | #2 | 0.21 | 0.11 |
| | #3 | -0.15 | 0.48 |
| $Al(OTf)_3$ | #1 | -0.21 | -0.06 |
| | #2 | -0.19 | 0.19 |
| | #3 | -0.23 | 0.25 |
| $In(OTf)_3$ | #1 | -0.13 | 0.14 |
| | #2 | -0.17 | -0.29 |
| | #3 | -0.17 | -0.05 |
| $Dy(OTf)_3$ | #1 | -0.22 | -0.24 |
| | #2 | -0.14 | -0.04 |
| | #3 | -0.11 | 0.08 |
| $Ce(OTf)_3$ | #1 | -0.24 | -0.22 |
| | #2 | -0.31 | -0.21 |
| | #3 | -0.27 | -0.24 |
| no triflate | #1 | -0.23 | 0.06 |
| | #2 | -0.27 | 0.15 |
| | #3 | -0.26 | 0.18 |

**Figure S24. PCA on $\alpha_2$, for R is measured at different time interval [50s;59s] | a,** PCA scores with 95% confidence ellipsoids. **b,** Individual variance for the different PC. **c,** PCA loadings of the different sensing elements' response for PC1 and PC2.

a)

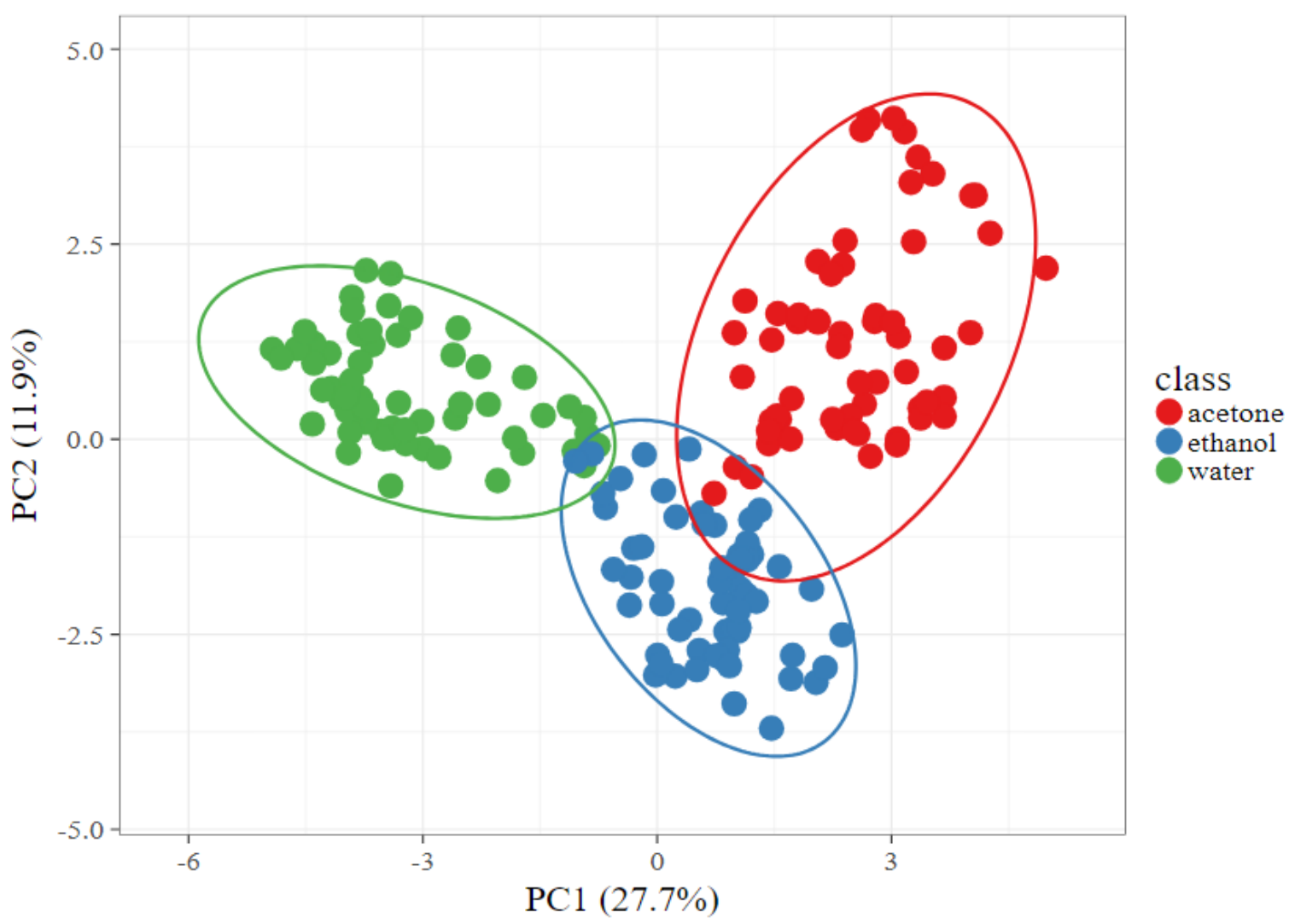

b)

| | Variance |
|---|---|
| **PC1** | 27.7% |
| **PC2** | 11.9% |
| **PC3** | 7.0% |
| **PC4** | 6.1% |
| **PC5** | 5.8% |
| **PC6** | 5.0% |
| **PC7** | 4.3% |
| **PC8** | 4.1% |
| **PC9** | 3.7% |
| **PC10** | 3.4% |
| **PC11** | 3.2% |
| **PC12** | 2.9% |
| **PC13** | 2.7% |
| **PC14** | 2.4% |
| **PC15** | 1.9% |
| **PC16** | 1.8% |
| **PC17** | 1.3% |
| **PC18** | 1.3% |
| **PC19** | 1.0% |
| **PC20** | 0.9% |
| **PC21** | 0.6% |
| **PC22** | 0.5% |
| **PC23** | 0.4% |
| **PC24** | 0.2% |

c)

| | | Loadings | |
|---|---|---|---|
| | | PC1 | PC2 |
| $Fe(OTf)_3$ | #1 | -0.24 | 0.01 |
| | #2 | -0.32 | -0.13 |
| | #3 | -0.11 | -0.40 |
| $Cu(OTf)_2$ | #1 | -0.06 | -0.10 |
| | #2 | -0.12 | 0.13 |
| | #3 | -0.06 | 0.12 |
| $Bi(OTf)_3$ | #1 | -0.22 | 0.42 |
| | #2 | 0.23 | 0.00 |
| | #3 | -0.14 | 0.45 |
| $Al(OTf)_3$ | #1 | -0.13 | -0.21 |
| | #2 | -0.18 | 0.05 |
| | #3 | -0.21 | 0.05 |
| $In(OTf)_3$ | #1 | -0.04 | 0.07 |
| | #2 | -0.06 | 0.13 |
| | #3 | -0.19 | 0.24 |
| $Dy(OTf)_3$ | #1 | -0.21 | -0.27 |
| | #2 | -0.18 | -0.17 |
| | #3 | -0.05 | -0.12 |
| $Ce(OTf)_3$ | #1 | -0.31 | -0.15 |
| | #2 | -0.34 | -0.16 |
| | #3 | -0.29 | -0.17 |
| no triflate | #1 | -0.26 | 0.04 |
| | #2 | -0.27 | 0.24 |
| | #3 | -0.20 | 0.13 |

**Figure S25. PCA on $\alpha_2$, for R is measured at different time interval [60s;69s] | a,** PCA scores with 95% confidence ellipsoids. **b,** Individual variance for the different PC. **c,** PCA loadings of the different sensing elements' response for PC1 and PC2.

a)

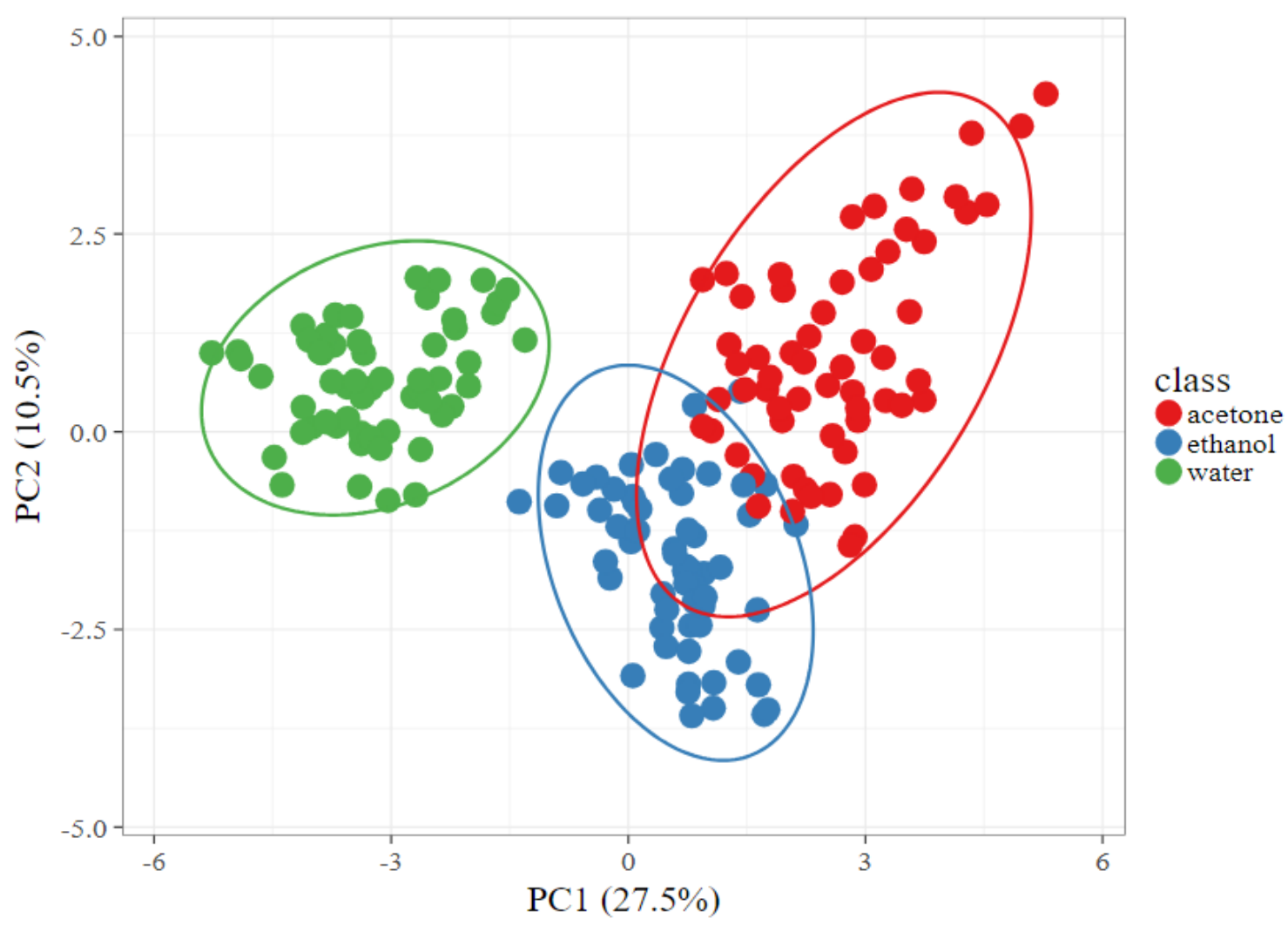

b)

| | Variance |
|---|---|
| **PC1** | 27.5% |
| **PC2** | 10.5% |
| **PC3** | 7.6% |
| **PC4** | 6.3% |
| **PC5** | 5.2% |
| **PC6** | 4.9% |
| **PC7** | 4.4% |
| **PC8** | 4.1% |
| **PC9** | 3.7% |
| **PC10** | 3.4% |
| **PC11** | 3.2% |
| **PC12** | 2.7% |
| **PC13** | 2.5% |
| **PC14** | 2.3% |
| **PC15** | 1.9% |
| **PC16** | 1.8% |
| **PC17** | 1.6% |
| **PC18** | 1.4% |
| **PC19** | 1.2% |
| **PC20** | 1.1% |
| **PC21** | 1.0% |
| **PC22** | 0.8% |
| **PC23** | 0.6% |
| **PC24** | 0.2% |

c)

| | | Loadings | |
|---|---|---|---|
| | | PC1 | PC2 |
| $Fe(OTf)_3$ | #1 | -0.28 | 0.09 |
| | #2 | -0.30 | 0.14 |
| | #3 | -0.13 | 0.38 |
| $Cu(OTf)_2$ | #1 | -0.08 | 0.14 |
| | #2 | -0.12 | 0.04 |
| | #3 | -0.01 | 0.25 |
| $Bi(OTf)_3$ | #1 | -0.24 | -0.34 |
| | #2 | 0.21 | -0.16 |
| | #3 | -0.10 | -0.44 |
| $Al(OTf)_3$ | #1 | -0.11 | -0.10 |
| | #2 | -0.15 | -0.30 |
| | #3 | -0.25 | -0.08 |
| $In(OTf)_3$ | #1 | -0.15 | 0.07 |
| | #2 | -0.09 | 0.05 |
| | #3 | -0.15 | -0.18 |
| $Dy(OTf)_3$ | #1 | -0.19 | 0.28 |
| | #2 | -0.04 | -0.09 |
| | #3 | 0.07 | -0.19 |
| $Ce(OTf)_3$ | #1 | -0.32 | 0.05 |
| | #2 | -0.34 | 0.11 |
| | #3 | -0.30 | 0.16 |
| no triflate | #1 | -0.25 | -0.05 |
| | #2 | -0.22 | -0.26 |
| | #3 | -0.26 | -0.18 |

**Figure S26. PCA on $\alpha_2$, for R is measured at different time interval [70s;79s] | a,** PCA scores with 95% confidence ellipsoids. **b,** Individual variance for the different PC. **c,** PCA loadings of the different sensing elements' response for PC1 and PC2.

a)

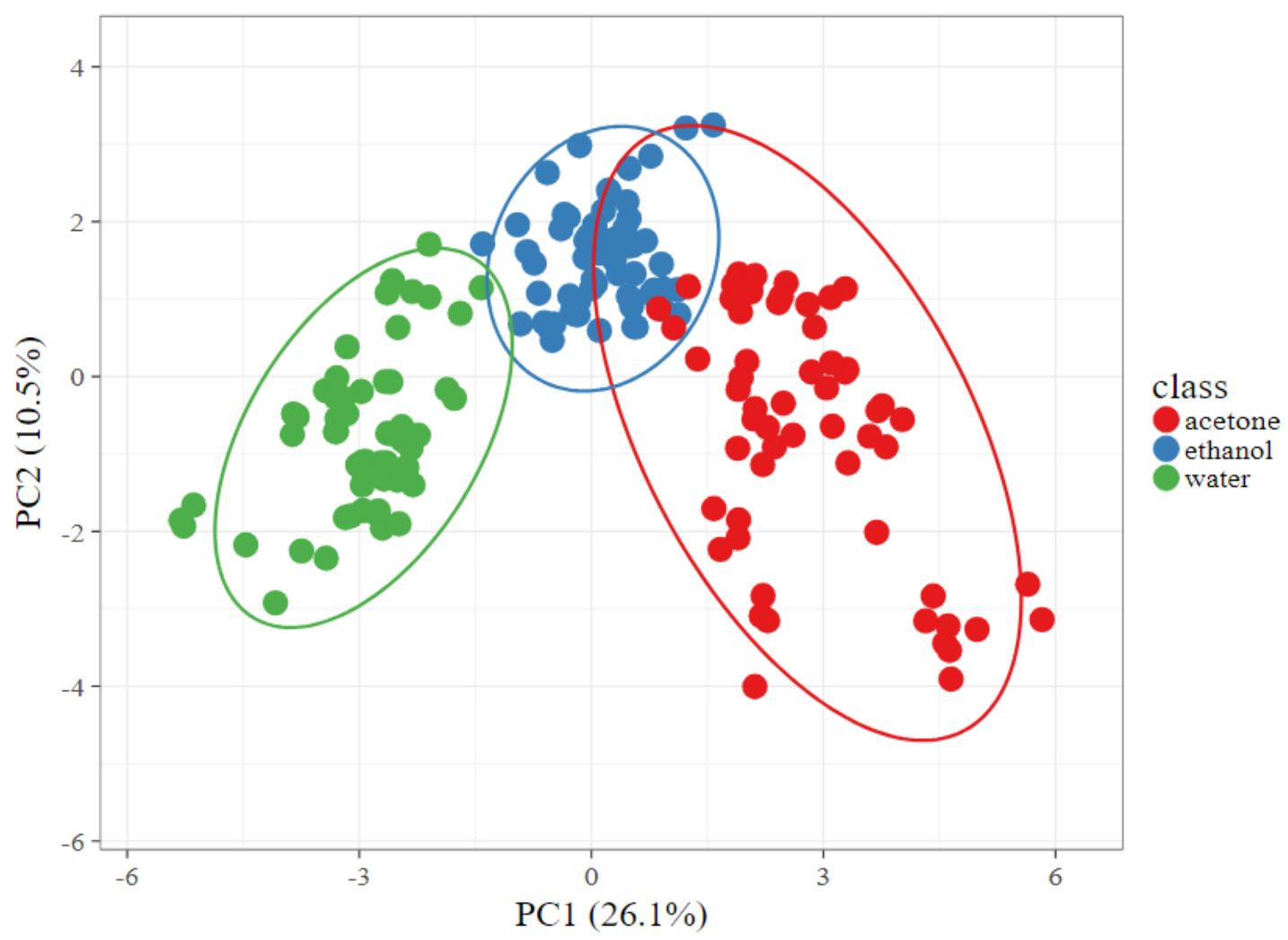

b)

| | Variance |
|---|---|
| **PC1** | 26.1% |
| **PC2** | 10.5% |
| **PC3** | 7.0% |
| **PC4** | 6.4% |
| **PC5** | 6.0% |
| **PC6** | 5.6% |
| **PC7** | 4.7% |
| **PC8** | 4.6% |
| **PC9** | 4.4% |
| **PC10** | 3.7% |
| **PC11** | 3.0% |
| **PC12** | 2.6% |
| **PC13** | 2.5% |
| **PC14** | 2.2% |
| **PC15** | 1.8% |
| **PC16** | 1.7% |
| **PC17** | 1.5% |
| **PC18** | 1.4% |
| **PC19** | 1.2% |
| **PC20** | 0.9% |
| **PC21** | 0.7% |
| **PC22** | 0.7% |
| **PC23** | 0.6% |
| **PC24** | 0.2% |

c)

| | | Loadings | |
|---|---|---|---|
| | | PC1 | PC2 |
| $Fe(OTf)_3$ | #1 | -0.19 | 0.33 |
| | #2 | -0.25 | -0.07 |
| | #3 | -0.01 | -0.31 |
| $Cu(OTf)_2$ | #1 | -0.17 | 0.14 |
| | #2 | -0.14 | 0.04 |
| | #3 | -0.03 | 0.00 |
| $Bi(OTf)_3$ | #1 | -0.29 | 0.34 |
| | #2 | 0.12 | 0.05 |
| | #3 | -0.19 | 0.28 |
| $Al(OTf)_3$ | #1 | -0.14 | -0.21 |
| | #2 | -0.18 | 0.12 |
| | #3 | -0.29 | 0.09 |
| $In(OTf)_3$ | #1 | -0.02 | 0.26 |
| | #2 | -0.05 | -0.07 |
| | #3 | -0.21 | 0.06 |
| $Dy(OTf)_3$ | #1 | -0.17 | -0.36 |
| | #2 | -0.05 | -0.36 |
| | #3 | -0.11 | -0.06 |
| $Ce(OTf)_3$ | #1 | -0.34 | -0.15 |
| | #2 | -0.33 | -0.25 |
| | #3 | -0.29 | -0.26 |
| no triflate | #1 | -0.30 | 0.06 |
| | #2 | -0.17 | -0.01 |
| | #3 | -0.25 | 0.12 |

**Figure S27. PCA on $\alpha_2$, for R is measured at different time interval [80s;89s] | a,** PCA scores with 95% confidence ellipsoids. **b,** Individual variance for the different PC. **c,** PCA loadings of the different sensing elements' response for PC1 and PC2.

a)

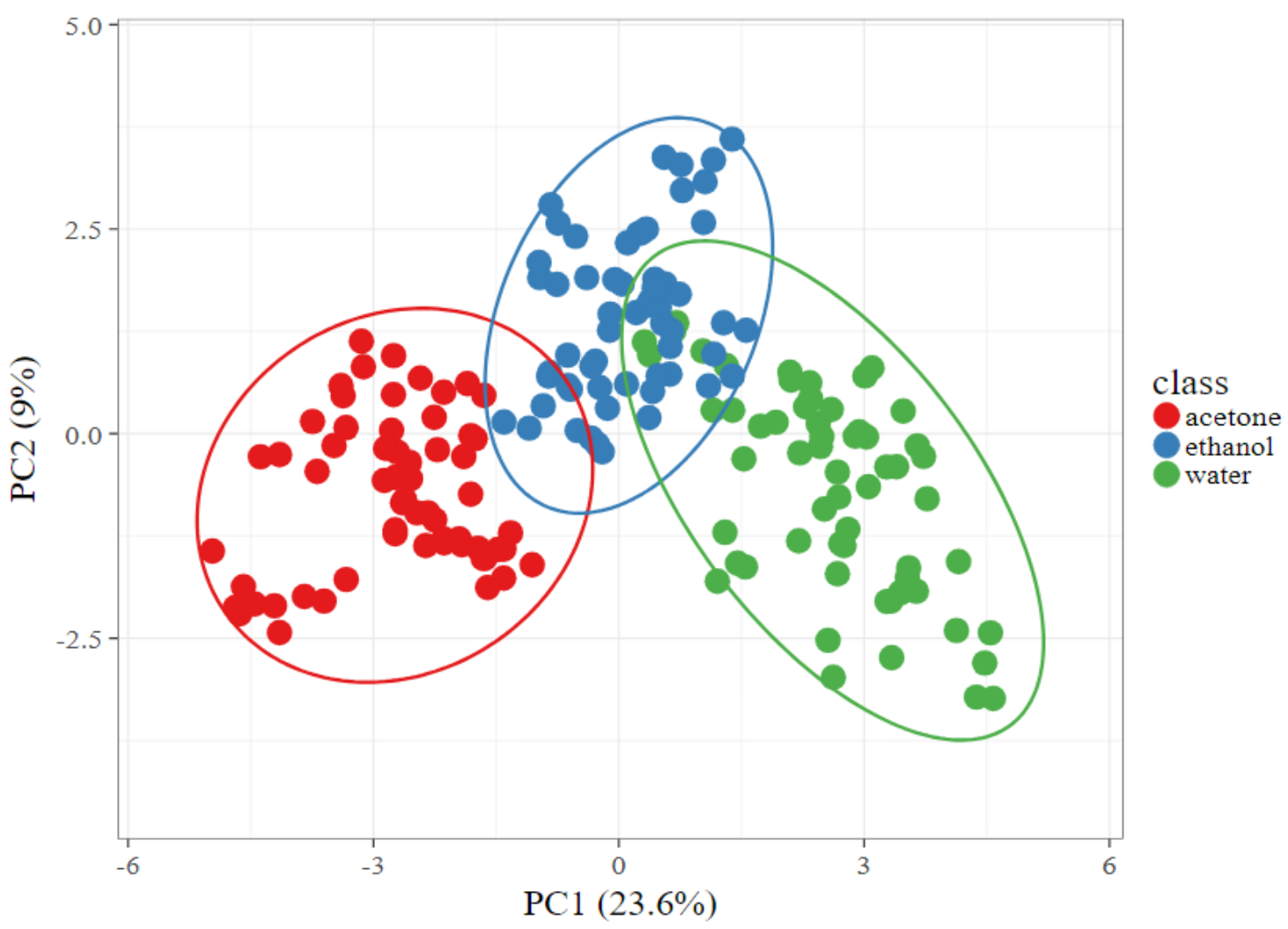

b)

| | Variance |
|---|---|
| **PC1** | 23.6% |
| **PC2** | 9.0% |
| **PC3** | 8.2% |
| **PC4** | 6.7% |
| **PC5** | 6.0% |
| **PC6** | 5.6% |
| **PC7** | 5.1% |
| **PC8** | 4.9% |
| **PC9** | 3.9% |
| **PC10** | 3.7% |
| **PC11** | 3.4% |
| **PC12** | 2.9% |
| **PC13** | 2.8% |
| **PC14** | 2.4% |
| **PC15** | 2.0% |
| **PC16** | 1.8% |
| **PC17** | 1.5% |
| **PC18** | 1.5% |
| **PC19** | 1.3% |
| **PC20** | 1.0% |
| **PC21** | 1.0% |
| **PC22** | 0.6% |
| **PC23** | 0.6% |
| **PC24** | 0.2% |

c)

| | | Loadings | |
|---|---|---|---|
| | | PC1 | PC2 |
| $Fe(OTf)_3$ | #1 | -0.21 | -0.30 |
| | #2 | -0.25 | 0.01 |
| | #3 | -0.19 | 0.20 |
| $Cu(OTf)_2$ | #1 | -0.07 | -0.11 |
| | #2 | -0.01 | -0.06 |
| | #3 | -0.14 | -0.34 |
| $Bi(OTf)_3$ | #1 | -0.29 | -0.32 |
| | #2 | 0.12 | -0.22 |
| | #3 | -0.29 | -0.31 |
| $Al(OTf)_3$ | #1 | -0.19 | -0.13 |
| | #2 | -0.17 | -0.21 |
| | #3 | -0.21 | 0.01 |
| $In(OTf)_3$ | #1 | -0.03 | 0.06 |
| | #2 | -0.02 | -0.26 |
| | #3 | -0.14 | -0.04 |
| $Dy(OTf)_3$ | #1 | -0.17 | 0.07 |
| | #2 | -0.09 | -0.07 |
| | #3 | 0.00 | -0.20 |
| $Ce(OTf)_3$ | #1 | -0.36 | 0.15 |
| | #2 | -0.35 | 0.28 |
| | #3 | -0.31 | 0.27 |
| no triflate | #1 | -0.25 | 0.00 |
| | #2 | -0.25 | 0.11 |
| | #3 | -0.08 | 0.36 |

**Figure S28. PCA on $\alpha_2$, for R is measured at different time interval [90s;99s] | a,** PCA scores with 95% confidence ellipsoids. **b,** Individual variance for the different PC. **c,** PCA loadings of the different sensing elements' response for PC1 and PC2.

a)

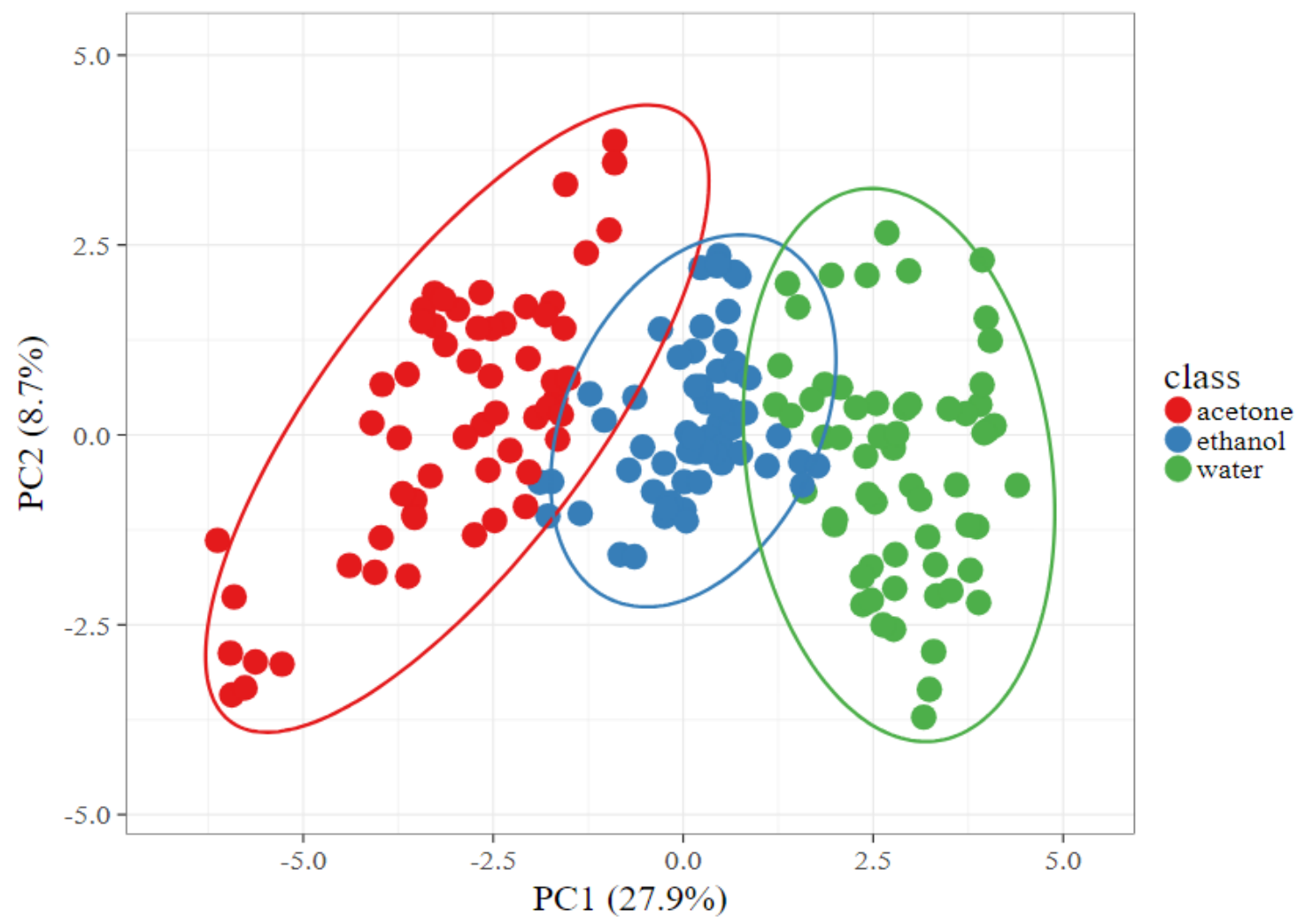

b)

| | Variance |
|---|---|
| **PC1** | 27.9% |
| **PC2** | 8.7% |
| **PC3** | 7.6% |
| **PC4** | 6.2% |
| **PC5** | 6.0% |
| **PC6** | 5.3% |
| **PC7** | 4.8% |
| **PC8** | 4.3% |
| **PC9** | 3.5% |
| **PC10** | 3.4% |
| **PC11** | 3.2% |
| **PC12** | 3.1% |
| **PC13** | 2.7% |
| **PC14** | 2.4% |
| **PC15** | 2.0% |
| **PC16** | 1.6% |
| **PC17** | 1.6% |
| **PC18** | 1.5% |
| **PC19** | 1.2% |
| **PC20** | 1.1% |
| **PC21** | 0.7% |
| **PC22** | 0.6% |
| **PC23** | 0.4% |
| **PC24** | 0.1% |

c)

| | | Loadings | |
|---|---|---|---|
| | | PC1 | PC2 |
| $Fe(OTf)_3$ | #1 | -0.29 | 0.13 |
| | #2 | -0.24 | 0.03 |
| | #3 | -0.20 | -0.31 |
| $Cu(OTf)_2$ | #1 | -0.06 | 0.26 |
| | #2 | -0.18 | 0.10 |
| | #3 | 0.10 | -0.26 |
| $Bi(OTf)_3$ | #1 | -0.28 | 0.17 |
| | #2 | 0.17 | -0.31 |
| | #3 | -0.27 | -0.03 |
| $Al(OTf)_3$ | #1 | -0.16 | -0.24 |
| | #2 | -0.17 | 0.26 |
| | #3 | -0.10 | 0.19 |
| $In(OTf)_3$ | #1 | -0.08 | 0.12 |
| | #2 | 0.01 | -0.44 |
| | #3 | -0.19 | 0.15 |
| $Dy(OTf)_3$ | #1 | -0.21 | 0.00 |
| | #2 | -0.02 | -0.16 |
| | #3 | 0.04 | -0.20 |
| $Ce(OTf)_3$ | #1 | -0.32 | -0.16 |
| | #2 | -0.31 | -0.21 |
| | #3 | -0.28 | -0.26 |
| no triflate | #1 | -0.28 | -0.10 |
| | #2 | -0.18 | -0.06 |
| | #3 | -0.23 | 0.09 |

**Figure S29. PCA on $\alpha_2$, for R is measured at different time interval [100s;109s] | a,** PCA scores with 95% confidence ellipsoids. **b,** Individual variance for the different PC. **c,** PCA loadings of the different sensing elements' response for PC1 and PC2.

a)

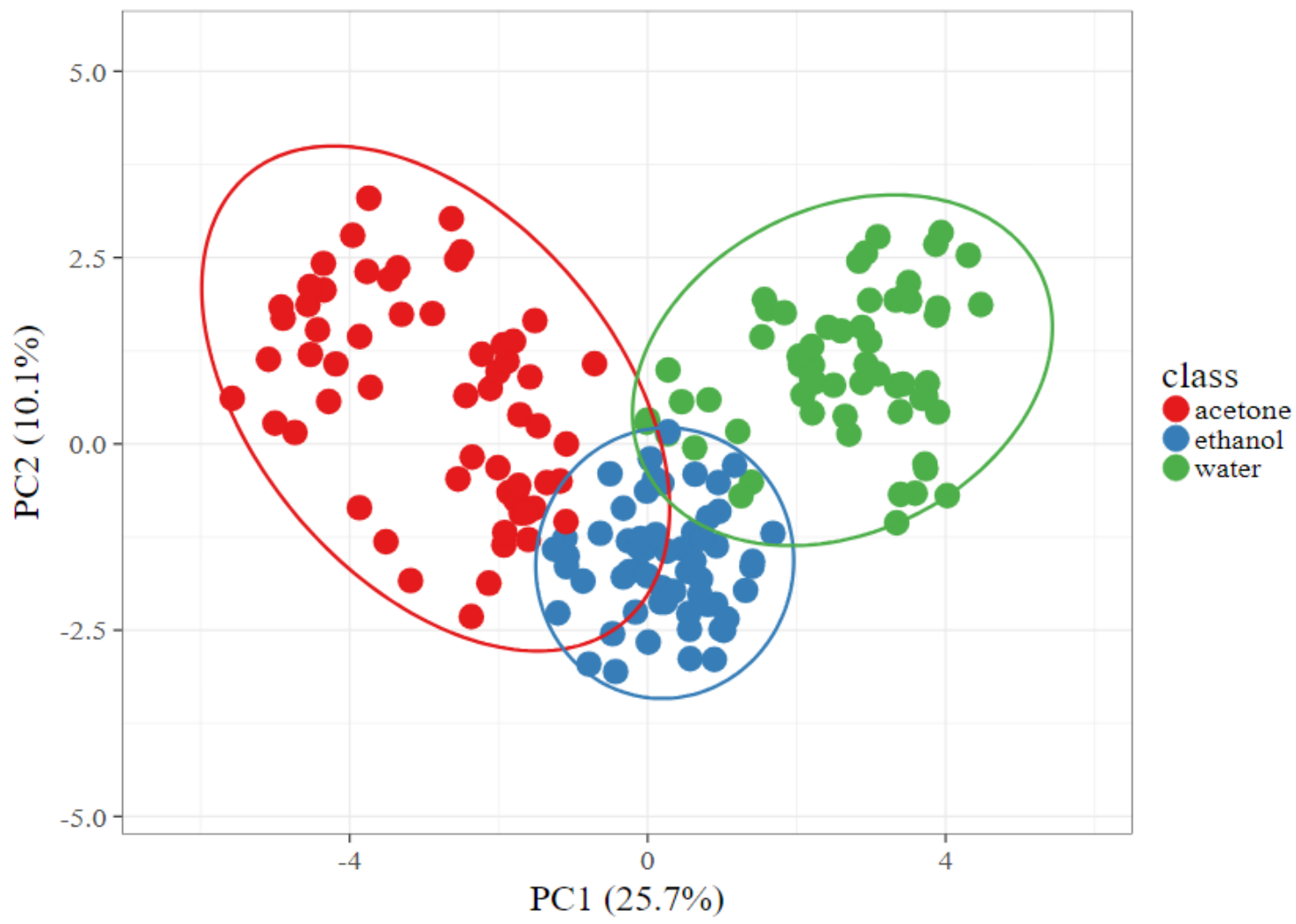

b)

| | Variance |
|---|---|
| **PC1** | 25.7% |
| **PC2** | 10.1% |
| **PC3** | 8.3% |
| **PC4** | 6.3% |
| **PC5** | 5.7% |
| **PC6** | 5.0% |
| **PC7** | 4.8% |
| **PC8** | 4.4% |
| **PC9** | 3.6% |
| **PC10** | 3.6% |
| **PC11** | 3.2% |
| **PC12** | 3.0% |
| **PC13** | 2.9% |
| **PC14** | 2.5% |
| **PC15** | 2.0% |
| **PC16** | 1.9% |
| **PC17** | 1.5% |
| **PC18** | 1.4% |
| **PC19** | 1.2% |
| **PC20** | 1.1% |
| **PC21** | 0.8% |
| **PC22** | 0.5% |
| **PC23** | 0.4% |
| **PC24** | 0.1% |

c)

| | | Loadings | |
|---|---|---|---|
| | | PC1 | PC2 |
| $Fe(OTf)_3$ | #1 | -0.16 | 0.21 |
| | #2 | -0.20 | -0.09 |
| | #3 | -0.05 | 0.16 |
| $Cu(OTf)_2$ | #1 | 0.04 | -0.03 |
| | #2 | -0.01 | -0.22 |
| | #3 | -0.30 | 0.03 |
| $Bi(OTf)_3$ | #1 | 0.22 | -0.31 |
| | #2 | -0.24 | 0.37 |
| | #3 | -0.22 | -0.36 |
| $Al(OTf)_3$ | #1 | -0.23 | 0.02 |
| | #2 | -0.18 | -0.07 |
| | #3 | -0.13 | 0.01 |
| $In(OTf)_3$ | #1 | -0.16 | -0.13 |
| | #2 | -0.07 | -0.03 |
| | #3 | -0.01 | 0.10 |
| $Dy(OTf)_3$ | #1 | -0.03 | 0.46 |
| | #2 | -0.06 | 0.07 |
| | #3 | -0.37 | 0.05 |
| $Ce(OTf)_3$ | #1 | -0.34 | 0.11 |
| | #2 | -0.30 | 0.26 |
| | #3 | -0.27 | 0.25 |
| no triflate | #1 | -0.17 | 0.22 |
| | #2 | -0.20 | -0.21 |
| | #3 | -0.19 | -0.16 |

**Figure S30. PCA on $\alpha_2$, for R is measured at different time interval [110s;119s] | a,** PCA scores with 95% confidence ellipsoids. **b,** Individual variance for the different PC. **c,** PCA loadings of the different sensing elements' response for PC1 and PC2.

a)

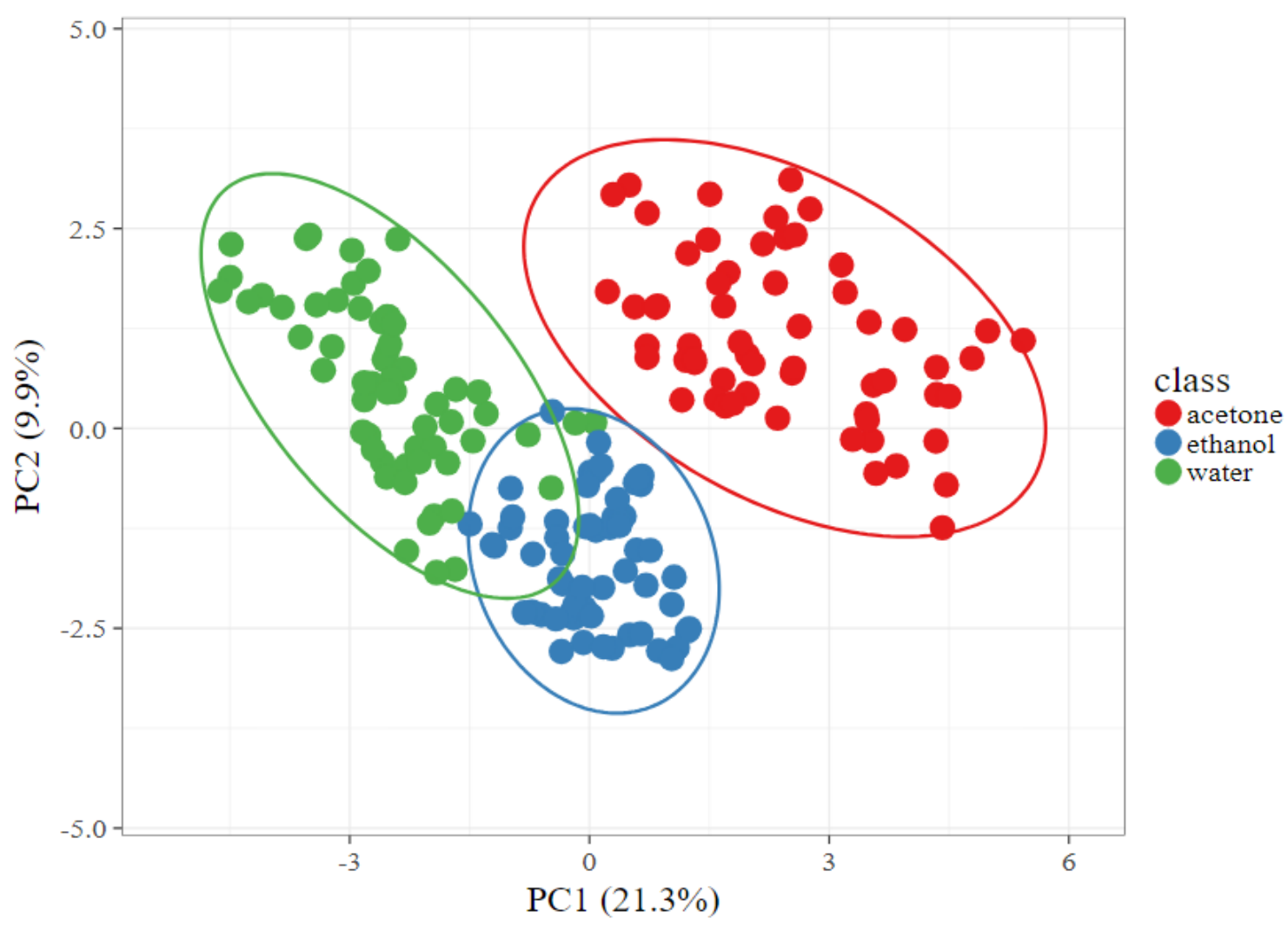

b)

| | Variance |
|---|---|
| **PC1** | 21.3% |
| **PC2** | 9.9% |
| **PC3** | 7.8% |
| **PC4** | 7.0% |
| **PC5** | 6.6% |
| **PC6** | 5.6% |
| **PC7** | 5.1% |
| **PC8** | 4.6% |
| **PC9** | 4.2% |
| **PC10** | 3.8% |
| **PC11** | 3.3% |
| **PC12** | 3.1% |
| **PC13** | 3.0% |
| **PC14** | 2.6% |
| **PC15** | 2.2% |
| **PC16** | 2.1% |
| **PC17** | 1.8% |
| **PC18** | 1.8% |
| **PC19** | 1.4% |
| **PC20** | 1.0% |
| **PC21** | 0.8% |
| **PC22** | 0.4% |
| **PC23** | 0.4% |
| **PC24** | 0.1% |

c)

| | | Loadings | |
|---|---|---|---|
| | | PC1 | PC2 |
| $Fe(OTf)_3$ | #1 | -0.10 | 0.38 |
| | #2 | -0.15 | 0.01 |
| | #3 | -0.27 | 0.04 |
| $Cu(OTf)_2$ | #1 | -0.05 | 0.29 |
| | #2 | -0.10 | -0.06 |
| | #3 | -0.02 | 0.12 |
| $Bi(OTf)_3$ | #1 | -0.32 | -0.27 |
| | #2 | 0.22 | -0.02 |
| | #3 | -0.23 | -0.45 |
| $Al(OTf)_3$ | #1 | -0.07 | -0.27 |
| | #2 | -0.08 | -0.06 |
| | #3 | -0.13 | -0.25 |
| $In(OTf)_3$ | #1 | -0.10 | 0.18 |
| | #2 | -0.02 | -0.06 |
| | #3 | -0.24 | 0.15 |
| $Dy(OTf)_3$ | #1 | -0.13 | 0.44 |
| | #2 | -0.08 | 0.18 |
| | #3 | 0.03 | -0.13 |
| $Ce(OTf)_3$ | #1 | -0.40 | 0.05 |
| | #2 | -0.38 | 0.10 |
| | #3 | -0.31 | 0.13 |
| no triflate | #1 | -0.20 | -0.04 |
| | #2 | -0.30 | -0.09 |
| | #3 | -0.15 | -0.05 |

**Figure S31. PCA on $\alpha_2$, for R is measured at different time interval [120s;129s] | a,** PCA scores with 95% confidence ellipsoids. **b,** Individual variance for the different PC. **c,** PCA loadings of the different sensing elements' response for PC1 and PC2.

a)

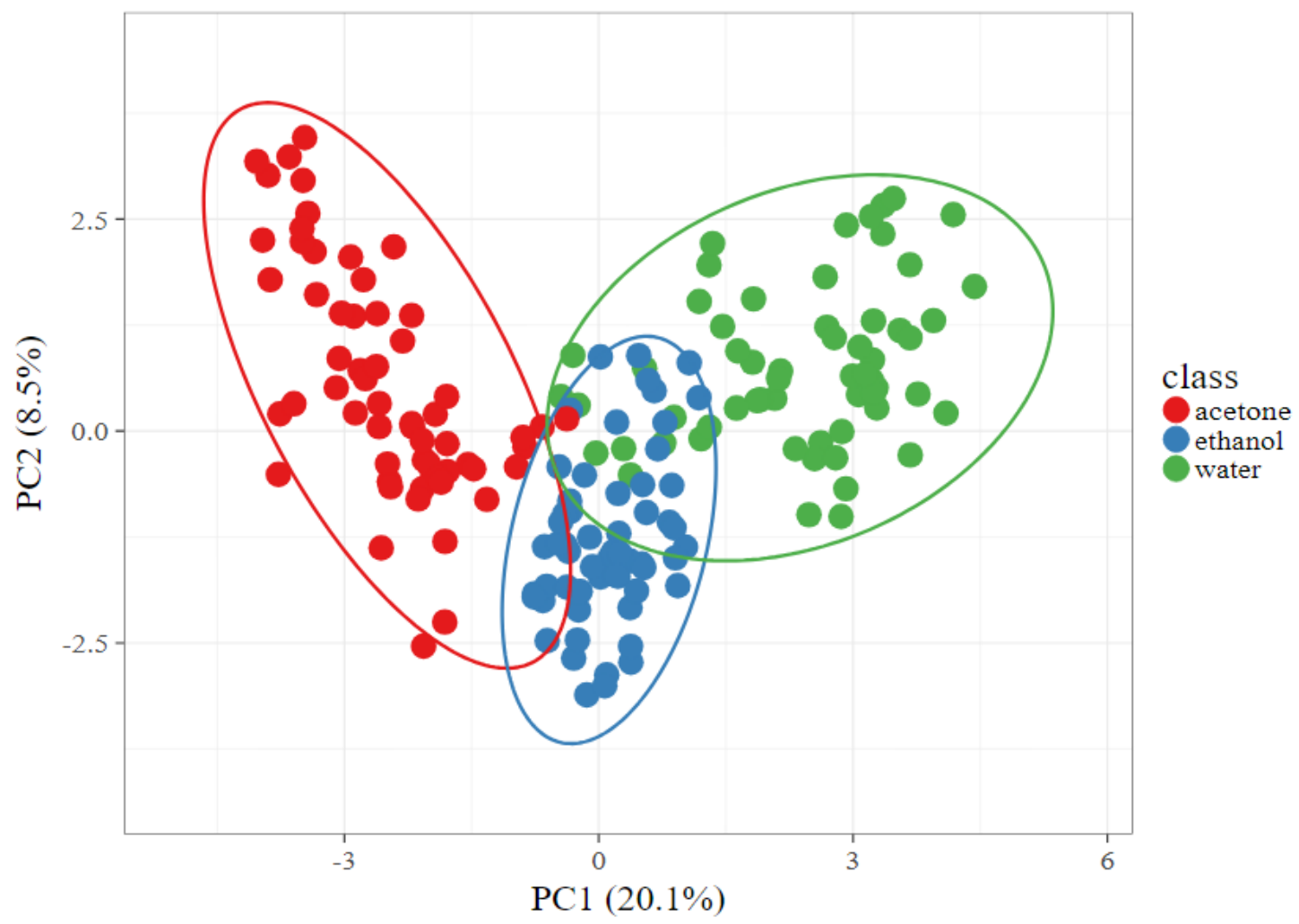

b)

| | Variance |
|---|---|
| **PC1** | 20.1% |
| **PC2** | 8.5% |
| **PC3** | 7.9% |
| **PC4** | 6.6% |
| **PC5** | 6.4% |
| **PC6** | 6.1% |
| **PC7** | 5.4% |
| **PC8** | 4.9% |
| **PC9** | 4.3% |
| **PC10** | 4.1% |
| **PC11** | 3.8% |
| **PC12** | 3.4% |
| **PC13** | 3.1% |
| **PC14** | 2.6% |
| **PC15** | 2.5% |
| **PC16** | 2.1% |
| **PC17** | 1.7% |
| **PC18** | 1.6% |
| **PC19** | 1.4% |
| **PC20** | 1.2% |
| **PC21** | 1.0% |
| **PC22** | 0.7% |
| **PC23** | 0.4% |
| **PC24** | 0.1% |

c)

| | | Loadings | |
|---|---|---|---|
| | | PC1 | PC2 |
| $Fe(OTf)_3$ | #1 | -0.07 | 0.35 |
| | #2 | -0.22 | -0.37 |
| | #3 | -0.10 | -0.16 |
| $Cu(OTf)_2$ | #1 | -0.01 | -0.21 |
| | #2 | -0.02 | -0.01 |
| | #3 | -0.08 | 0.14 |
| $Bi(OTf)_3$ | #1 | -0.29 | -0.37 |
| | #2 | 0.02 | 0.30 |
| | #3 | -0.28 | -0.30 |
| $Al(OTf)_3$ | #1 | -0.12 | 0.03 |
| | #2 | -0.16 | -0.20 |
| | #3 | -0.14 | -0.08 |
| $In(OTf)_3$ | #1 | -0.08 | 0.29 |
| | #2 | -0.12 | 0.02 |
| | #3 | -0.12 | 0.04 |
| $Dy(OTf)_3$ | #1 | -0.04 | 0.26 |
| | #2 | -0.11 | -0.07 |
| | #3 | -0.03 | 0.04 |
| $Ce(OTf)_3$ | #1 | -0.42 | 0.09 |
| | #2 | -0.41 | 0.17 |
| | #3 | -0.40 | 0.12 |
| no triflate | #1 | -0.18 | 0.21 |
| | #2 | -0.29 | 0.14 |
| | #3 | -0.19 | 0.13 |

**Figure S32. PCA on $\alpha_2$, for R is measured at different time interval [130s;139s] | a,** PCA scores with 95% confidence ellipsoids. **b,** Individual variance for the different PC. **c,** PCA loadings of the different sensing elements' response for PC1 and PC2.

a)

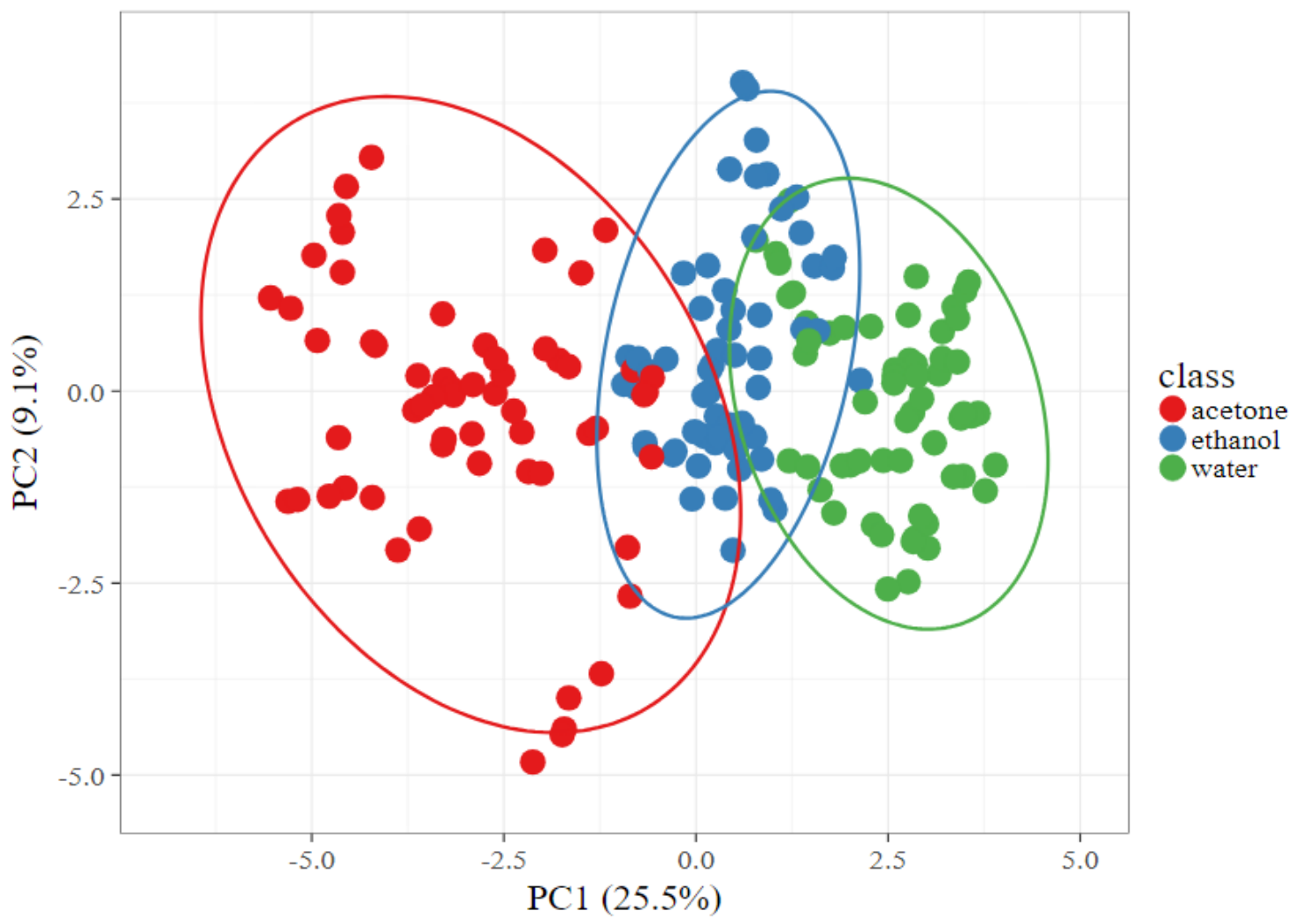

b)

| | Variance |
|---|---|
| **PC1** | 25.5% |
| **PC2** | 9.1% |
| **PC3** | 8.0% |
| **PC4** | 7.3% |
| **PC5** | 5.8% |
| **PC6** | 5.4% |
| **PC7** | 4.6% |
| **PC8** | 4.1% |
| **PC9** | 3.8% |
| **PC10** | 3.5% |
| **PC11** | 3.2% |
| **PC12** | 2.8% |
| **PC13** | 2.7% |
| **PC14** | 2.5% |
| **PC15** | 2.4% |
| **PC16** | 1.9% |
| **PC17** | 1.6% |
| **PC18** | 1.5% |
| **PC19** | 1.4% |
| **PC20** | 1.2% |
| **PC21** | 0.7% |
| **PC22** | 0.4% |
| **PC23** | 0.4% |
| **PC24** | 0.1% |

c)

| | | Loadings | |
|---|---|---|---|
| | | PC1 | PC2 |
| $Fe(OTf)_3$ | #1 | -0.20 | 0.33 |
| | #2 | -0.11 | 0.01 |
| | #3 | -0.10 | -0.35 |
| $Cu(OTf)_2$ | #1 | -0.17 | -0.14 |
| | #2 | -0.12 | -0.10 |
| | #3 | -0.09 | 0.26 |
| $Bi(OTf)_3$ | #1 | -0.31 | -0.02 |
| | #2 | -0.03 | -0.42 |
| | #3 | -0.29 | 0.00 |
| $Al(OTf)_3$ | #1 | -0.25 | -0.09 |
| | #2 | -0.19 | 0.15 |
| | #3 | -0.19 | 0.22 |
| $In(OTf)_3$ | #1 | -0.07 | -0.01 |
| | #2 | 0.06 | 0.28 |
| | #3 | -0.13 | -0.26 |
| $Dy(OTf)_3$ | #1 | -0.07 | 0.41 |
| | #2 | 0.07 | -0.04 |
| | #3 | -0.13 | -0.27 |
| $Ce(OTf)_3$ | #1 | -0.35 | 0.04 |
| | #2 | -0.33 | 0.06 |
| | #3 | -0.35 | 0.07 |
| no triflate | #1 | -0.26 | -0.11 |
| | #2 | -0.23 | -0.06 |
| | #3 | -0.20 | -0.06 |

**Figure S33. PCA on $\alpha_2$, for R is measured at different time interval [140s;149s] | a,** PCA scores with 95% confidence ellipsoids. **b,** Individual variance for the different PC. **c,** PCA loadings of the different sensing elements' response for PC1 and PC2.

a)

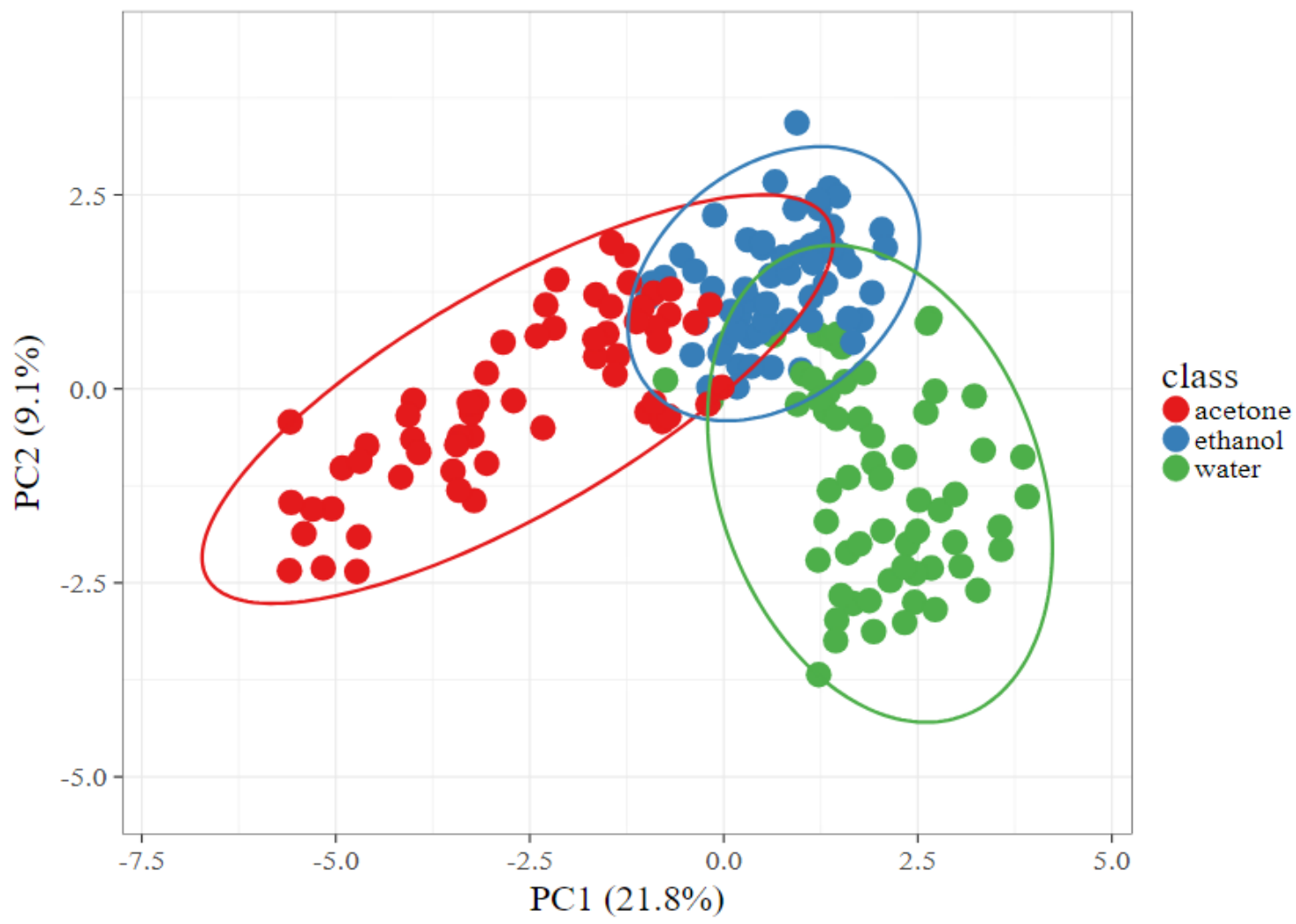

b)

| | Variance |
|---|---|
| **PC1** | 21.8% |
| **PC2** | 9.1% |
| **PC3** | 8.3% |
| **PC4** | 7.2% |
| **PC5** | 6.0% |
| **PC6** | 5.5% |
| **PC7** | 4.9% |
| **PC8** | 4.7% |
| **PC9** | 4.2% |
| **PC10** | 3.9% |
| **PC11** | 3.7% |
| **PC12** | 3.5% |
| **PC13** | 2.8% |
| **PC14** | 2.4% |
| **PC15** | 2.1% |
| **PC16** | 1.8% |
| **PC17** | 1.7% |
| **PC18** | 1.6% |
| **PC19** | 1.5% |
| **PC20** | 1.3% |
| **PC21** | 1.0% |
| **PC22** | 0.5% |
| **PC23** | 0.4% |
| **PC24** | 0.2% |

c)

| | | Loadings | |
|---|---|---|---|
| | | PC1 | PC2 |
| $Fe(OTf)_3$ | #1 | -0.29 | 0.06 |
| | #2 | -0.23 | -0.26 |
| | #3 | -0.17 | -0.36 |
| $Cu(OTf)_2$ | #1 | -0.16 | -0.03 |
| | #2 | 0.02 | 0.09 |
| | #3 | -0.10 | -0.01 |
| $Bi(OTf)_3$ | #1 | -0.33 | -0.28 |
| | #2 | -0.04 | -0.12 |
| | #3 | -0.33 | -0.24 |
| $Al(OTf)_3$ | #1 | -0.10 | -0.08 |
| | #2 | -0.22 | -0.14 |
| | #3 | 0.05 | 0.16 |
| $In(OTf)_3$ | #1 | 0.03 | 0.06 |
| | #2 | -0.09 | -0.20 |
| | #3 | -0.03 | -0.06 |
| $Dy(OTf)_3$ | #1 | -0.11 | 0.29 |
| | #2 | 0.04 | 0.31 |
| | #3 | -0.08 | -0.18 |
| $Ce(OTf)_3$ | #1 | -0.36 | 0.22 |
| | #2 | -0.33 | 0.36 |
| | #3 | -0.39 | 0.18 |
| no triflate | #1 | -0.22 | 0.18 |
| | #2 | -0.14 | 0.31 |
| | #3 | -0.19 | -0.01 |

**Figure S34. PCA on $\alpha_2$, for R is measured at different time interval [150s;159s] | a,** PCA scores with 95% confidence ellipsoids. **b,** Individual variance for the different PC. **c,** PCA loadings of the different sensing elements' response for PC1 and PC2.

a)

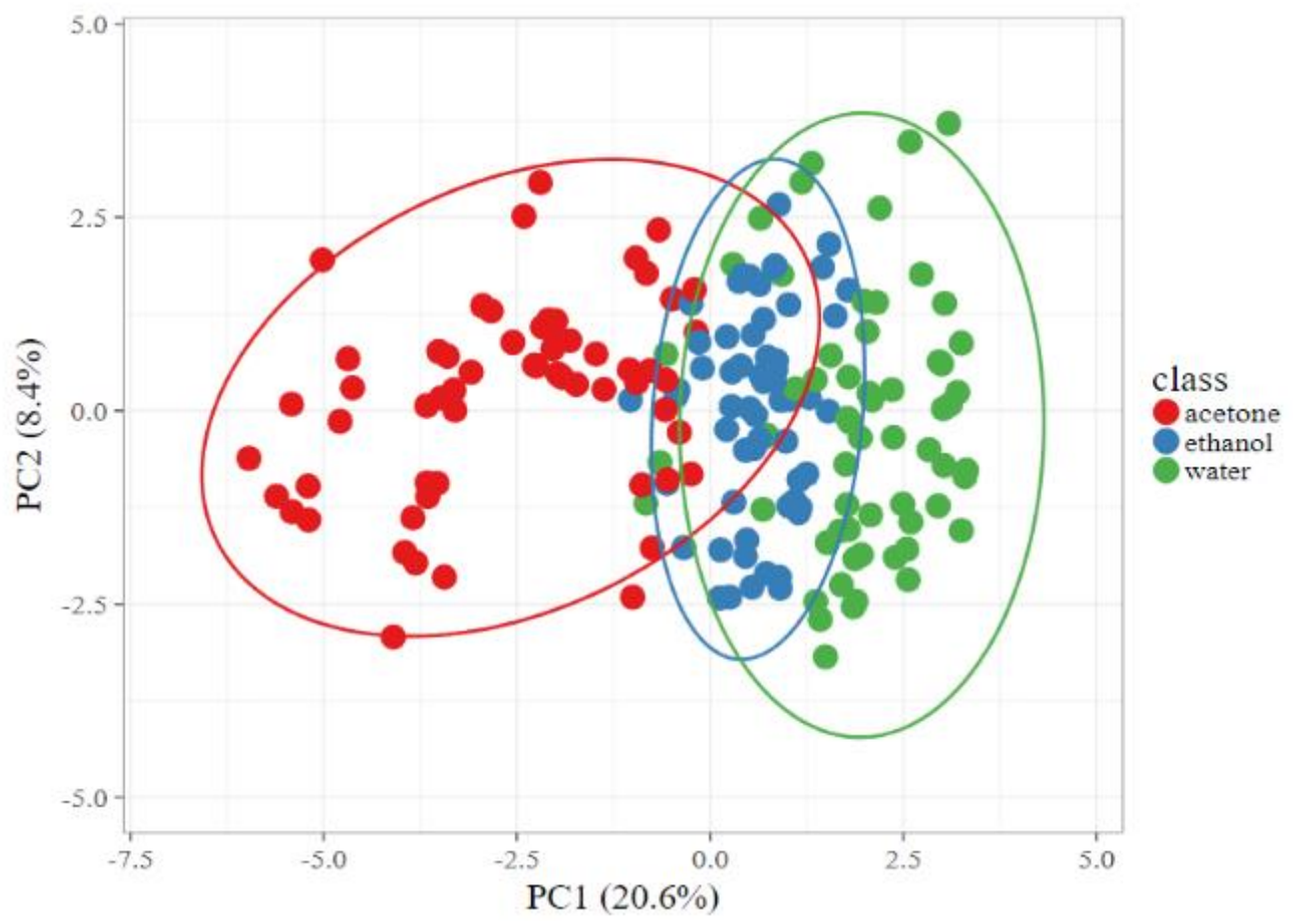

b)

| | Variance |
|---|---|
| **PC1** | 20.6% |
| **PC2** | 8.4% |
| **PC3** | 8.2% |
| **PC4** | 7.0% |
| **PC5** | 6.0% |
| **PC6** | 5.4% |
| **PC7** | 5.2% |
| **PC8** | 5.0% |
| **PC9** | 4.3% |
| **PC10** | 3.8% |
| **PC11** | 3.5% |
| **PC12** | 3.1% |
| **PC13** | 3.0% |
| **PC14** | 2.6% |
| **PC15** | 2.3% |
| **PC16** | 2.2% |
| **PC17** | 2.1% |
| **PC18** | 1.8% |
| **PC19** | 1.6% |
| **PC20** | 1.4% |
| **PC21** | 1.2% |
| **PC22** | 0.6% |
| **PC23** | 0.4% |
| **PC24** | 0.3% |

c)

| | | Loadings | |
|---|---|---|---|
| | | PC1 | PC2 |
| $Fe(OTf)_3$ | #1 | 0.18 | -0.21 |
| | #2 | 0.16 | 0.14 |
| | #3 | 0.23 | 0.30 |
| $Cu(OTf)_2$ | #1 | 0.08 | 0.34 |
| | #2 | -0.09 | -0.30 |
| | #3 | 0.13 | -0.07 |
| $Bi(OTf)_3$ | #1 | 0.34 | 0.01 |
| | #2 | 0.09 | 0.12 |
| | #3 | 0.30 | -0.08 |
| $Al(OTf)_3$ | #1 | 0.10 | -0.32 |
| | #2 | 0.17 | -0.16 |
| | #3 | 0.15 | 0.18 |
| $In(OTf)_3$ | #1 | 0.08 | -0.04 |
| | #2 | 0.08 | -0.40 |
| | #3 | 0.13 | -0.07 |
| $Dy(OTf)_3$ | #1 | 0.07 | -0.04 |
| | #2 | 0.06 | -0.20 |
| | #3 | -0.02 | -0.20 |
| $Ce(OTf)_3$ | #1 | 0.39 | -0.06 |
| | #2 | 0.38 | -0.14 |
| | #3 | 0.39 | -0.10 |
| no triflate | #1 | 0.17 | 0.01 |
| | #2 | 0.20 | 0.37 |
| | #3 | 0.18 | 0.19 |

**Figure S35. PCA on $\alpha_2$, for R is measured at different time interval [160s;169s] | a,** PCA scores with 95% confidence ellipsoids. **b,** Individual variance for the different PC. **c,** PCA loadings of the different sensing elements' response for PC1 and PC2.

a)

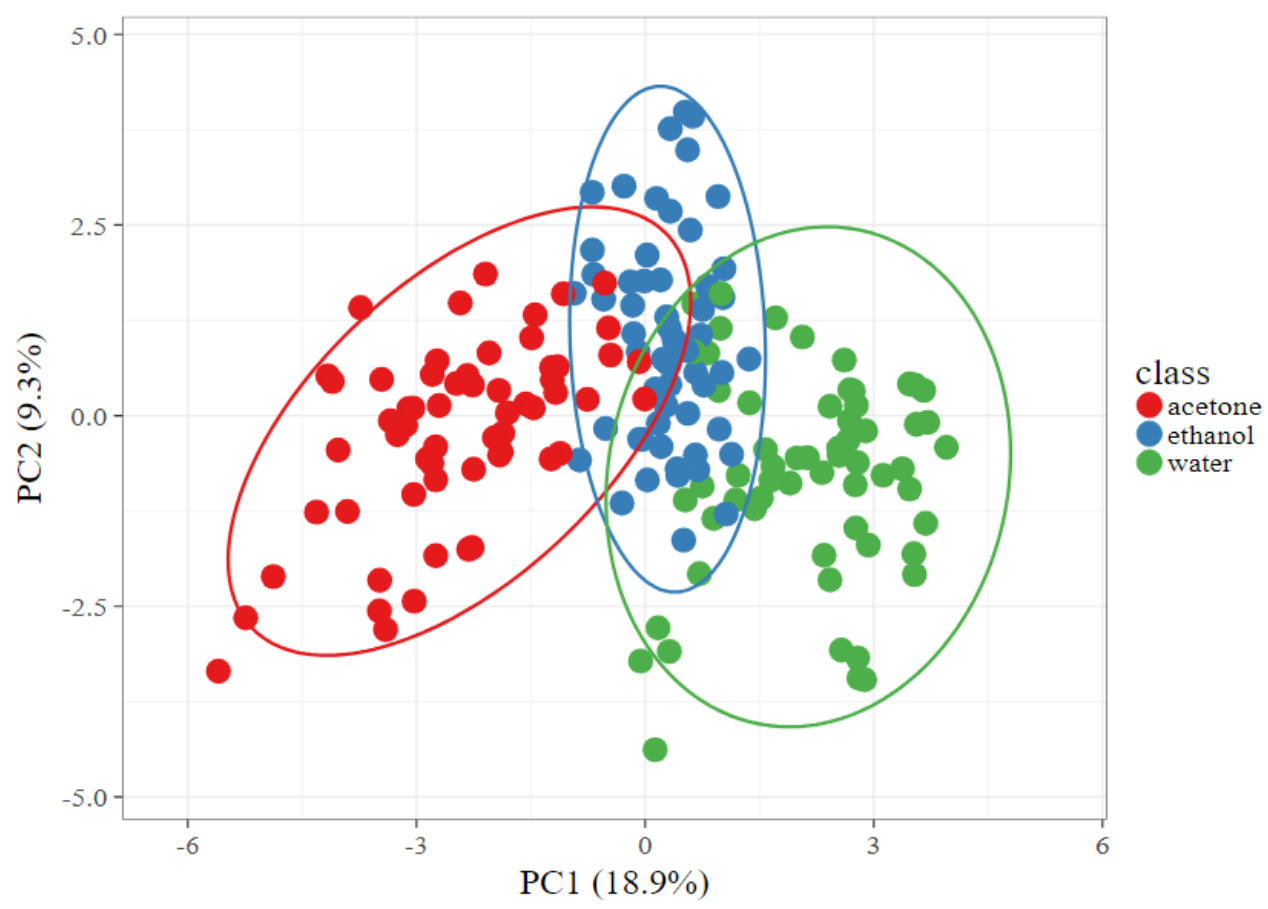

b)

| | Variance |
|---|---|
| **PC1** | 18.9% |
| **PC2** | 9.3% |
| **PC3** | 8.0% |
| **PC4** | 6.8% |
| **PC5** | 6.6% |
| **PC6** | 6.0% |
| **PC7** | 5.2% |
| **PC8** | 5.1% |
| **PC9** | 4.3% |
| **PC10** | 4.1% |
| **PC11** | 3.9% |
| **PC12** | 3.3% |
| **PC13** | 3.1% |
| **PC14** | 2.6% |
| **PC15** | 2.2% |
| **PC16** | 2.0% |
| **PC17** | 1.8% |
| **PC18** | 1.7% |
| **PC19** | 1.4% |
| **PC20** | 1.2% |
| **PC21** | 1.0% |
| **PC22** | 0.6% |
| **PC23** | 0.5% |
| **PC24** | 0.3% |

c)

| | | Loadings | |
|---|---|---|---|
| | | PC1 | PC2 |
| $Fe(OTf)_3$ | #1 | -0.20 | -0.06 |
| | #2 | -0.10 | 0.40 |
| | #3 | -0.17 | 0.16 |
| $Cu(OTf)_2$ | #1 | -0.12 | 0.33 |
| | #2 | -0.07 | -0.11 |
| | #3 | -0.11 | 0.02 |
| $Bi(OTf)_3$ | #1 | -0.33 | -0.18 |
| | #2 | -0.18 | -0.32 |
| | #3 | -0.35 | -0.12 |
| $Al(OTf)_3$ | #1 | -0.03 | -0.13 |
| | #2 | -0.06 | -0.28 |
| | #3 | -0.11 | -0.32 |
| $In(OTf)_3$ | #1 | 0.03 | 0.27 |
| | #2 | -0.10 | 0.00 |
| | #3 | 0.01 | -0.33 |
| $Dy(OTf)_3$ | #1 | -0.14 | 0.09 |
| | #2 | -0.14 | 0.02 |
| | #3 | 0.07 | 0.02 |
| $Ce(OTf)_3$ | #1 | -0.42 | 0.02 |
| | #2 | -0.42 | 0.11 |
| | #3 | -0.40 | 0.00 |
| no triflate | #1 | -0.04 | -0.15 |
| | #2 | -0.06 | 0.08 |
| | #3 | -0.18 | 0.34 |

**Figure S36. PCA on $\alpha_2$, for R is measured at different time interval [170s;179s] | a,** PCA scores with 95% confidence ellipsoids. **b,** Individual variance for the different PC. **c,** PCA loadings of the different sensing elements' response for PC1 and PC2.